\documentclass{aastex}
\usepackage{spr-astr-addons}
\usepackage{epsf}
\usepackage{psfig}
\usepackage{lscape}
\usepackage{graphicx}
\usepackage{rotating}
\newcommand{\LZ}{\mathcal{L}}
\newcommand{\BVR}{\mathcal{(B\!-\!R)/(B\!+\!V\!+\!R)}}
\newcommand{\BUO}{\mathcal{(B{\rm 2}\!-\!R)/(B\!+\!V\!+\!R)}}
\newcommand{\msait}{Mem. Soc. Astr. Italiana}
\newcommand{\aca}{Acta Astron.}
\newcommand{\bast}{Baltic Astron.}
\bibpunct[]{(}{)}{;}{a}{}{,} 

\begin{document}
\title{Horizontal Branch Stars: The Interplay between Observations 
       and Theory, and Insights into the \\Formation of the Galaxy}    
\author{M. Catelan\altaffilmark{1,2}}   
\altaffiltext{1}{John Simon Guggenheim Memorial Foundation Fellow}
\altaffiltext{2}{Pontificia Universidad Cat\'olica de Chile, Departamento de Astronom\'\i a 
       y Astrof\'\i sica, Av. Vicu\~na Mackenna 4860, 
       782-0436 Santiago, Chile; e-mail: {\tt mcatelan@astro.puc.cl}}    
\shorttitle{Horizontal Branch Stars: Observations and Theory}
\shortauthors{M. Catelan}

\begin{abstract} 
We review and discuss horizontal branch (HB) stars in a broad 
astrophysical context, including both variable and non-variable 
stars. A reassessment of the Oosterhoff dichotomy is 
presented, which provides unprecedented detail regarding its origin
and systematics. We show that the Oosterhoff dichotomy and the 
distribution of globular clusters in the HB morphology-metallicity 
plane both exclude, with high statistical significance, the possibility 
that the Galactic halo may have formed from the accretion of dwarf 
galaxies resembling present-day Milky Way satellites such as 
Fornax, Sagittarius, and the LMC---an 
argument which, due to its strong reliance on the ancient RR Lyrae 
stars, is essentially independent of the chemical evolution of these 
systems after the very earliest epochs in the Galaxy's history. 
Convenient analytical fits to isochrones in the HB type--[Fe/H] 
plane are also provided. In this sense, a rediscussion of the 
second-parameter problem is also presented, focusing on the cases of
NGC~288/NGC~362, M13/M3, the extreme outer-halo globular clusters 
with predominantly red HBs, and the metal-rich globular clusters
NGC~6388 and NGC~6441. 
The recently revived possibility that the 
helium abundance may play an important role as a second parameter is 
also addressed, and possible constraints on this scenario discussed.
We critically discuss the possibility that the observed properties of 
HB stars in NGC~6388 and NGC~6441 might be accounted for if these 
clusters possess a relatively minor population of helium-enriched 
stars. 
A technique is proposed to estimate the HB types of extragalactic 
globular clusters on the basis of integrated far-UV photometry. The 
importance of bright type II Cepheids as tracers of faint blue HB 
stars in distant systems is also emphasized. The relationship 
between the absolute $V$ magnitude of the HB 
at the RR Lyrae level and metallicity, as obtained on the basis of 
trigonometric parallax measurements for the star RR Lyr, is also 
revisited. Taking into due account the evolutionary status of 
RR Lyr, the derived relation implies a true distance modulus to 
the LMC of $(m\!-\!M)_0 = 18.44\pm 0.11$. Techniques providing discrepant 
slopes and zero points for the $M_V({\rm RRL})-{\rm [Fe/H]}$ relation
are briefly discussed. We provide a convenient analytical fit to 
theoretical model predictions for the period change 
rates of RR Lyrae stars in globular 
clusters, and compare the model results with the available data. 
Finally, the conductive opacities used in evolutionary 
calculations of low-mass stars are also investigated. 
\end{abstract}


\section{Introduction}                           

\subsection{A Bit of History} 

In her beautiful review of (hot) horizontal-branch (HB) stars, \citet{m01} 
notes that \citet{b1900} was the first to detect the presence of (blue) 
horizontal-branch stars in globular clusters. The term {\em horizontal branch} 
appears to have been coined by \citet{tB27}, to whom \citet{m04} assigns the 
discovery of the horizontal branch---which he noticed when plotting the 
color-magnitude data obtained by \citet{s15} in the latter's study 
of NGC~5272 (M3). 
Of course, with the development of nuclear astrophysics and the establishment 
of modern stellar evolution theory still several years away, it was 
not until three decades later that \citet{hs55} first correctly identified HB
stars as the progeny of low-mass red giant branch (RGB) stars, burning helium 
in their center and hydrogen in a shell around the core. 

The first successful HB models were 
actually computed by \citet{f66}, and \citet{cr68} and \citet{ir70} 
were the first to recognize 
that substantial mass loss on the RGB phase was needed to explain the observed 
colors of HB stars in globular clusters, with moreover a significant spread in 
mass loss amounts from star to star in any given globular cluster being needed 
to explain their observed color ranges---blue HB stars losing, on average, more 
mass than red HB stars. The distribution of masses along the HB often resembles 
a normal or Gaussian distribution 
(\citeauthor{rc89} \citeyear{rc89}; \citeauthor{dea96} \citeyear{dea96};
\citeauthor{vc08} \citeyear{vc08}), and normal deviates 
are accordingly often adopted in the construction of ``synthetic horizontal 
branches'' \citep*[e.g.,][]{r73,ct81,cea87,c93,l90,ldz90,scea04}. The presence of 
mass distributions that resemble Gaussian deviates strongly suggests the presence 
of {\em stochastic mass loss processes} on the RGB. However, deviations from a 
Gaussian shape are also not uncommon among globular clusters, particularly in the 
cases of those having bimodal HBs and/or long blue tails with gaps \citep{cea98,
ffea98,gpea99,ymea04}.

\subsection{The Complexity of the ``HB Phenomenon''} 

It is virtually impossible to write a short review paper on HB 
stars covering ``observations'', ``theory'', and ``implications for the formation 
of the Galaxy'': each one of these subjects covers so much material that one 
could rather write separate review papers for each one of them. 
Moreover, a review of HB stars 
cannot be complete without looking into their progenitors and their progeny. The 
task of a reviewer of HB stars is accordingly a daunting one, and it is virtually 
impossible to aim at completeness. In the present paper, while attempting to 
cover a broad spectrum of HB-related topics, we again hold no hope of providing 
a complete review of the literature on these subjects. Recent reviews 
focusing on several more or less specific topics related to HB stars have been 
provided by \citet{cc99,car03}, \citet{bc99}, \citet{dB99}, 
\citet{m01,m04}, \citet{as97b,as99}, \citet{pd99}, \citet{wl99}, \citet{ywlea99}, 
\citet{aw00}, \citet*{egea01}, \citet{cc03}, \citet{gb03}, \citet{rdm03}, 
\citet{gp03}, \citet{pm04a,pm04b}, \citet{sc05}, \citet{jst06}, 
\citet{uh08}, and \citet*{rrea08}; and very instructive 
earlier reviews, covering diverse astrophysical contexts, include those by 
\citet{as85,as90,as94}, \citet{ap94}, \citet{ac95}, \citet{bd95}, \citet{hs95}, 
\citet{tb96}, \citet*{psea96}, \citet*{scd97}, \citet{fpb97}, \citet*{rrea97}, 
and \citet{r98}. 
Similarly, excellent sections focused on HB stars can be found in the reviews 
on the evolution of low-mass stars, Population~II stars, globular clusters, and
related topics by \citet{ar77,ar83}, \citet{ir84}, \citet{vc85,vc99}, \citet{fc85,fc98}, 
\citet{rfp88}, \citet{rc89}, \citet{ii91}, \citet{rz93a,z93}, \citet{fd99},
\citet{mf99}, \citet{bc01}, \citet{wh01}, and 
\citet*{rgea04b}, among others. Other recent reviews by the present 
author on the subject of HB stars include 
\citet{c04b,mc06,mc08a,mc08b,mc09}.  

\begin{figure}[t]
  \plotone{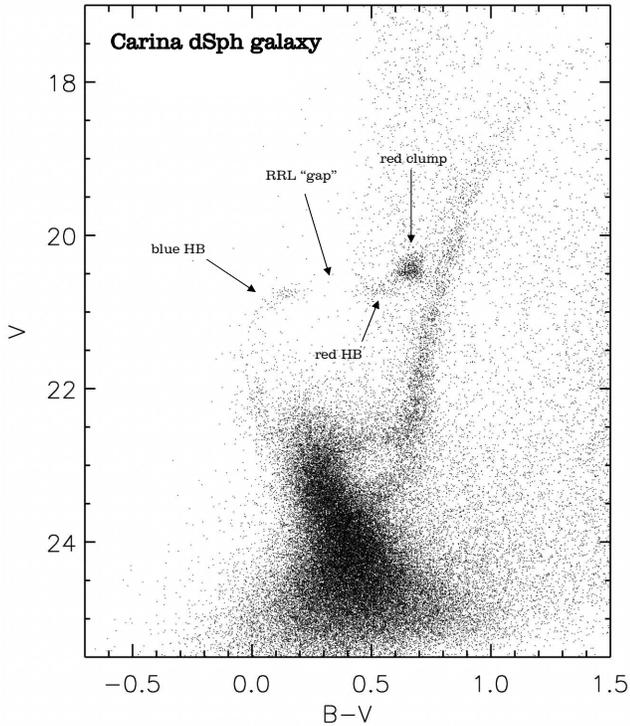}
  \caption{$V$, $\bv$ CMD for the Carina dSph galaxy, with the positions of red 
    clump stars, red HB stars, the RR Lyrae ``gap,'' and the blue HB indicated. 
    The red clump is associated with the younger turnoff, 
    at $V \approx 23$~mag, $\bv \approx 0.25$~mag. 
    The other components derive from the 
    old turnoff whose presence is indicated by the faint subgiant branch at
    $V \approx 23.25$~mag, $\bv \approx 0.55$~mag. 
    Adapted from \citet{mmea03}.\label{fig:CMD}  
}
\end{figure}

\begin{figure}[ht]
  \plotone{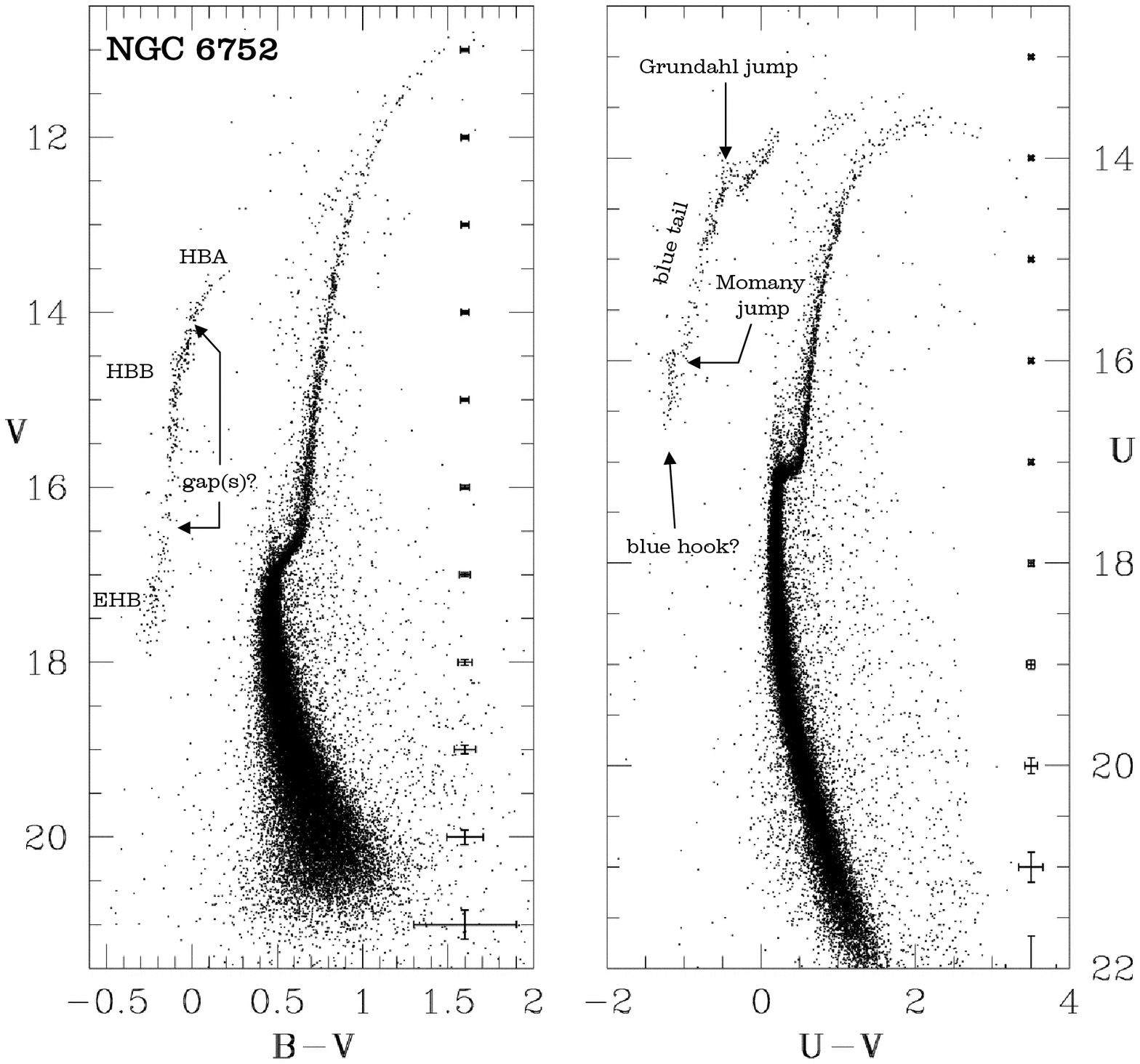}
  \caption{$V$, $\bv$ ({\em left panel}) and $U$, $U\!-\!V$ ({\em right panel}) 
    CMDs for NGC~6752. In the left panel, the positions of HBA, HBB, and EHB
    stars (collectively forming the ``blue tail'') are indicated, along with 
    the positions of possible gaps along the distribution. In the right panel, 
    besides the blue tail, the positions of the \citet{gea99} and
    \citet{ymea02,ymea04} ``jumps'' are also indicated, along with the 
    approximate position where ``blue hook'' stars would be found if they 
    were present in the cluster. 
    Adapted from \citet{ymea02}.\label{fig:NGC}  
}
\end{figure}

In Figure~\ref{fig:CMD} we show a visual color-magnitude diagram (CMD) 
for the Carina dwarf spheroidal (dSph) satellite of the Milky Way, with 
several different HB components indicated, including both a {\em red
clump} and a {\em red HB}. The {\em RR Lyrae ``gap''} and the 
{\em blue HB} are
also indicated. Similarly, Figure~\ref{fig:NGC} shows CMDs for the
Galactic globular cluster NGC~6752, in the visual ({\em left panel})
and in the $U$, $U\!-\!V$ plane ({\em right panel}). These plots reveal
the complexity of the blue tail phenomenon, with the positions of 
the HBA, HBB, and EHB components (see below) indicated, along with those 
of the \citet{gea99} and \citet{ymea02,ymea04} ``jumps'' and of possible 
blue HB gaps. The place where blue hook stars would be found, if 
present in the cluster (NGC~6752 actually appears to lack blue hook 
stars, according to \citeauthor{ymea04}), is also schematically 
indicated. 

In the next section, we briefly discuss each of these components in turn.

\section{The Different Constituents of the Horizontal Branch} 

\subsection{\em The red clump} 

Red clumps originate from red giants that undergo the helium flash at the 
tip of the RGB while still having a total mass 
of more than $\sim 1\, M_{\odot}$ (but less than $\sim 2-2.5\, M_{\odot}$; 
see, e.g., Fig.~2 in \citeauthor{lg99} \citeyear{lg99}); 
accordingly, they are commonly present in intermediate-age systems and 
old open clusters \citep*[e.g.,][]{rc70,ebea01,gs02}. They are physically 
related to, but should not be confused with, the so-called 
``red HB stars,'' which are less massive, significantly fainter and usually 
fall along the ``horizontal'' part of the HB, as clearly shown in 
Figure~\ref{fig:CMD}. 

A common mistake is to confuse 
``red clump''  and ``RGB bump'' stars, but they are totally different types 
of stars: the latter are bona-fide red giants with a partially degenerate 
helium core that burn hydrogen in a thin shell surrounding the core
\citep{t67}. This 
hydrogen-burning shell moves outward in mass as the star evolves up the RGB, 
and eventually reaches a chemical discontinuity left behind by the previous, 
maximum inward penetration of the hydrogen-rich convective envelope, leading 
to a sudden replenishment of the shell with fresh nuclear fuel and a momentary 
reversal of evolutionary path on the RGB \citep*[see][ for a review and 
additional references]{msea02}. Recent papers discussing several 
different aspects of red clump stars include those by \citet{lg99} and 
\citet*{msea03}.

\subsection{\em The red HB} 

As just stated, red HB stars are the lower-mass analogues 
of the red clump stars. They are commonly present in metal-rich  
\citep[e.g.,][]{ta88,oea95} as well as relatively young globular clusters 
\citep[e.g.,][]{psea89,rbea93}. 
However, red HB stars can also be found in metal-poor systems with ``normal'' 
ages, where they are usually interpreted as either the progeny of red giants 
that lost very little mass on their ascent of the RGB, or as former blue HB stars 
or RR Lyrae variables that have evolved to the right on the color-magnitude 
diagram, now approaching the end of their evolution as HB stars and the 
beginning of the asymptotic giant branch (AGB) phase. 
Another route towards producing (admittedly fewer) 
red HB stars is through the evolution of blue straggler stars 
\citep[BSS;][]{rfp88,fpea92}. \citeauthor{fpea92} suggest, in fact, 
that a red HB star ``of BSS origin'' should be present for every $\sim 6$ BSS
present in a given globular cluster. (Note that, depending on the amount of 
mass lost by the BSS on the RGB phase, at least some of these might better 
classify as red clump stars.)

\subsection{\em The RR Lyrae ``gap''}\label{sec:RRL} 

This is the part of the HB that crosses the Cepheid 
instability strip. The term ``gap'' is very inadequate, but is still commonly 
used. The reason for this is that, in order to properly place 
an RR Lyrae variable in the color-magnitude diagram, one needs to follow its 
whole pulsation cycle and thereby obtain reliable mean colors and magnitudes. 
Since most color-magnitude diagram studies lack the sufficient time coverage, 
these stars are often simply omitted from the published CMDs, thereby 
leading to an artificial ``gap'' at the instability strip level. However, there 
are indeed several globular clusters---the so-called {\em bimodal-HB globular 
clusters}---which do seem to have relatively few RR Lyrae stars compared to the 
nonvariable HB stars to 
their right and left in the CMD \citep[see][ for a review and extensive 
references]{cea98}. The most famous such cluster is 
NGC~2808 \citep{wh74}, which in spite of the recent discovery of a sizeable 
population of RR Lyrae stars, still remains firmly classified as having a 
bimodal HB \citep{mcea04}.

\subsection{\em The blue HB}\label{sec:BHB} 

As clearly revealed by its name, blue HB stars are HB
stars falling to the blue of the RR Lyrae instability strip. There have been 
numerous subdivisions of the blue HB, the most common including HBA (or A-type 
HB), HBB (B-type HB), and EHB (extreme or extended HB) stars
(see Fig.~\ref{fig:NGC}). 
HBA stars are blue HB stars cooler than about 12,000~K; HBB stars include those 
with temperatures in the range between 12,000~K and 20,000~K; and EHB stars 
include HB stars hotter than 20,000~K. The latter cover a remarkably small 
range in envelope masses, generally less than $0.02\,M_{\odot}$ 
\citep*[e.g.,][]{bdea93}---and therefore also in total masses, since the
He-core mass is essentially the same for all low-mass stars with a given 
chemical composition \citep[e.g.,][]{cdi95}. 
In a visual CMD, the blue HB component 
may contain a ``horizontal'' part---the canonical 
blue HB (see Fig.~\ref{fig:CMD})---and an effectively 
``vertical'' component, commonly referred to as the {\em the blue HB tail}
(see Fig.~\ref{fig:NGC}). 

Many authors have suggested that the blue HB proper and the blue HB tail are 
separated by a ``gap'' which is indeed seen in several globular clusters 
around $(\bv)_0 \simeq 0$ \citep*[e.g.,][]{rbea85,c99}---but perhaps not in 
{\em all} clusters containing blue tails \citep{cea98,lc99}. On the other 
hand, several additional ``gaps'' may be present along the 
blue tail \citep{bn73,ng76,ns78,lc80,ffea98,gpea99,ymea04}; 
a recent, detailed description of these gaps has been provided 
by \citeauthor{ymea04}.  Several of these gaps, as discussed  
by \citeauthor{cea98}, may still require more sophisticated statistical 
analyses to establish their reality beyond reasonable doubt, particularly 
when well-observed clusters seem to lack such gaps altogether, as in the case
of M2 \citep[NGC~7089;][]{lc99}. Most impressively, \citet{bb03b} has recently 
argued that the field ``blue HB star'' sample studied by \citeauthor{bn73} and 
\citeauthor{ng76}---and which gave rise to the very concept of gaps along the 
blue HB---is mostly comprised of stars in {\em different} evolutionary 
phases. We quote \citeauthor{bb03b}:

\begin{quotation}
We expected to find [based on detailed
spectral analysis] a high fraction of HB candidates among the faint blue 
high-latitude stars listed by \citet{bn73} and \citet{ng76}, especially in 
light of the `gaps' in the color distribution of these stars... But fewer than 
half (11 of 27) of the Newell stars that we observed were clearly HB objects, 
with another 11 stars classified as Population~I dwarfs, and the remaining 
five stars marked as pAGB [post-AGB], subgiants, and such.
\end{quotation}

The {\em EHB component} is the globular cluster analog of the field {\em blue 
subdwarf} (or sdB) stars \citep{c72}. The majority of the (nearby) field sdB 
stars, according to \citet*{maltea04} and \citet{miaea05}, appear to have disk 
kinematics, unlike HBA stars, 
which they find to be mostly halo stars \citep*[see also][]{aea05}. 
Accordingly, \citet{maltea04} pose the question 
whether field HBA and EHB stars truly have a similar physical origin.  

Detailed studies of the color-magnitude diagrams of globular clusters in several 
different bandpasses indicate the presence of additional ``fine structure'' on 
the blue HB, most notaly the {\em Grundahl jump} \citep{gea99} at a temperature 
around 11,500~K, which is seen as a sudden deviation, towards brighter 
magnitudes, of HB stars hotter than this point, particularly evident when using 
the Str\"omgren $u$ band \citep{gea99}. However, 
it is also present when using the Johnson $U$ passband 
\citep*{bn70,msea99,bea00,pbea01,hmea01,mcea02a,ymea02,ymea04}, as can be 
clearly seen from Fig.~\ref{fig:NGC}. This ``jump'' 
has been interpreted by \citeauthor{gea99}
as due to {\em radiative levitation and diffusion effects}, which lead to the 
metal-poor HB stars hotter than 11,500~K possessing strongly {\em metal-enhanced 
and helium-depleted atmospheres}.\footnote{Fortunately, and as emphasized by
\citet{wl99} and \citet{bb03a}, Mg appears to be 
virtually unaffected by radiative levitation and diffusion effects, thus 
allowing a ``window'' into the original metallicity of the star---which is 
of extreme importance in the case of field stars in particular.} 
The earlier literature was extensively reviewed by \citeauthor{gea99}; 
empirical evidence supporting their arguments was provided in spectroscopic 
studies of blue HB stars in globular clusters by  
\citet{bbea99}, \citet{smea99,smea00,smea03}, \citet{bb03a}, and 
\citet{dfea05}. Diffusion calculations that showed how some of the observed 
spectroscopic and photometric patterns come about were 
provided by \citet*{gmea83}, \citet*{gmea07,gmea08}, and 
\citet*{hbhea00}. The ``gap'' 
at $(\bv)_0 \approx 0$ has been tentatively associated with this phenomenon
\citep{c99}, but such a color appears much redder than would be expected 
for the observed 
Grundahl jump temperature of 11,500~K. In addition, the theoretical calculations 
by \citeauthor{gea99} suggest that other bandpasses in the Str\"omgren and 
Johnson systems besides $u$ and $U$ should
not be dramatically affected by this phenomenon. 

Most interestingly, 
recent observations have shown that the Grundahl jump phenomenon is also 
accompanied by a sharp drop in measured rotation velocities, blue HB stars 
hotter than 11,500~K presenting essentially no rotation, in contrast with 
cooler stars which may show quite significant rotation velocities, up to 
about $40\,{\rm km}\,{\rm s}^{-1}$  
\citep*[][]{rpea83,rp83,rp85,rpea95,bbea00a,bbea00b,
tkea00,arbea02,arbea04,bb03a,bb03b,bcea03}. 
While \citeauthor{bcea03} note that the presence of rotation in both red and 
blue HB stars presents a problem for models in which rotation is a candidate 
second parameter, it must be noted that the abrupt disappearance of 
rotation exactly at the Grundahl jump temperature indicates that diffusion 
and levitation patterns interfere with the star's observed surface rotation, 
somehow quickly damping the latter for temperatures higher than about 
11,500~K \citep{sp00,as02}. 
Therefore, we have at present no means to check whether stars 
hotter than this temperature may have arrived on the ZAHB with surface 
rotation velocities faster than those observed for red HB and cooler HB 
stars. Asteroseismology may provide a very useful probe of the (internal) 
rotation velocities of hot HB stars in the near future \citep{kh05}. 
A most intriguing piece of the puzzle 
is provided by the RR Lyrae stars, for which no evidence of rotation has 
been observed so far (see \citeauthor{bcea03} \citeyear{bcea03} 
for extensive references). Why 
would cool blue HB stars and red HB stars show clear signatures of rotation, 
but not the intermediate-temperature RR Lyrae variables? More extensive 
spectroscopic observations of RR Lyrae stars are clearly needed to settle 
this issue. 

What is the physical origin of the sudden onset of radiative levitation 
and diffusion patterns that lead to the \citet{gea99} jump at 11,500~K? 
The answer 
to this question is not entirely clear at present, but \citet{as02} noted 
that this temperature is very close to the one corresponding to the 
disappearance of surface convection in HB stars, thus strongly suggesting
a link between the disappearance of convection and the onset of radiative 
levitation and gravitational diffusion effects in these stars. In addition, 
\citeauthor{as02} suggests that the low rotation velocities of stars hotter 
than the Grundahl jump may be due to the spin down of the surface layers 
by a weak stellar wind induced by the radiative levitation of Fe. As shown  
by \citet{vc02}, such winds are indeed predicted by theory.

\subsection{\em The extended (or extreme) HB} 

Before a star arrives on the HB, it must undergo a {\em helium flash}---the onset 
of helium burning, under partially degenerate conditions, which takes place at 
the tip of the RGB \citep{sh62}. 
It is interesting to note that several early hydrodynamic 
studies of the He-flash indicated the occurrence of an explosive event and 
actual {\em disruption} of the star \citep[e.g.,][]{ae69,aw77,cd80}. Given 
that this prediction is in clear conflict with the very {\em existence} of HB 
stars, significant effort has also been devoted towards the computation of 
{\em hydrostatic} flashes \citep[e.g.,][]{mg76,kd80,ms81}. The hydrostatic 
approximation, with some numerical and physical refinements, is still often 
used in modern calculations 
\citep*[e.g.,][]{tbea01,scea03,tlea04,lpea04,sw05}---and 
these have indeed been vindicated by modern hydrodynamic studies 
\citep*[e.g.,][]{rd96,ddea06} 
which revealed that the aforementioned hydrodynamical predictions were indeed 
incorrect. It is perhaps not a conclusively settled matter whether any of 
the material that is 
nuclearly processed during the He-flash may reach the surface of the star, 
and also whether one or more mass loss episodes may be triggered by the 
primary and secondary core flashes; however, and to the extent that the 
latest He-flash calculations provide a realistic description of the process, 
no such phenomena should be expected. Indeed, \citeauthor{ddea06} have recently 
pointed out that, due to expansion velocities that are much lower than the 
local sound speed, hydrostatic modeling should indeed capture the essence 
of the helium flash process. On the other hand, the 3D models by 
\citeauthor{ddea06} intriguingly reveal motions ``of an apparently convective 
nature'' {\em beyond} the H-burning shell, which these authors claim not to be 
directly associated with the flash, but which may bring the products of 
H-burning to the surface of the star. 

On the other hand, 
the situation can become much more complex when the star loses so much mass on 
its ascent of the RGB that the He-flash ends up taking place not at the RGB tip, 
but rather during the helium white dwarf cooling curve---a so-called 
{\em late hot flasher} \citep*{tbea01,scea03,mcea06}. 
In this case, extensive mixing between the envelope and regions that underwent 
significant hydrogen and helium burning is indeed expected, and these stars, 
when they finally manage to settle on the zero-age HB (ZAHB), may end up as the 
so-called {\em blue hook stars} \citep*[e.g.,][]{dcea96,dcea00,jwea98,area04,gbea07,vrea07}, 
which have temperatures in excess of 35,000~K \citep{smea02,smea04,smea07}.   
This is to be compared with the case in which 
the star undergoes an {\em early hot flash}---i.e., {\em prior} to arriving on 
the white dwarf cooling sequence---in which case \citeauthor{tbea01} find that no 
mixing takes place and a ``canonical'' EHB star results. In this scenario, 
the reason why the blue hook stars appear fainter than its peers on the EHB
is twofold. First, they tend to have smaller helium-core masses, due 
to the fact that they leave the RGB {\em before} igniting helium in their 
core, as a consequence of extreme mass loss on the 
RGB \citep[e.g.,][]{cc93,dcea96,tbea01}. Second, and most importantly in 
the late flasher scenario, their atmospheres present
large enhancements in both helium and carbon as a consequence of mixing, 
thus affecting the bolometric corrections \citep{tbea01}. 

Spectroscopic evidence 
largely favoring this ``late flash mixing'' scenario for the origin of 
blue hook stars has recently been presented  
by \citet{tlea04} and \citet{smea02,smea04,smea07}. 
An intriguing indication from these spectroscopic studies for at least some of 
the hotter EHB stars is that some hydrogen still remains in their atmospheres, 
which is not expected on the basis of the theoretical models. As argued by 
\citeauthor{smea04}, some residual hydrogen may perhaps survive a late hot 
flash and later diffuse to the surface during the HB phase. Note also that, 
in analogy with the \citet{gea99} ``jump,'' another ``jump'' has 
recently been suggested to be present in blue-tail globulars, namely the 
Momany jump (see Fig.~\ref{fig:NGC}), 
at a temperature around 21,000~K \citep{ymea02,ymea04}---which 
these authors conjecture to be related to the early helium flashers
\citep[but see][ for a critical discussion]{mc08a}. 

Recently, a 
sub-population of cluster stars with a large, primordial enhancement in the 
helium abundance has also been suggested as a possible channel producing 
the blue hook stars \citep{lea05,sjyea08}. However, it is unclear how the abundance 
patterns observed in blue hook stars would be accounted for in this 
scenario \citep[e.g.,][]{smea07}. 
In like vein, \citet{vcea05} have recently advanced the 
intriguing suggestion that at least some of the very hot and underluminous 
blue hook stars might actually be more straightforwardly explained as 
photometric blends and/or binary stars.

\section{Variable Stars on the HB} 

Until recently, it was thought that RR Lyrae stars were the only class of 
variable stars on the HB. The situation has changed recently, with the 
discovery (in a beautiful example of theoretical modelling preceding the 
observations) of {\em non-radial pulsation} among sdB stars. These 
non-radial pulsators are now divided into two different groups. Therefore, 
there are currently three known classes of variable HB stars, namely:  

\begin{itemize}

\item {\em RR Lyrae stars}: These are radial pulsators with periods in 
the range between about 0.2~d and 1.0~d, known to pulsate primarily in the 
{\em fundamental mode} (RR Lyrae stars of type a and b, now lumped together 
as RRab or RR0 stars) and in the {\em first overtone} (RR Lyrae stars of type c, 
nowadays also referred to as RR1 stars). They have long been known to be 
present in large numbers in Galactic globular clusters \citep{pb1895}, and
were first correctly identified as radially pulsating stars by \citet{hs14}. 
The letters ``a,'' ``b,'' ``c'' were first used by \citet{b02} 
(see his p.~132) to classify RR Lyrae stars as a function of 
their light curve shapes; \citeauthor{b02} also noted that the RRb subclass 
is 

\begin{quotation}
... similar to (the RR)a (subclass), of which it may be regarded as a 
modification.
\end{quotation}

\noindent Indeed, both are now known to be fundamental-mode pulsators; in 
fact, the identification of RRab's with fundamental pulsation and of the RRc's
with first overtone pulsation seems to have first been clearly made by \citet{ms40}. 
The notation in which the numbers 0 and 1 are used as opposed to the letters ``ab'' 
and ``c,'' respectively, was first introduced by \citet{caea00}. 

In addition, there are also {\em double-mode} pulsators (RRd or RR01 stars), 
pulsating simultaneously in the fundamental and first-overtone modes, as  
first recognized by \citet{jw77} among field stars and by \citet*{asea81} 
(see their Sect.~IIIa) in globular clusters
(see also \citeauthor{mgea07} \citeyear{mgea07} for an impressively 
detailed frequency analysis, using MOST [Microvariability Oscillations 
of Stars] satellite observations, of the RRd star AQ~Leo, the prototype 
of this class). 
The notation ``RRd'' for these
stars appears to first have been used by \citet{jn84,jn85}. 
The ratio between the first-overtone and fundamental period 
for the RRd stars is quite well defined; indeed, on the basis of the data 
for M68 (NGC~4590), IC~4499, and M15 (NGC~7078) compiled in Table~8 of 
\citet{kw99}, we obtain $\langle P_1/P_0\rangle = 0.7454$ for 38 
stars,\footnote{It is basically this ratio that allows one to compute 
the ``fundamentalized'' period of an RRc star by using the relation 
$\log P_{0} = \log P_{\rm 1} + 0.128$.} 
with a minimum value of $P_1/P_0 = 0.7433$ and a maximum value 
$P_1/P_0 = 0.7481$. For the LMC RRd, \citet{caea00} find comparable 
period ratios, but with upper and lower values shifted downward by 
$\approx 0.0015$ (compare with their Fig.~5). An impressive summary of 
period ratios for a variety of stellar systems (the LMC, the SMC, Draco, 
Sculptor, M15, M68, IC~4499, M3, and the Galactic field) is provided in 
Figure~1 of \citet*{pdc00}, which basically confirms the above range in 
period ratios. Recently, \citet{gcea04} have reported on the 
discovery of two RRd variables in M3 whose period ratios, in the range
0.738--0.739, fall well below those for previously known RRd stars. 

It has also been suggested that RR Lyrae stars may pulsate in the 
{\em second} overtone \citep[e.g.,][]{dw77,caea96,wn96,lkea99,cr00}; 
accordingly, short-period, 
low-amplitude variables which are suspected of being second-overtone pulsators
are now often classified as RRe or RR2 stars (but see \citeauthor{gk98}
\citeyear{gk98}, \citeauthor{c04b} \citeyear{c04b}, 
\citeauthor{gbea97a} \citeyear{gbea97a,gbea97b}, 
for arguments 
suggesting that at least some of them may rather represent the short-period 
end of the RRc distribution). In fact, \citet{caea00} suggest that 
double-mode variables pulsating simultaneously in the first {\em and second} 
overtones may also exist, and tentatively assign them an RR12 subclass. 

Non-radial modes have also been 
suggested to be present in a fraction of the RR Lyrae stars \citep[e.g.,]
[; and references therein]{gk95,kk02,mgea07}, primarily as a means to 
explain the so-called \citet{b07} effect \citep[e.g.,][]{dm04}. 
The Blazhko effect consists in a 
periodic modulation, on a much longer timescale than the primary period, 
of the light curve shape.\footnote{In their review of 100 years of 
observations of the star RR Lyrae, \citet{sk00} point out that the 
phenomenon might more appropriately be called the {\em Blazhko-Shapley 
effect}, since \citet{s16} was actually the first to demonstrate that 
the oscillations in the maxima of an RR Lyrae star can be described as a
periodic variation in the shape of the light curve and in the height of 
the maxima.} The modulation (or Blazhko) period falls in the range between 
5.309~d \citep[as found by][ for the field star SS~Cnc]{jjea06} and 
530~d \citep[as found by][ for the field star RS~Boo]{an98}.
A variety of other multiperiodic phenomena may also take 
place in RR Lyrae stars \citep[see, e.g.,][]{caea00,caea04,tm04,mgea07}. 
We note, in passing, 
that the Fourier decomposition ($D_m$) method \citep{jk96} has recently 
been criticized by \citet*{ccea05} as a diagnostic of the Blazhko effect 
in RR Lyrae stars. 

\item {\em V361~Hya variables}: Also known as EC~14026 or sdBV stars, these 
are non-radial, short-period, low-order p-mode pulsators, whose periods fall 
in the range 80--400~s, and whose amplitudes bracket the interval between 
4 and 25~mmag. Their temperatures are found in the range between 29,000~K 
and 36,000~K, and their gravities in the range $5.2 \leq \log g \leq 6.1$ 
\citep{dk02,gfea04,gfea06a,gfea06b}. 
Note that this class of variable star had been 
predicted \citep{scea96} {\em before} it was first observed \citep{dkea97}.

\item {\em V1093~Her variables}: Also known as 
PG1716+426 or ``Betsy'' variables (after Elizabeth Green, the 
discoverer of the group), these are non-radial, relatively long-period, 
high-order g-mode pulsators \citep[e.g.,][]{egea03,gfea03,mrea04}. 
Their periods fall in the range 2000--9000~s, and 
their amplitudes are very small, generally being smaller than 0.5~mmag. 
Their temperatures 
bracket the interval 25,000--30,000~K, and their gravities are in the range
$5.1 \leq \log g \leq 5.8$ \citep{gfea04,gfea06a,gfea06b,mrea04}. 

\item {\em Mixed-mode variables}: Variable stars showing both V361~Hya 
{\em and} V1093~Her characteristics have recently been discovered as well 
\citep{abea05,ssea06}.    
\end{itemize}

For a recent 
discussion of the pulsation mechanism (the ``Fe-bump opacity mechanism'') 
driving the pulsations in V361~Hya and V1093~Her stars, see \citet{js06}.

It is important to emphasize that the presence of variable stars on the HB phase 
provides us with a unique opportunity to utilize {\em stellar pulsation observations 
and theory} to improve our understanding of HB stars in general. 
As well known, the pulsation properties of stars are fundamentally 
related to their mean densities through the so-called {\em period-mean density
relation}---an expression originally due to \citet{r1879}. This extremely 
important equation specifies that the period is inversely proportional to 
the square root of the mean density of the star, which in turn can be expressed
in terms of the star's global physical parameters, such as mass $M$, luminosity 
$L$, and effective temperature $T_{\rm eff}$ 
\citep*[e.g.,][]{vab71,gbea97b,fcea98,mdcea04}. As a matter of fact, Ritter's
relation is applicable, to first order, over an impressively wide range of stellar 
parameters \citep[e.g.,][]{jc74}.  
In addition, the observed periods---and 
especially so in the case of non-radial pulsators---depend on the star's detailed 
structural profile. This is a natural consequence of the fact that the measured
periods are directly related to the speed of travel of the sound waves across
the stellar interior. 
In this sense, the importance of the non-radial variables 
on the HB to constrain the internal structures of HB stars, including the amount 
of internal rotation, has recently been emphasized by \citet{sk05}, 
\citet{kh05}, and \citet{gfea06a}, and 
nicely demonstrated in practice by \citet{scetal05,scea06}, \citet{srea07}, 
and \citet{vvgea08}. 
It is interesting to note that the high rotation velocity recently derived by 
\citet*{vdea04} for the post-AGB star ZNG~1 in M5 might be accounted for, as 
noted by those authors, if RGB and HB stars are able to maintain rapid rotating 
cores, as may also be required \citep{sp00}
to explain the presence of fast rotators on the HB
(see Sect.~\ref{sec:BHB} for extensive references).

It is worth noting that several authors have called 
attention to the possible presence of non-variable HB stars {\em inside} the 
RR Lyrae instability strip
\citep*[e.g.,][; and references therein]{sk68,aw90,dxea98,sea05}. 
The recent discovery of 
Cepheid variables presenting pulsation amplitudes at the mmag level 
\citep{jbea05} suggests that at least some of these ``non-variable'' stars 
may indeed be varying, though with much smaller amplitudes than would  
have been possible to detect with more traditional techniques---as nicely 
demonstrated by \citet{cr01} in the case of NGC~5897. 
According to  
\citeauthor{jbea05}, ``ultra-low amplitude'' RR Lyrae stars are 
predicted to exist close to both the blue and red edges of the instability
strip. Note, on the other hand, that the  {\em General Catalog of Variable 
Stars} \citep{pkea98} defines RR Lyrae stars as having amplitudes in the 
range 0.2 to 2~mag in 
$V$,\footnote{{\tiny{See \texttt{http://www.sai.msu.su/groups/cluster/gcvs/gcvs/iii/vartype.txt}.}}} 
thus showing that RR Lyrae with $V$ amplitudes smaller than 0.2~mag, though 
known to exist even among the normally higher-amplitude RRab's
\citep[e.g.,][]{as81,gw86,cr01,isea02,isea03,ccea05}, are indeed quite rare. 
Note that variable stars falling on the red HB---i.e., 
{\em outside} the formal boundaries of the RR Lyrae instability strip---have 
also been suggested to exist \citep*[][ and references therein]{dxea98}.

Still in regard to pulsation amplitudes, an interesting recent result 
worth mentioning are the very large amplitudes---up to $\sim 5$~mag---found for 
RR Lyrae stars in the far UV \citep[][; see also Fig.~1 in \citeauthor{jbea82} 
\citeyear{jbea82} for an example of an earlier RR Lyrae light curve in the 
far UV, already indicative of large amplitudes]{rdea04,adea05,adea07,bwea05,jwea05}, 
and up to $\sim 2.6$~mag in the near UV \citep{sb05}. Conversely, in the near-IR, 
RR Lyrae amplitudes become quite small \citep*[e.g.,][]{rjea88,lj90a}. Thus, 
RR Lyrae stars usually reveal themselves more easily when time-series surveys 
are conducted using bluer bandpasses, whereas average magnitudes can be more 
easily computed using near-IR observations (see also \S\ref{sec:M31halo} below). 

Before closing this section, we would like to comment on the possibility that some 
stars in the rapid phase of evolution between the RGB tip and the 
ZAHB do in fact become variables. Evolutionary paths for these stars 
are shown, for instance, in Fig.~4 in \citet{ms81} and Fig.~4 in \citet{tbea01}, 
where it can be seen that many of these stars do cross the instability strip before 
settling on the ZAHB. While distinguishing them from more ordinary RGB, AGB, and 
RR Lyrae stars is obviously far from straightforward, both due to the very small 
expected number of stars in this phase and because of their overlapping with 
them on the CMD, their extremely fast evolutionary timescale ($\sim 10^6$~yr) hints 
at the possibility of identifying them by appropriate monitoring in wide-field 
surveys. This is because much more extreme {\em period change rates} might obtain 
for them than for most other variables in a similar position on the CMD. To the best 
of our knowledge, no other technique has been proposed in the literature for the 
detection of these elusive post-RGB tip/pre-ZAHB stars. The reader is referred
to the recent paper by \citet{vsaea08} for an application of this method to
the case of the Galactic globular cluster M3.

\section{Detecting HB Stars in Distant Systems: Methods}

While HB stars are relatively bright, their direct detection in distant systems 
(i.e., at the distance of M31 and beyond) is a challenge. This is particularly so 
in the case of blue HB stars, which can be several magnitudes 
fainter than the ``horizontal'' level in $V$ (Fig.~\ref{fig:NGC}), 
due to the increase in the $V$-band bolometric correction at 
high temperatures. While the blue HB stars 
become brighter towards the far ultraviolet, which could make them 
candidates for detection from space, the required exposure times, at the 
distance of M31, are prohibitively large. Therefore, while extremely long-exposure 
{\em HST} observations have demonstrated that it is possible to reliably detect 
blue HB stars down to at least the main sequence turnoff level in M31 
\citep[e.g.,][]{tbea03,tbea04,tbea06}, 
alternative approaches are needed at present to obtain a more complete 
assessment of the HB morphologies in extragalactic systems.

\subsection{Far-UV Observations}

The far-UV output from globular clusters is strongly dependent on the 
temperature of the sources. Since hot HB stars are important UV 
emitters \citep*[e.g.,][; see also \citeauthor{mc08b} \citeyear{mc08b} for 
a recent review]{bdea95}, this is expected to 
translate into a far UV (integrated) color--HB type correlation, 
especially when the contribution of bright, individual far-UV sources (such as 
post-AGB stars), which are present in clusters in non-statistically significant 
numbers, is removed from estimates of the integrated far UV colors.

\begin{figure}[t]
  \plotone{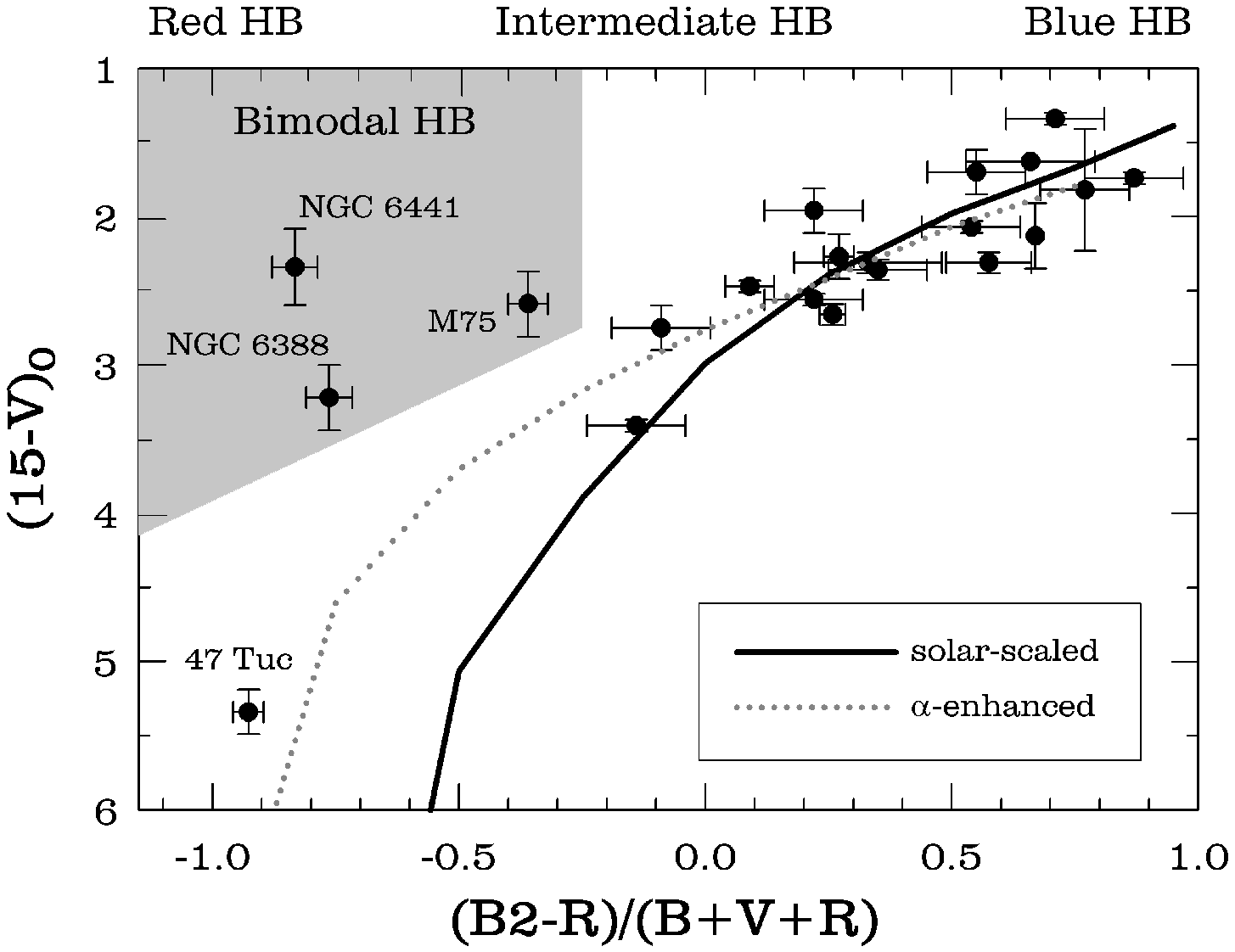}
  \caption{Variation in the integrated $(15\!-\!V)_0$ color of Galactic globular 
  clusters as a function of the HB morphology parameter $\BUO$
  \citep{rb93}. The shaded area (labeled ``bimodal HB'') indicates a region 
  occupied by clusters which have sizeable blue HB components, but which at 
  the same time have sizeable red HB components as well. The lines indicate 
  theoretical predictions based on solar-scaled (continuous lines) and 
  $\alpha$-enhanced (dotted gray lines) abundances. From \citet{wlea01} 
  and Landsman \& Catelan (2009, in preparation). \label{fig:FUV}
}
\end{figure}

\citet{wlea01} and Landsman \& Catelan (2009, in preparation) 
tested this prediction, using integrated far UV 
fluxes for Galactic globular clusters as summarized by \citet{bdea95}, and 
including revisions to the far UV photometry based on images taken with the 
{\em Ultraviolet Imaging Telescope} (UIT). The result is shown in 
Figure~\ref{fig:FUV}, which depicts the trend 
of variation in the $(15\!-\!V)_0$ color (where ``15'' stands for 
a bandpass centered at 1500~\AA) as 
a function of the HB morphology parameter $\BUO$ \citep{rb93,rbea97}, 
where $\mathcal{B}2$ is the number of blue HB stars bluer than $(\bv)_0 = -0.02$, and 
$\mathcal{B}$, $\mathcal{V}$, $\mathcal{R}$ are the numbers of blue, variable (RR Lyrae), and red HB stars, 
respectively. The latter number counts are also widely used in constructing 
the so-called {\em Lee-Zinn parameter}, $\LZ = \BVR$ \citep{rz86,l90,ldz90}. 
The reason why we use the Buonanno as opposed to the Lee-Zinn parameter 
is that it is more successful at 
breaking the degeneracy that characterizes $\LZ$ in the case of clusters with 
completely blue HBs. More specifically, $\LZ$ is not able to distinguish between 
a long blue tail cluster such as M13 or NGC~6752 and a stubby blue HB cluster 
such as NGC~288 (all having $\LZ \simeq +1.0$), whereas $\BUO$ (as 
well as several other HB morphology parameters, including some of those defined 
by \citeauthor{fpea93} \citeyear{fpea93} and \citeauthor*{cea01a} \citeyear{cea01a})
are better suited for the task. Indeed, Figure~\ref{fig:FUV} shows that, 
apart from the 
clusters with bimodal HBs (NGC~6388, NGC~6441, M75), there is a good correlation 
between the Buonanno parameter and $(15\!-\!V)_0$, characterized by a high correlation 
coefficient $r = 0.93$. Using $\LZ$, one finds an $r = 0.82$ instead, but this is 
primarily due to the large difference in far UV color between 47~Tucanae 
(NGC~104) and the group 
of blue HB clusters: removing 47~Tuc, the correlation coefficient, when using $\LZ$, 
drops to $r = 0.57$, whereas the one obtained using the Buonanno parameters still
remains fairly high at $r = 0.84$. These correlations are in good agreement with 
the theoretical predictions by \citeauthor{wlea01} for an $\alpha$-enhanced 
composition, as can clearly be seen from the dotted line in Figure~\ref{fig:FUV}. 
In summary, 
one should be able to estimate, at least as a first approximation, the HB type 
of an old extragalactic globular cluster by measuring its {\em integrated} far 
UV light.  

Indeed, that the far UV light is a powerful diagnostic of the presence of hot HB 
stars has been demonstrated by the observation of a peculiarly high far UV flux 
from the moderately metal-rich Galactic globular clusters NGC~6388 and NGC~6441
\citep*{mrea93}, which led the authors to suggest the existence of long blue tails 
in these clusters several years {\em before} these were first directly observed 
with {\em HST} \citep{gpea97,mrea97}. Recently, \citet{rpea03} have performed a detailed 
analysis of the mid-UV flux from the extremely massive M31 globular cluster G1, 
finding that it likely presents both regular and extreme blue HB stars.  
Its recent detection in far-UV observations with GALEX \citep{screa05,screa07} lends  
support to their conclusion. The {\em HST} color-magnitude diagram for the 
cluster \citep{rea05} cannot reveal the presence of hot HB stars since it 
just reaches the HB level, but it is not clearly inconsistent with the presence 
of a few blue HB stars at the ``horizontal'' level.\footnote{While \citet{rea05}
provide $\LZ$ values for their observed M31 globular clusters, no completeness 
corrections were applied in deriving them, which may lead to important 
underestimates of the number of blue HB stars---and therefore of $\LZ$---for 
those among them possessing well-populated blue HB tails.} 

To close, it is important to note that the H$\beta$ and H$\delta$ integrated 
indices can be sensitive indicators of the presence and temperature of HB stars 
in unresolved stellar systems 
\citep*[e.g.,][; and references therein]{hclea00,cmea03,rsea04,spea09}. 
The dependence of these and several other photometric and spectroscopic indicators 
on HB morphology has recently also been discussed by \citet*{hclea02},
\citet*{rpea04}, \citet{sac07}, and \citet{mkea08}, among others.

\subsection{Type II Cepheids: Bright Indicators of the Presence of 
            Faint Blue HB Stars}

Another technique to infer the presence of a blue HB component in extragalactic 
systems without the need to obtain exceedingly deep photometry could use fairly 
bright {\em type II Cepheids} (including BL Herculis, W Virginis, and RV Tauri 
stars) as an indicator. Indeed, type II Cepheids are widely believed to be the 
immediate progeny of HB stars with little envelope mass (but with still 
enough to reach the AGB stage)---i.e., blue HB stars. This is supported both by 
the observational record, which shows that type II Cepheids are present only when 
a sizeable blue HB component is also present \citep{gw70,sw85}, seemingly 
irrespective of the metallicity \citep{pea02,pea03};\footnote{There is a single known
exception to this rule, provided by the type II Cepheid in Palomar~3 \citep*{jbea00}.}
and by theoretical models 
(\citeauthor{sh70} \citeyear{sh70}; 
\citeauthor{rg76} \citeyear{rg76,rg85}; 
see also 
\citeauthor{mdcea07} \citeyear{mdcea07} for a recent discussion and 
additional references). 
Accordingly, detection of type II 
Cepheids should immediately imply the presence of a sizeable blue HB component. 
In this sense, it is worth noting that \citet{adea04} and \citet{bw05} found 
possible type II
Cepheids in variability studies of small M31 halo fields; if this classification 
is confirmed (the authors note that the stars could also be Anomalous Cepheids, 
for instance), 
this would be consistent with the deep M31 CMDs obtained by \citet{tbea03,tbea06}, 
which clearly reveal the presence of a sizeable (though seemingly not very 
extended) blue HB component. Likewise, \citet{bpea05a} suggest that the M31 
dSph satellites And~I and And~III may also contain a small number of type II 
Cepheids, which should also imply the presence of blue HB components. Their 
{\em HST} CMDs for these galaxies do indeed suggest the presence of blue HB 
stars, particularly evident in the case of And~I. Last but not least, we note 
that \citet{jfea06} and \citet*{fvea07} have recently reported the detection of 
type II Cepheids in M31, including many in the direction of the M31 bulge---which, 
if confirmed, implies the presence of a sizeable blue HB component in that 
direction as well.

\begin{table*}[t]
\caption{Formulae to Compute the Mass Loss Rate in Red Giant Stars\tablenotemark{a}   \label{tb:DM}}
\smallskip
\begin{center}
{\small
\begin{tabular}{cc}
\tableline
\noalign{\smallskip}
Formula name
&   Formula for $\frac{dM_{\rm RGB}}{dt}$ (in $M_{\odot}{\rm yr}^{-1}$)    \\
\noalign{\smallskip}
\tableline
\noalign{\smallskip}
Reimers  & 
   $5.5\times 10^{-13} \, \left(\frac{L}{g R}\right)$     \\ \\
Modified Reimers  & 
   $8.5\times 10^{-10} \, \left(\frac{L}{g R}\right)^{+1.4}$     \\ \\
Mullan     & 
   $2.4\times 10^{-11} \, \left(\frac{g}{R^{3/2}}\right)^{-0.9}$ \\ \\
Goldberg   & 
   $1.2\times 10^{-15} \, R^{+3.2}$                              \\ \\
Judge-Stencel & 
   $6.3\times 10^{-8} \, g^{-1.6}$                               \\ \\
VandenBerg & 
   $3.4\times 10^{-12} \,L^{+1.1} g^{-0.9}$                               \\

\tableline
\tablenotetext{a}{Reimers formula as given in \citet{kr78}. 
All other expressions derived by \citet[][see his Appendix~A for details]{c00} 
on the basis of the 
data provided by \citet{js91}. The gravity $g$ is in cgs units, and luminosity 
$L$ and radius $R$ in solar units. } 
\end{tabular}
}
\end{center}
\end{table*}

\section{Mass Loss on the RGB}\label{sec:DM} 

Knowledge of how red giants lose mass is one of the indispensable conditions for 
understanding the properties of HB stars, including their color distributions, the 
production of RR Lyrae and sdB stars, the variation in HB morphology with metallicity, 
and the (in)famous second-parameter phenomenon. 
Several previous reviewers have emphasized the need for an advancement 
in our knowledge of RGB mass loss for a consensus on a variety of problems involving 
HB morphology to be achieved \citep[e.g.,][]{r73,r98,rc89,rrea98}. 
Unfortunately, even though more than three decades have passed since a very 
convenient mass loss formula was advanced by \citet{r75a,r75b}, his expression 
is still widely used as a ``law,'' even though it is now known that the Reimers 
mass loss formula does not properly account for the mass loss rates that have 
become available in the more recent literature. In fact, several other formulae 
have been proposed that do provide a better description of these empirical mass 
loss rates, some of which are summarized in Table~\ref{tb:DM}. Additional 
formulations and references are provided in \citet{loea07}, \citet{sc07}, and 
\citet{mcdvl07}. \citet{mbea08} provide a critical discussion of recent results
based on {\em Spitzer Space Telescope} infrared observations of Galactic globular
clusters, and caution that some of the results by \citeauthor{loea07} may
actually be due to blends in their infrared data.

\begin{figure}[t]
  \plotone{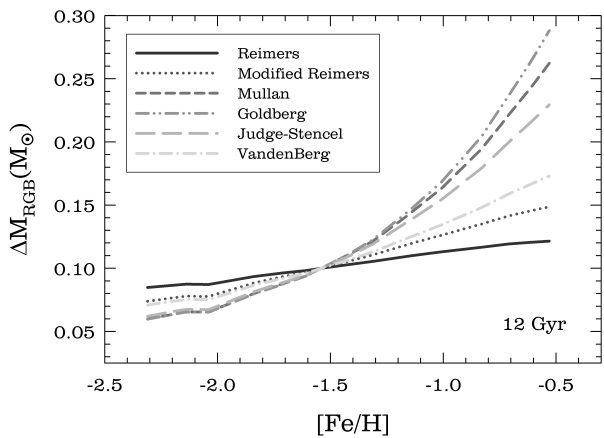}
  \caption{Variation in integrated mass loss along the RGB with metallicity, for an
  age of 12~Gyr and several different prescriptions for mass loss 
  (see Table~\ref{tb:DM}). For all equations,
  mass loss values were normalized to a value of $\Delta M_{\rm RGB} = 0.10\, M_{\odot}$
  at ${\rm [Fe/H]} = -1.5$. 
  \label{fig:DM}
}
\end{figure}

Figure~\ref{fig:DM} shows the variation in the integrated mass loss along the RGB, 
on the basis of different prescription for the mass loss rate, 
for an age of 12~Gyr and several different metallicities (see Sect.~\ref{sec:MC}
below for more information on the RGB models used in these calculations, and 
\citeauthor{mc09} \citeyear{mc09} for an extension to the 
\citeauthor{loea07} \citeyear{loea07} and \citeauthor{sc07} \citeyear{sc07}
cases). Clearly, the metallicity dependence differs 
substantially among the different expressions. This can have important 
consequences for the prediction of the temperature distribution of hot HB 
stars in metal-rich systems, with implications not only for our understanding
of the origin and nature of hot HB stars in the Galactic bulge 
\citep[e.g.,][]{gbea05,bm08}, but also for our modelling and 
interpretation of the UV-upturn phenomenon in elliptical galaxies and  
bulges of spirals (e.g., \citeauthor*{ehea92} \citeyear{ehea92}; 
\citeauthor*{abea94} \citeyear{abea94}; 
\citeauthor{bdea95} \citeyear{bdea95}; 
\citeauthor*{syea97} \citeyear{syea97}; \citeauthor{syea99} \citeyear{syea99}), 
as well as for our understanding 
of the relatively red HB morphologies 
of the most metal-poor globular clusters 
in our galaxy (see Fig.~\ref{fig:HBR} below). 
In Sect.~\ref{sec:TPR}, we 
shall also address the possible impact of several different mass loss formulae
upon analyses of specific pairs of second-parameter globular clusters. 

On the other hand, the reader should bear in mind that it remains very 
unclear at present whether {\em any} analytical 
formula can be reliably used to compute the mass loss in red giant stars. 
For instance, \citet{loea02} find, 
using ISOCAM near-IR data for six globular clusters with a range in metallicities, 
that: i)~None of the formulae given in Table~\ref{tb:DM}
reproduces their derived mass 
loss rates, the latter being in general larger than the former by at least one 
order of magnitude; ii)~There is no clear dependence between mass loss 
rate and any of the basic stellar parameters $L$, $g$, or $R$; 
iii)~Mass loss takes place near the RGB tip only (more specifically, within the 
very final $10^6$ years of evolution on the RGB---compare with the timescale 
shown in Fig.~\ref{fig:TRHOM} below), and is not 
constant but rather episodic; iv)~The mass loss episodes must last longer than a 
few days, but less than $10^6$~yr; v)~There is no clear correlation between mass 
loss and metallicity. We will discuss the implications of their results for 
the evolutionary interpretation of the second parameter problem in 
Sect.~\ref{sec:TPR}; see also Sect.~\ref{sec:ORIG} below for a discussion of 
the impact of different mass loss recipes upon theoretical isochrones in the 
$\LZ - {\rm [Fe/H]}$ plane. 

Assuming their results are able to withstand the test of time \citep[see][ for 
some recently raised caveats]{mbea08},
how does one reconcile the \citet{loea02} study with the empirical results upon 
which the formulae in Table~\ref{tb:DM} are based? The answer to this question 
is not clear at present, but it is possible that the latter provide a better 
description of mass loss on the AGB, the former being instead more suitable 
for first-ascent giants. The connection between mass loss and stellar 
variability, which has frequently been addressed in the literature 
\citep[e.g.,][]{wb84,gb88,lw88,w00,rj01,lw05}, may provide the key to  
the riddle. \citet{tlea05} concluded that, at a similar luminosity, 
red giants with higher pulsation amplitudes present higher mass loss 
rates. At the same time, while variability in AGB stars (which become 
more easily enshrouded in dust) has long been established, 
that in first-ascent red giants is a fairly recent result \citep{yiea02}. 
Importantly, it appears rather clear now that AGB and RGB stars have 
different pulsation properties \citep[e.g.,][]{kb04,isea04}, first-ascent 
variables having smaller pulsation amplitudes and falling on a different
period-luminosity relation than AGB variables. On the other  
hand, \citeauthor{isea04} suggest, based on the coincidence of the RGB 
pulsators belonging to the LMC, the SMC, and the Galactic bulge in a 
\citet{jp73} (or period ratio vs. period) diagram, that RGB pulsators with 
different metallicities do not have significantly different pulsation 
properties. If confirmed, these results may provide very important 
constraints on the extent to which mass loss on the RGB may vary with 
metallicity. 


\section{The Oosterhoff Dichotomy: Constraints on the Galaxy's 
         Formation History}\label{sec:OO}

\subsection{The Oosterhoff Dichotomy: Systematics}\label{sec:SYS}

It has recently been argued that the \citet{o39,o44} dichotomy may very 
well hold the key to the formation history of the Galactic halo. The 
argument goes as follows: the Galactic halo shows a sharp division 
between Oosterhoff type I (OoI, average periods of the ab-type RR Lyrae 
variables $\langle P_{\rm ab} \rangle \approx 0.55$~d) and Oosterhoff 
type II (OoII, with $\langle P_{\rm ab} \rangle \approx 0.65$~d) globular 
clusters, with very few clusters in the range 
$0.58 \leq \langle P_{\rm ab}{\rm (d)} \rangle \leq 0.62$ 
(the ``Oosterhoff gap'').\footnote{The question whether the 
Galactic halo {\em field} 
also presents the Oosterhoff dichotomy, as found by \citet*{nsea91}, or 
not, as suggested by the QUEST \citep{vz03} and ROTSE \citep{kk04} 
surveys \citep[see also][]{as06}, 
remains an open issue at present \citep[see also][]{c04b}. 
We note, however, that inspection of Figure~9 in 
\citet{avea04} reveals that many of the candidate Oosterhoff-intermediate 
stars in the QUEST survey (e.g., their stars 1, 9, 27, 95) have extremely 
uncertain amplitudes, thus rendering their position in the Bailey 
diagram---and thus their Oosterhoff status---similarly uncertain
\citep{mc06}. The 
recently released results of the LONEOS-I survey also strongly support 
the reality of the Oosterhoff dichotomy among field halo stars: in this  
sense, Figures~19 and 20 in \citet{amea08} constitute especially striking 
evidence.} On the other hand, 
the dwarf spheroidal (dSph) satellite galaxies of the Milky Way, as well 
as their respective globular clusters, fall {\em preferentially} on the 
``Oosterhoff gap'' region. One of the main scenarios for the formation 
of the Galactic halo envisages the build-up of the halo from the accretion 
of smaller ``protogalactic fragments'' not unlike the present-day Milky 
Way dSph satellite galaxies \citep[e.g.,][]{sz78,z93}. However, if this 
were the case, the present-day halo should {\em not} display the Oosterhoff 
dichotomy, since the dSph galaxies and their globular clusters are 
predominantly intermediate between the two Oosterhoff classes. Therefore, 
the Galactic halo cannot have been assembled by the accretion of dwarf 
galaxies resembling the present-day Milky Way satellites---including, 
in fact, the LMC dIrr \citep{c04b}. 

One criticism that might perhaps be drawn against this argument is related 
to the fact that not too many globular clusters satisfied the fairly 
strict selection criteria established by \citet{c04b}, which restricted 
his sample to globular clusters containing at least 10 known RRab variables
with measured periods. 
With this selection criterion, \citeauthor{c04b} found that 19 globular 
clusters were OoI, 9 were OoII, and two (or 6.7\%) were 
Oosterhoff-intermediate. Two additional globular clusters, NGC~6388 and 
NGC~6441, were tentatively assigned to a new class, OoIII 
\citep{pea00,pea01,pea02,pea03}. As one can easily see, only a relatively 
small fraction of the Galactic globular clusters, of order 20\%, were 
included. 

\begin{figure*}[ht]
  \includegraphics[scale=0.56]{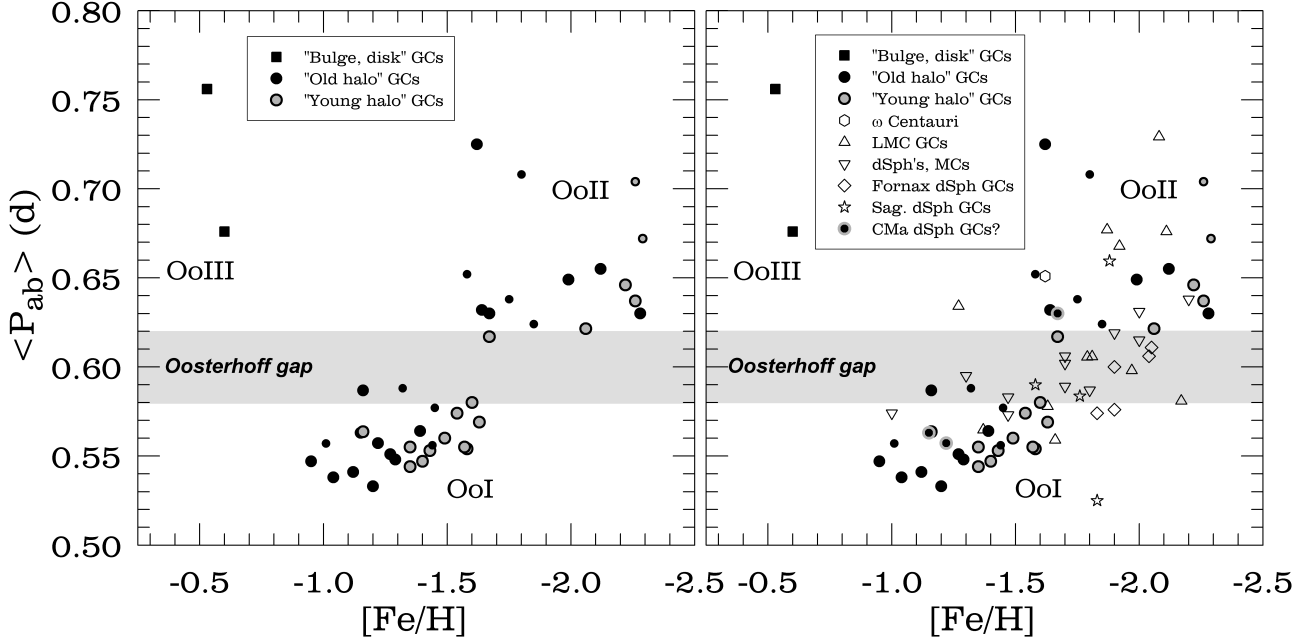} 
  \caption{({\em Left panel}) The Oosterhoff dichotomy among Galactic 
  globular clusters. 
  Clusters belonging to the bulge or disk are shown as squares; those  
  belonging to the ``young halo'' in the \citet{mvdb05} classification 
  scheme are shown as filled gray circles; and clusters belonging to 
  their ``old halo'' are shown as filled black circles. Clusters with 
  at least 10 RRab stars with measured periods are shown as large 
  symbols, while those with 5 to 9 RRab's are shown with smaller 
  symbols. 
  ({\em Right panel}) Same as in the previous plot, but now including dwarf 
  satellite 
  galaxies of the Milky Way and their associated globular clusters.
  The dSph galaxies and the Magellanic 
  Clouds are shown as open inverted triangles; LMC globulars are shown 
  as open triangles; globular clusters associated with the Fornax dSph 
  are shown as open losanges; those associated with the Sagittarius dSph 
  are shown as open stars; possible Canis Major dSph globulars are shown 
  as black circles with gray contours; 
  and $\omega$~Cen is shown as an open 
  hexagon. It is obvious from this plot that the dwarf galaxies orbiting 
  the Milky Way and their associated globular clusters preferentially 
  {\em occupy} the Oosterhoff gap region, in stark contrast with the 
  Galactic globular clusters.\label{fig:OO}
}
\end{figure*}

In order to improve the statistics, we add to the sample originally 
studied by \citet{c04b} all Galactic globular clusters with at least 5 
known RRab variables with measured periods. The result is shown in 
Figure~\ref{fig:OO} ({\em left panel}), 
where clusters with 5 to 10 RRab's are shown with smaller 
symbols than those with 10 or more fundamental-mode variables. 
While the main source is the extensive compilation and catalog of 
variable stars in Galactic globular clusters by 
\citet{cea01},\footnote{\texttt{http://www.astro.utoronto.ca/$\sim$cclement/read.html}} 
this plot includes several updates compared to similar previous plots, 
such as the new measurements by \citet{psea05} for NGC~362, 
by \citet{cea05} for NGC~6266 (M62), and by \citet{tmcea05} for 
NGC~6388 and NGC~6441. 

In addition, we 
separate the clusters into ``bulge/ disk,'' ``young halo'' and ``old halo'' 
subsystems, following the classification scheme by \citet{mvdb05} (based 
on the positions of globular clusters in the HB morphology-metallicity 
plane). Except for the peculiar positions of the clusters labeled as 
OoIII in this plot, it is very clear that both the ``young'' and the 
``old'' halo components present the Oosterhoff dichotomy, perhaps the 
only apparent systematic difference between the two being 
a systematic shift of the ``young'' clusters towards metallicities 
that are lower than for the ``old'' clusters, by $\approx 0.25$~dex. 
Importantly, the sample size is now significantly larger, thus 
clearly improving the statistics: we find that, out of a total of 
41 Galactic globulars with periods measured for at least 5 RRab stars, 
only 4 (or less than 10\%) are Oosterhoff-intermediate.

In Figure~\ref{fig:OO} ({\em right panel}), 
dwarf satellite galaxies of the Milky Way 
and their respective globular cluster systems are added to the previous 
plot. Shown are not only the LMC, SMC, Sagittarius dSph and Fornax dSph 
systems (along with their globular clusters), but also the recently 
discovered Canis Major dSph and its suggested globular cluster system 
\citep{nfmea04}---the galaxy itself (if it indeed exists) 
appears to be extremely poor in 
RR Lyrae variables \citep*{kea04}---as well as $\omega$~Centauri.
The latter, according to many 
authors, is also the remnant of a dwarf galaxy that was accreted by the 
main body of the Milky Way long ago 
\citep[e.g.,][; and references therein]{jnea96,dd02,aea05,mea05,dcc08}. 
This plot dramatically illustrates the contrast between the Oosterhoff 
behavior of Galactic globular clusters and the systems associated with 
the dwarf satellite galaxies of the Milky Way. 
Note that the ``external'' systems shown in 
the figure fall more naturally along the trend defined by the ``young 
halo'' globular clusters---which, as already pointed out, are shifted 
with respect to the ``old halo'' globulars by about 0.25~dex, towards 
lower [Fe/H].

The globular cluster 
data used in Figure~~\ref{fig:OO} are given in Table~\ref{tb:MW} 
(Galactic globulars) 
and Table~\ref{tb:DSPH} (globular clusters associated with the dwarf 
galaxy satellites of the Milky Way). The information contained therein 
comes from the following sources: 

\begin{itemize}
\item For the Fornax dSph globular clusters 3, 4, and 5, we adopt the 
$\langle P_{\rm ab}\rangle$ values from 
\citet[][; see also \citeauthor{cgea07} \citeyear{cgea07} for a recent, 
detailed discussion of the case of Fornax~4]{cgea05}. For clusters 
1 and 2, we adopt the values from the entry ``RRab with good $P$'' 
in Table~4 of \citet{mg03}. Metallicity values are the ones provided 
by \citet{mg04}. 

\item For globular clusters that have been associated with the Sagittarius 
dSph and which have at least 5 RR Lyrae variables, we have adopted 
$\langle P_{\rm ab}\rangle$ values from \citet*{cea02} (M54~=~NGC~6715), 
\citet{rsea05} (Arp~2, NGC~5634), and \citet{sea05} (NGC~4147). While 
the association of M54 and Arp~2 to the Sagittarius dSph dates back to 
early studies of the system \citep[e.g.,][]{iea95,da95}, that of the 
other two quoted globular clusters has only recently been advanced
(NGC~4147: \citeauthor*{mbea03a} \citeyear{mbea03a}; 
\citeauthor*{mbea03b} \citeyear{mbea03b}; NGC~5634: 
\citeauthor*{mbea02} \citeyear{mbea02}).\footnote{Note that the 
possibility has also been raised that the massive 
outer-halo globular cluster NGC~2419 has once been 
associated with the Sagittarius dSph \citep{hnea03}, 
or perhaps been the stripped 
nucleus of a dwarf galaxy \citep{vm04}---but we 
consider it a bona fide Galactic globular cluster in the present analysis
\citep[see also][]{vrea07}.
There are a few additional clusters whose association with Sagittarius 
has been proposed \citep*[e.g.,][]{ddea01,cpea02,mea03,gcea07}. 
However, it is well known 
that the Fornax dSph, with its five globular clusters, has an anomalously 
high globular cluster specific frequency \citep[e.g.,][ and references 
therein]{vdb98}; accordingly, Sagittarius, with a total luminosity 
that seems comparable to that of Fornax \citep{vdb00,mea03}, would have 
an even more anomalous specific frequency of globular clusters if all these 
candidates (in addition to the five fairly clear-cut cases of M54, Arp~2, 
Ter~7, Ter~8, and Pal~12) were indeed associated to it. 
Therefore, caution recommends not to attribute final membership status 
to all the new candidates before their association to the galaxy has 
been conclusively established. 
Likewise, we do not include NGC~6388 and NGC~6441 as extragalactic 
globular clusters, in spite of their suggested association with dwarf
galaxies that were long ago captured by the Milky Way \citep{crea02}.
Finally, we note that \citet*{ywlea07} recently raised the intriguing 
possibility that pretty much {\em all} globular clusters with extended 
HBs are of extragalactic origin (being the former nuclei of dwarf 
galaxies), unlike most other globulars in the halo; the latter would 
be more consistent with a dissipational collapse scenario.} 
NGC~5634 is an interesting case, since the preliminary results of 
\citet{rsea05} indicate that, in spite of a $\langle P_{\rm ab}\rangle$ 
value indicative of OoII status, its Bailey (or period-amplitude) 
diagram\footnote{The widespread usage of the term ``Bailey diagram'' to 
refer to the period-amplitude diagram appears to be relatively recent: 
for instance, it is not used either in 
\citeauthor{as58}'s (\citeyear{as58}) {\em Vatican Conference} 
paper or in \citeauthor{hs95}'s (\citeyear{hs95}) monograph. 
To be sure, \citet{sb13,sb19} did produce plots showing, 
among many other quantities, ``range'' (i.e., amplitude) vs. period (his 
Figs.~4 and 7, respectively), but no great emphasis appears to have been 
placed by him on this specific diagram, other than noting that ``there 
appears to be... a fairly well marked relation between the period and the 
maximum magnitude, the mimimum magnitude, and the range of variation'' 
\citep[][, his p.~87]{sb13}.}  
suggests 
instead that the cluster is Oosterhoff-intermediate \citep[see][ for a 
discussion of the importance of the Bailey diagram in defining Oosterhoff 
status]{c04b}. 
Metallicity values are from the \citet{h96} 
catalog\footnote{{\tiny\texttt{http://www.physics.mcmaster.ca/Globular.html}}}
(Feb. 2003 update).

\begin{table*}[!htbp]
\caption{Galactic Globular Clusters with at Least 5 Known RRab Variables\label{tb:MW}}
\smallskip
\begin{center}
{\small
\begin{tabular}{cccccccc}
\tableline
\noalign{\smallskip}
Name   &   Other  & [Fe/H]  & $\langle P_{\rm ab}\rangle$ & $N_{\rm ab}$ & Type & $\LZ$ & Population\tablenotemark{a} \\
       &          &         &             (d)             &              &      &   & \\
\noalign{\smallskip}
\tableline
\noalign{\smallskip}
NGC 362  &        & $-1.16$ & 0.564 &  13 & OoI     & $-0.87$ & YH \\
NGC 1261 &        & $-1.35$ & 0.555 &  13 & OoI     & $-0.71$ & YH \\
NGC 2419 &        & $-2.12$ & 0.655 &  24 & OoII    & $+0.86$ & OH \\
NGC 3201 &        & $-1.58$ & 0.554 &  72 & OoI     & $+0.08$ & YH \\
NGC 4590 & M68    & $-2.06$ & 0.622 &  14 & OoII    & $+0.17$ & YH \\
NGC 4833 &        & $-1.88$ & 0.708 &   7 & OoII    & $+0.93$ & OH \\
NGC 5024 & M53    & $-1.99$ & 0.649 &  29 & OoII    & $+0.81$ & OH \\
NGC 5053 &        & $-2.29$ & 0.672 &   5 & OoII    & $+0.52$ & YH \\
NGC 5139 & $\omega$ Cen  & $-1.62$ & 0.651 &  76 & OoII  & $+0.89$ & $\omega$C \\
NGC 5272 & M3     & $-1.57$ & 0.555 & 145 & OoI    & $+0.18$ & YH \\
NGC 5466 &        & $-2.22$ & 0.646 &  13 & OoII   & $+0.58$ & YH \\
NGC 5824 &        & $-1.85$ & 0.624 &   7 & OoII   & $+0.79$ & OH \\
NGC 5904 & M5     & $-1.27$ & 0.551 &  91 & OoI    & $+0.31$ & OH \\
NGC 5986 &        & $-1.58$ & 0.652 &   7 & OoII    & $+0.97$ & OH \\
NGC 6121 & M4     & $-1.20$ & 0.553 &  31 & OoI     & $-0.06$ & OH \\
NGC 6171 & M107   & $-1.04$ & 0.538 &  15 & OoI    & $-0.73$ & OH \\
NGC 6229 &        & $-1.43$ & 0.553 &  30 & OoI     & $+0.24$ & YH \\
NGC 6266 & M62    & $-1.29$ & 0.548 & 131 & OoI     & $+0.55$ & OH \\
NGC 6284 &        & $-1.32$ & 0.588 &   6 & Oo-Int  & $+0.88$ & OH \\
NGC 6333 & M9     & $-1.75$ & 0.638 &   8 & OoII    & $+0.87$ & OH \\
NGC 6341 & M92    & $-2.28$ & 0.630 &  11 & OoII    & $+0.91$ & OH \\
NGC 6362 &        & $-0.95$ & 0.547 &  18 & OoI     & $-0.58$ & OH \\
NGC 6388 &        & $-0.60$ & 0.676 &   9 & OoIII   & $-0.69$ & BD \\
NGC 6402 & M14    & $-1.39$ & 0.564 &  39 & OoI    & $+0.65$ & OH \\
NGC 6426 &        & $-2.26$ & 0.704 &   9 & OoII    & $+0.58$ & YH \\
NGC 6441 &        & $-0.53$ & 0.756 &  43 & OoIII   & $-0.73$ & BD \\
NGC 6558 &        & $-1.44$ & 0.556 &   6 & OoI     & $+0.70$ & OH \\
NGC 6584 &        & $-1.49$ & 0.560 &  34 & OoI     & $-0.15$ & YH \\
NGC 6626 & M28    & $-1.45$ & 0.577 &   8 & OoI    & $+0.90$ & OH \\
NGC 6642 &        & $-1.35$ & 0.544 &  10 & OoI     & $-0.04$ & YH \\
NGC 6656 & M22    & $-1.64$ & 0.632 &  10 & OoII   & $+0.91$ & OH \\
NGC 6712 &        & $-1.01$ & 0.557 &   7 & OoI     & $-0.62$ & OH \\
NGC 6723 &        & $-1.12$ & 0.541 &  23 & OoI     & $-0.08$ & OH \\
NGC 6864 & M75    & $-1.16$ & 0.587 &  25 & Oo-Int  & $-0.07$ & OH \\
NGC 6934 &        & $-1.54$ & 0.574 &  68 & OoI    & $+0.25$ & YH \\
NGC 6981 & M72    & $-1.40$ & 0.547 &  24 & OoI     & $+0.14$ & YH \\
NGC 7006 &        & $-1.63$ & 0.569 &  53 & OoI    & $-0.28$ & YH \\
NGC 7078 & M15    & $-2.26$ & 0.637 &  39 & OoII   & $+0.67$ & YH \\
NGC 7089 & M2     & $-1.62$ & 0.725 &  17 & OoII   & $+0.92$ & OH \\
IC 4499  &        & $-1.60$ & 0.580 &  63 & Oo-Int &  $+0.11$ & YH \\
Ruprecht 106 &    & $-1.67$ & 0.617 &  13 & Oo-Int &  $-0.82$ & YH \\
\tableline
\end{tabular}
}
\tablenotetext{a}{YH = ``Young Halo''; OH = ``Old Halo''; BD = ``Bulge/Disk''; $\omega$C = $\omega$~Centauri.
                  From \citet{mvdb05}.}
\end{center}
\end{table*}

\begin{table*}[!htbp]
\caption{Milky Way Dwarf Galaxy Satellite Globular Clusters with at Least 
         5 RRab Variables\label{tb:DSPH}}
\smallskip
\begin{center}
{\small
\begin{tabular}{cccccccc}
\tableline
\noalign{\smallskip}
Name   &   Other  & [Fe/H]  & $\langle P_{\rm ab}\rangle$ & $N_{\rm ab}$ & Type & $\LZ$ & Population \\
       &          &         &             (d)             &              &      &       & \\
\noalign{\smallskip}
\tableline
\noalign{\smallskip}
\multicolumn{8}{c}{\em LMC Globular Clusters} \\
\tableline
\noalign{\smallskip}
NGC 1466 &        & $-2.17$ & 0.581 &  19 & Oo-Int &  $+0.42$ & LMC \\
NGC 1786 &        & $-1.87$ & 0.677 &  17 & OoII   &  $+0.39$ & LMC \\
NGC 1835 &        & $-1.79$ & 0.606 &  55 & Oo-Int &  $+0.57$ & LMC \\
NGC 1841 &        & $-2.11$ & 0.676 &  18 & OoII   &  $+0.71$ & LMC \\
NGC 1898 &        & $-1.37$ & 0.565 &  17 & OoI    &  $+0.03$ & LMC \\
NGC 1916 &        & $-2.08$ & 0.729 &   6 & OoII   &  $+0.97$ & LMC \\
NGC 1928 &        & $-1.27$ & 0.634 &   6 & OoII   &  $+0.94$ & LMC \\
NGC 2005 &        & $-1.92$ & 0.668 &   6 & OoII   &  $+0.90$ & LMC \\
NGC 2019 &        & $-1.81$ & 0.606 &  24 & Oo-Int &  $+0.66$ & LMC \\
NGC 2210 &        & $-1.97$ & 0.598 &  13 & Oo-Int &  $+0.65$ & LMC \\
NGC 2257 &        & $-1.63$ & 0.578 &  17 & OoI    &  $+0.42$ & LMC \\
Reticulum  &      & $-1.66$ & 0.559 &  16 & OoI    &  $+0.00$ & LMC \\
\tableline
\noalign{\smallskip}
\multicolumn{8}{c}{\em Fornax dSph Globular Clusters} \\
\tableline
\noalign{\smallskip}
Fornax GC1  &     & $-2.05$ & 0.611 &   5 & Oo-Int & $-0.30$ & Fornax \\
Fornax GC2  &     & $-1.83$ & 0.574 &  15 & OoI    & $+0.50$ & Fornax \\
Fornax GC3  & NGC 1049    & $-2.04$ & 0.606 &  13 & Oo-Int  & $+0.44$ & Fornax \\
Fornax GC4  &     & $-1.90$ & 0.600 &  11 & Oo-Int &  $-0.42$ & Fornax \\
Fornax GC5  &     & $-1.90$ & 0.576 &   7 & OoI    &  $+0.52$ & Fornax \\
\tableline
\noalign{\smallskip}
\multicolumn{8}{c}{\em Sagittarius dSph (Candidate) Globular Clusters} \\
\tableline
\noalign{\smallskip}
NGC 4147 &        & $-1.83$ & 0.525 &  12 & OoI    & $+0.55$ & Sag? \\
NGC 5634 &        & $-1.88$ & 0.660 &  12 & OoII/Int  & $+0.91$ & Sag/OH? \\
NGC 6715 & M54    & $-1.58$ & 0.590 &  55 & Oo-Int &  $+0.54$ & Sag \\
Arp 2    &        & $-1.76$ & 0.584 &   8 & Oo-Int &  $+0.53$ & Sag \\
\tableline
\noalign{\smallskip}
\multicolumn{8}{c}{\em CMa dSph (Candidate) Globular Clusters} \\
\tableline
\noalign{\smallskip}
NGC 1851 &        & $-1.22$ & 0.571 &  21 & OoI    &  $-0.32$ & CMa/OH? \\
NGC 2808 &        & $-1.15$ & 0.563 &  10 & OoI    &  $-0.49$ & CMa/OH? \\
NGC 5286 &        & $-1.67$ & 0.630 &  29 & OoII   &  $+0.80$ & CMa/OH? \\
\tableline
\end{tabular}
}
\end{center}
\end{table*}

\item For globular clusters which have been associated with the CMa dSph
\citep{fea04,nfmea04}, 
the data come from \citet{mcea04} (NGC~2808), \citet{w98} (NGC~1851, with 
the addition of five new confirmed RRab's from Sumerel et al. 2004), and 
Zorotovic et al. (2009, in preparation) (NGC~5286). Metallicity values 
are from the \citet{h96} catalog (Feb. 2003 update). 

\item As to the LMC globular clusters, we retrieved OGLE data 
\citep{isea03} from their online 
catalog,\footnote{{\tiny\texttt{ftp://ftp.astrouw.edu.pl/ogle/ogle2/var$_{-}$stars/lmc/rrlyr/}}} 
and derived mean periods therefrom. The clusters that were included in
their survey turned out to be  
NGC~1835, 
NGC~1898, 
NGC~1916, 
NGC~1928, 
NGC~2005, and 
NGC~2019. 
For NGC~1466, we used the value from \citet{w92b}. For the remaining 
clusters---namely, Reticulum, NGC~1786, NGC~1841, NGC~2210, 
and NGC~2257---we utilized the values summarized in Table~6 of 
\citet{w92a}. Metallicity values for all LMC clusters were taken from 
the recent \citet{mg04} compilation. 

An interesting characteristic found in the OGLE data for NGC~1835 is the 
presence of a bimodal period distribution for the RRab variables, as if 
the cluster actually presented an {\em internal} Oosterhoff dichotomy
\citep[see Fig.~4 in][]{isea03}. A similar phenomenon has recently been 
discovered in the Fornax dSph globular cluster 4 \citep{cgea05}.  
There are preliminary indications (Vidal et al. 2009, in preparation) 
that NGC~6284
may, in spite of the cluster's Oosterhoff-intermediate classification 
indicated by $\langle P_{\rm ab}\rangle$, lack RR Lyrae stars in the 
Oosterhoff gap region as well. 

\item In the special case of the OoIII globular clusters NGC~6388 and 
NGC~6441, the newly revised $\langle P_{\rm ab}\rangle$ values from 
\citet{tmcea05} were adopted. 

\item For the dSph galaxies, the values come from Table~1 in \citet{c04b}. 

\item For the SMC, we adopted the $\langle P_{\rm ab}\rangle$ value from 
\citet{isea02}, and a metallicity that is an average between the 
\citet{sea92} and \citet*{dbea82} values. 

\item For the LMC, the adopted RR Lyrae metallicity represents an average 
between the spectroscopic measurements of \citet{jbea04} and \citet{gea04a}. 
The question regarding the LMC's $\langle P_{\rm ab}\rangle$ value is 
more complicated: as it happens, the MACHO \citep{caea96} and OGLE 
\citep{isea03} surveys provide results that differ by 0.01~d from one 
another, the MACHO result, $\langle P_{\rm ab}\rangle = 0.583$~d, 
placing the galaxy inside the Oosterhoff gap region in 
Figure~\ref{fig:OO}, but 
the OGLE result, $\langle P_{\rm ab}\rangle = 0.573$~d, indicating an 
OoI classification instead. Given the 
large databases used in both studies---5455 RRab's in the case of OGLE 
\citep{isea03} and 6158 in the case of MACHO \citep{aea03}---this 
difference appears to be intrinsic. This could be due to population gradients 
in the LMC, since whereas the OGLE fields are concentrated along the bar 
of the LMC, the MACHO survey includes many LMC halo 
fields.\footnote{For a comparison 
between the two teams' area coverage, see the maps available at 
{\tiny\texttt{http://bulge.princeton.edu/$\sim$ogle/ogle2/rrlyr$_{-}$lmc$_{-}$map.html}} 
(OGLE team) and 
{\tiny\texttt{http://wwwmacho.mcmaster.ca/Systems/Coords/LMC$_{-}$Fields.gif}} 
(MACHO project).} A similar segregation has been suggested to exist in 
our own galaxy, in the sense that OoII (i.e., longer-period) variables 
would be more confined 
to relatively small distances from the Galactic plane, whereas OoI 
variables would lie farther from the plane on 
average\footnote{This presumably  
implies that the HB morphology of the field is bluer near the plane 
than farther away, which is consistent with the results from \citet*{pea91}
and \citet*{kea94}. Interestingly, \citet{lc99} note that the results 
of \citet{l96} for RR Lyrae stars and \citet{wea96} for HB stars in 
general support an accompanying change in kinematic behavior, from 
prograde systemic rotation to retrograde rotation as one moves away 
from the plane.}---only that, in the case of the LMC, the longer-period 
variables would be located preferentially farther away from the bar.  
For our present purposes, we decided to include two separate datapoints 
for the LMC RR Lyrae in Figure~\ref{fig:OO}, to reflect the possible presence 
of two different populations. We note, in addition, that \citet{caea96} 
already called attention to the Oosterhoff-intermediate nature of the 
RR Lyrae period distribution in that galaxy. For a recent discussion of the 
evidence for a genuine halo component in the LMC (as indicated primarily 
by the velocity dispersion of the RR Lyrae), see \citet{dmea03} and 
\citet{jbea04}, but also \citet{cgea04}, \citet{z04}, and \citet{rcea08} 
(and references therein) for possible alternative interpretations. 

\end{itemize}

\begin{figure*}[ht]
  \includegraphics[angle=0,scale=0.56]{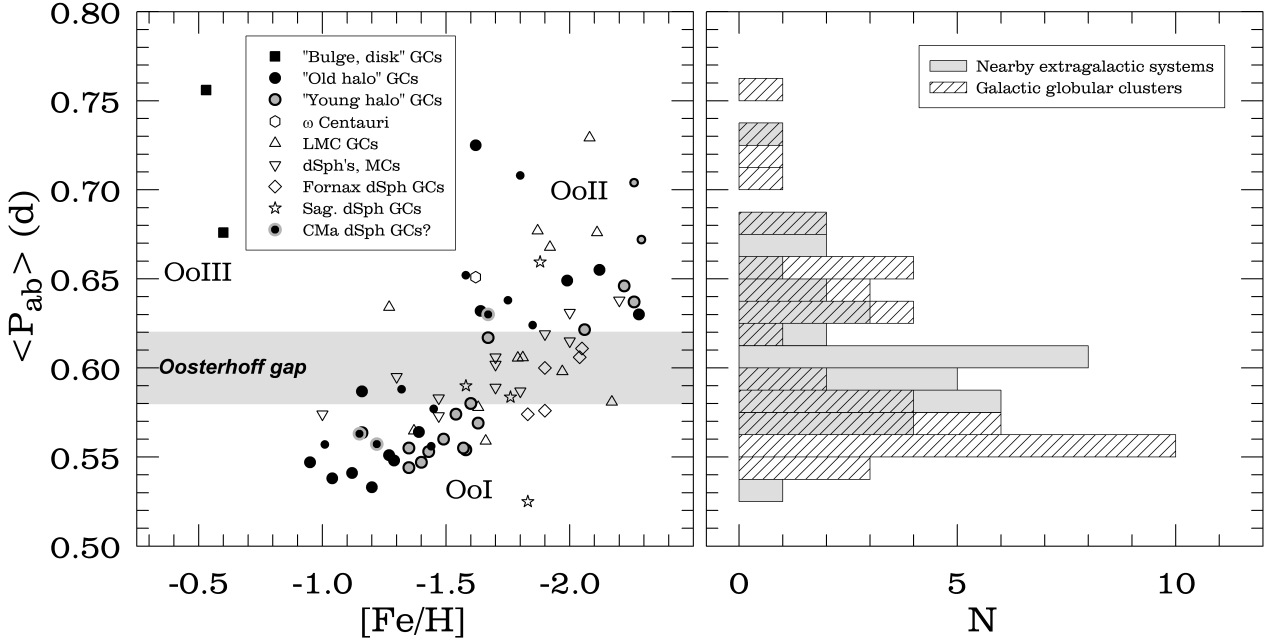} 
  \caption{({\em Left panel}) The right panel of Figure~\ref{fig:OO} is
  repeated here for convenience. 
  ({\em Right panel}) Histogram of $\langle P_{\rm ab}\rangle$ values
  for Galactic globular clusters (hatched bars) and for the nearby 
  extragalactic sysems (gray bars). There is a marked difference between 
  the two histograms, with the objects of extragalactic origin showing a 
  peak at $\langle P_{\rm ab}\rangle \simeq 0.6$~d, which however corresponds 
  to a {\em minimum} in the corresponding Galactic globular cluster 
  distribution.\label{fig:OOH} 
}
\end{figure*}

Figure~\ref{fig:OOH} (right panel) compares the 
$\langle P_{\rm ab}\rangle$ distributions for Galactic globular clusters 
and nearby extragalactic systems. Simple eye inspection clearly reveals that 
the two distributions are remarkably different, with the one for extragalactic 
systems strongly peaking at a $\langle P_{\rm ab}\rangle \simeq 0.6$~d, which 
however corresponds to the {\em minimum} of the Galactic globular cluster 
distribution. A Kolmogorov-Smirnov test confirms that the difference is highly 
significant, with a $D_{\rm KS} = 0.3338$, implying a probability that the two 
sets are derived from the same parent distribution of only $P_{\rm KS} = 1.8\%$. 
Removing the dwarf galaxies (but not their associated globular clusters) from 
the sample changes these figures slightly; we now find $D_{\rm KS} = 0.2856$,
implying a $P_{\rm KS} = 12.5\%$. (These figures do not change significantly if 
we keep only the ``young halo'' component among the Galactic globulars.) 
Removing NGC~6388 and NGC~6441 from the Galactic 
globular cluster sample so that the comparison is limited to globular cluster
systems having comparable metallicities, we find $D_{\rm KS} = 0.3034$,
implying a $P_{\rm KS} = 9.2\%$. Finally, by assuming that $\omega$~Cen, NGC~1851, 
NGC~2808, and NGC~5286 are all bona-fide members of the Galactic globular cluster 
system---which is equivalent to assuming that the CMa dSph does not possess its 
own globular cluster system, which might perhaps be supported by the fact that 
the suggested CMa galaxy appears to be much too young and metal-rich 
\citep*[e.g.,][]{lsea05b,gcea08} to host  
several old and metal-poor globulars---we find $D_{\rm KS} = 0.3714$, with a corresponding $P_{\rm KS} = 2.8\%$. 

In summary, it appears quite clear that the distribution of mean RRab periods 
for the Galactic and nearby extragalactic systems have only a very small 
probability of being compatible with one another. 


\subsection{On the Origin of the Oosterhoff Dichotomy}\label{sec:ORIG}

\begin{figure*}[ht]
  \includegraphics[angle=0,scale=0.56]{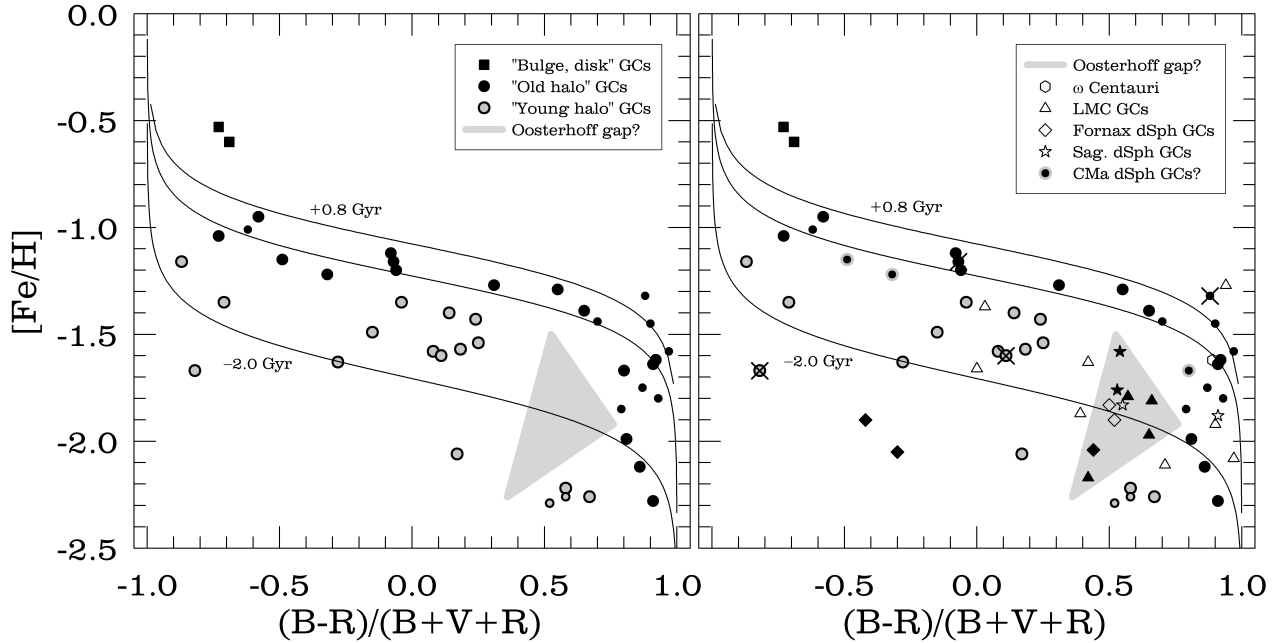} 
  \caption{({\em Left panel}) Position of the Galactic globular clusters 
  with a defined Oosterhoff type in the metallicity--``HB type'' plane. 
  The symbols are the same as in Figure~\ref{fig:OO}. The region marked as a triangle
  and termed ``Oosterhoff gap?'' represents a seemingly ``forbidden region'' 
  for bona-fide Galactic globular clusters. The overplotted lines are 
  isochrones from \citet{cfp93}. ({\em Right panel}) To the previous plot 
  the position of the globular clusters which have been associated with 
  dwarf satellite galaxies of the Milky Way are added. Only globulars 
  with determined HB types are shown. Filled symbols for the extragalactic 
  systems indicate an {\em Oosterhoff-intermediate} status. Of the 9 
  extragalactic globulars classified as Oosterhoff-intermediate, 7 fall 
  inside the triangular region, which we tentatively call ``Oosterhoff gap'' 
  region in analogy with Figure~\ref{fig:OO}. Bona-fide Galactic globular 
  clusters with Oosterhoff-intermediate status are marked with crosses.\label{fig:HBR}
}  
\end{figure*}

In Figure~\ref{fig:HBR}, we show the variation in the $\LZ$ parameter with 
metallicity for Galactic globular clusters. 
To produce this figure, the main source of HB morphology parameters was 
the compilation by \citet{mvdb05} for Galactic globulars, and the one 
by \citet{mg04} for nearby extralactic
globular clusters. The [Fe/H] values came 
from the Feb.~2003 edition of the \citet{h96} catalog. We note, however, 
that some of the values provided in the \citeauthor{mvdb05} compilation 
appear to be incorrect. 
For NGC~6388 and NGC~6441, in particular, they quote values of $\LZ = -1.0$, 
which cannot be correct given the well-known presence of extended blue HB 
components \citep{gpea97,mrea97} and RR Lyrae variables 
\citep{nsea94,lea99,pea00,
pea01,pea02,pea03,tmcea05} in both clusters. Accordingly, we computed 
new $\LZ$ values for these globulars, from the photometry 
provided in the quoted papers and in \citet{gpea02}. Number counts for 
non-variable stars in the \citeauthor{gpea02} study were kindly provided 
by M. Zoccali (2003, priv. comm.), whereas the number counts for RR Lyrae 
variables were taken from the \citeauthor{pea01} and \citeauthor{tmcea05}
studies. We obtain $\LZ \simeq -0.69$ for NGC~6388, and $\LZ \simeq -0.73$ 
for NGC~6441. 

In the case of M3, \citet*{cea01b} and \citet*{mcea02b} have 
shown that the HB morphology is not constant with radius, the HB type 
closer to the center being significantly bluer than the HB type farther 
away. Previous values of the $\LZ$ parameter for the cluster tended to 
be based on photographic measurements that were carried out for the outer 
parts, so that the HB type was accordingly underestimated. When properly 
taking into account HB stars from all radial regions in the cluster in 
the number counts, a value $\LZ = 0.18$ obtains \citep{c04a}---which 
is about 0.1 bluer than provided in the available compilations 
\citep[e.g.,][]{h96,mvdb05}. 

$\omega$~Centauri (NGC~5139) lacks an HB type measurement in both the 
\citet{h96} and \citet{mvdb05} compilations. We accordingly adopt the 
HB type provided by \citet{bm00}, namely $\LZ = 0.89$. 

In Figure~\ref{fig:HBR} are also displayed isochrones based on the 
work of \citet{cfp93}. These are practically identical to the ones 
presented in their Figure~1a, the difference in age with respect to the 
middle (reference) isochrone being $+0.8$~Gyr (upper isochrone) and 
$-2.0$~Gyr (lower isochrone).\footnote{Note, however, 
that those authors also pointed out that there is a ``degeneracy 
effect'' associated with the ages of these isochrones, since several 
different ages can provide morphologically indistinguishable isochrones, 
depending on the assumptions regarding the amount of mass loss on the 
RGB and the chemical composition. Conversely, different assumptions 
for the chemical composition and RGB mass loss can be adopted to mimic 
a favored relative age scale on the $\LZ - {\rm [Fe/H]}$ plane 
[which is actually the reason why this particular set of 
isochrones was chosen in the present study: it appears to 
provide a reasonable analog of Fig.~9b in \citet{screa01} in the 
scenario in which mass loss does not depend on metallicity
\citep{loea02}]. Accordingly, we expect that different ``age labels'' 
may be assigned to these same isochrones in the future, reflecting 
(for instance) changes in the absolute age and/or RGB mass loss scales 
for Galactic globular clusters. Note also that the WMAP and SDSS 
results on the age of the Universe \citep{dsea03,dsea07,mtea04,mtea06} 
strongly constrain the range of possible solutions among those discussed 
by \citet{cfp93}.} These isochrones were computed assuming that the 
total mass loss on the RGB does not depend on metallicity, as recently 
suggested by \citet{loea02}. On the other hand, an RGB mass loss 
dependence on metallicity, as suggested by the expressions 
shown in Table~\ref{tb:DM} (see Fig.~\ref{fig:DM}), 
would lead to somewhat distorted isochrones 
in the $\LZ - {\rm [Fe/H]}$ plane, predicting redder HB types at lower 
[Fe/H] (and vice-versa) than would be the case for a similar, 
constant-$\Delta M$ isochrone. 
This effect is clearly shown in Fig.~9 of 
\citet{screa01}, which compares isochrones computed with constant mass 
loss and isochrones computed assuming a mass loss rate as given by the 
\citet{r75a,r75b} mass loss formula.   

Note that these isochrones can be well described by a 
{\em modified Fermi-Dirac profile} as follows: 

\begin{equation}
  \LZ = \frac{2}{1 + \gamma \, \exp\left(\frac{{\rm [Fe/H]}+\alpha}{\beta}\right)} - 1.
  \label{eq:LZ}
\end{equation}

\noindent The middle isochrone provides a reasonable approximation 
to the inner-halo globular clusters, which in fact comprise 
a large fraction of the ``old halo'' globulars shown in 
Figure~\ref{fig:OO}. It is also very similar to the isochrones 
labeled ``$\Delta t = 0.0$'' in \citet{screa01} (their Fig.~9). For 
this isochrone, the 
parameters used are $\alpha = 1.227$, $\beta = 0.130$, and $\gamma = 1$. 
For the lower isochrone, which in \citet{cfp93} corresponded to the LMC 
globular clusters, one has $\alpha = 1.707$, $\beta = 0.140$, and 
$\gamma = 1$; the relative age scale is fairly similar to that in 
Fig.~9b of \citeauthor{screa01}, except (as expected) for the 
most metal-poor clusters with the redder HB types. 
Finally, the upper isochrone represents the RR Lyrae 
stars in the Galactic bulge, and is well described by  
$\alpha = 1.027$, $\beta = 0.130$, and $\gamma = 1$. 

Note that the plots in Figure~\ref{fig:HBR} 
actually use an inverted 
form of eq.~(\ref{eq:LZ}), which we also provide for convenience: 

\begin{equation}
  {\rm [Fe/H]} = \beta \, \ln\left[\left(\frac{2}{1+\LZ}-1\right)\,\frac{1}{\gamma}\right] - \alpha.
  \label{eq:LZ2}
\end{equation} 

Figure~\ref{fig:HBR} ({\em left panel}) 
shows very clearly that there is a {\em zone of 
avoidance} for RR Lyrae-rich globular clusters in our galaxy, defined by a 
rectangle located in the region $-2.18 \la {\rm [Fe/H]} \la -1.5$, 
$0.3 \la \LZ \la 0.75$. However, in the right panel, when globular 
clusters associated with the neighboring galaxies are plotted, one 
sees that this region is actually {\em preferentially occupied} by 
the external globulars. The situation immediately brings to mind  
the similar phenomenon that was found in our analysis of 
Figure~\ref{fig:OO}; in 
that case, the ``zone of avoidance'' for the Galactic globulars---the 
{\em Oosterhoff gap}---was also the zone preferentially 
occupied by the globulars associated with neighboring galaxies. 
Therefore, one cannot help but suspect that the noted ``avoidance 
region'' in the left panel of 
Figure~\ref{fig:HBR} will {\em also} be related to 
the Oosterhoff-intermediate globulars, as was the Oosterhoff gap 
region in Figure~\ref{fig:OO}. Indeed, 
Figure~\ref{fig:HBR} ({\em right panel}) clearly shows that 
nearby extragalactic globulars do not only preferentially fall in 
the quoted region of this diagram: there is in fact a smaller, 
well-defined, triangular-shaped region in the plot where most 
Oosterhoff-intermediate globulars of extragalactic origin 
(7 out of 9) can be 
found. The vertices of this triangle, in the ${\rm [Fe/H]}-\LZ$ 
plane, are given by the following coordinates: $(-1.5, \, 0.525)$; 
$(-1.92, \, 0.76)$; and $(-2.26, \, 0.36)$. 

Note that the four Oosterhoff-intermediate Galactic globulars 
(NGC~6284; M75~= NGC~6864; IC~4499; Ruprecht~106), indicated by large 
``$\times$'' symbols in the right-hand plot, are located very far 
away from the ``Oosterhoff gap'' region in this diagram, and may 
accordingly represent different phenomena, or even be due to statistical 
fluctuations \citep[see also][]{c04b}. Note, in this sense, that 
NGC~6284 has a mere 6 known RRab's \citep{cea01}, {\em none of which} 
with a period that falls in the Oosterhoff gap range (i.e., 0.58--0.62~d); 
that M75 has a multimodal HB; and that the RRab's in Rup~106 (which 
entirely lacks RRc's) all fall very close to the red edge of the instability 
strip. Finally, IC~4499 is at the very edge of the Oosterhoff gap region, 
with a mean ab-type period of exactly 0.580~d, and might as well have 
been classified as OoI. 

\begin{figure}[t]
  \plotone{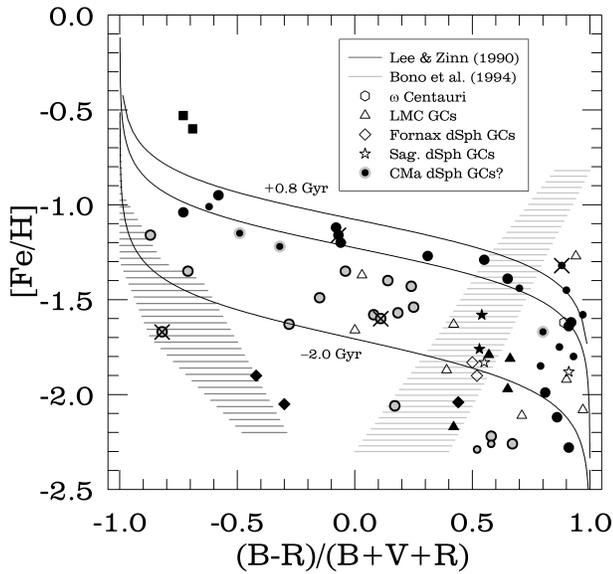} 
  \caption{Same as in the right panel of the previous figure, but now 
  showing, as hatched regions, the predicted Oosterhoff-intermediate 
  region, according to the model calculations by 
  \citet[][, right, in light gray]{bea94} and 
  \citet[][, left, in dark gray]{lz90}. \label{fig:HBRB}
}  
\end{figure}

This scenario provides unprecedented detail about the origin of the 
Oosterhoff dichotomy, going significantly beyond the original and 
insightful early analyses of the problem by \citet{vc83}, \citet{ar83}, 
\citet{lz90}, and \citet*{bea94}---who previously 
suggested that the origin of the Oosterhoff dichotomy among 
Galactic globular clusters was the lack of RR Lyrae-rich globulars (or, 
more specifically, the presence of a majority of clusters with exclusively 
blue HBs) at metallicities intermediate between the bulk of the OoI and 
the OoII clusters. Indeed, theoretical predictions of an 
``Oosterhoff-intermediate area'' in the ${\rm [Fe/H]} - \LZ$
plane have previously been provided by 
\citeauthor{lz90} and \citeauthor{bea94}; 
the results of these studies are summarized in Figure~\ref{fig:HBRB}, 
which shows reasonable success in predicting which globulars should 
be Oosterhoff-intermediate. It is interesting to note that the two 
Oosterhoff-intermediate Fornax dSph globular clusters which lie 
farther away from the triangular-shaped region in Figure~\ref{fig:HBRB}, 
as well as Rup~106, fall very close to the Oosterhoff-intermediate 
region located towards red HB types in this plane, in agreement with
the theoretical predictions by \citeauthor{lz90}. (A region equivalent  
to the area labeled ``\citeauthor{bea94}'' in this plot was also predicted, 
though over a somewhat more limited $\LZ$ range for a given [Fe/H], by 
\citeauthor{lz90}---see their Fig.~2.)

An important question that 
remains unanswered and which will certainly require additional work is 
the following: {\em why} did the Galactic globular cluster system avoid 
the triangular-shaped region in 
Figure~\ref{fig:HBR}, whereas the nearby extragalactic 
globulars were instead preferentially ``attracted'' to it? According to 
Figure~\ref{fig:HBR}, part of the explanation could be that the 
Galactic ``old halo'' and nearby 
extragalactic globular clusters differ in age by $\sim 2$~Gyr (on 
average). However, little (if any) evidence for an age difference 
has been found so far: in the case of the Fornax dSph, \citet{rbea98} 
show that all globular clusters have similar ages, comparable to those 
found in Galactic globular clusters of similar metallicity. The exception 
appears to be cluster 4, which is $\sim 3$~Gyr younger than the 
other Fornax dSph globulars \citep{rbea99}, and which indeed appears to
have the redder HB type (see Table~\ref{tb:DSPH}). 
In the case of the LMC, 
virtually all ``old'' clusters have ages comparable to those of the 
oldest Galactic globular clusters, a point which has been repeatedly 
emphasized in the recent literature 
\citep[e.g.,][]{ebea96,jjea99,jjea02,mg04a}. 
Among the Oosterhoff-intermediate globular clusters in the Sagittarius
dSph, only Arp~2 was previously suggested to be ``young'' (but note 
that it has a predominantly blue HB), although the recent \citet{dv00} 
and \citet{sw02} studies suggest that its age is probably not much 
lower than those of Galactic globular clusters of similar 
metallicity.\footnote{Among the Oosterhoff-intermediate Galactic globular 
clusters, both IC~4499 \citep{ifea95} and Rup~106 \citep{rbea93} were 
indeed previously found to be young globular clusters, although the 
case for a young IC~4499 is not supported by \citet{dv00}, \citet{sw02}, 
or \citet{fdaea05}, 
Rup~106 being instead more consistent, according to these studies, with 
an age slightly lower than found for the bulk of the Galactic globular 
clusters.} Also, \citet{ls00} argue that the Sagittarius dSph globulars 
M54 and Arp~2 (as well as Ter~8) are basically coeval with the bulk 
of the Galactic halo globular clusters. Therefore, it appears difficult, 
given the available age determinations for Oosterhoff-intermediate 
globular clusters, to argue that they differ in age significantly 
with respect to OoII and OoI globulars. In a similar vein, it should 
be noted that the separation of the clusters into ``young halo'' and 
``old halo'' groups by \citet{mvdb05} is somewhat artificial, since 
there is clearly significant overlap in turnoff ages between the two 
groups, as can be seen from Figure~9 in \citet{mg04}: according to 
the quoted figure, the oldest ``young halo'' globular clusters are at 
least as old as the {\em oldest} ``old halo'' clusters, and likewise, 
the youngest ``old halo'' clusters have ages that are similar to those 
of the youngest ``young halo'' clusters. This clearly shows the perils 
of using HB morphology as an age indicator for individual globular 
clusters. We note, in passing, that \citet{fdaea05} find no clear
correlation between the ages of Galactic globular clusters and their 
galactocentric distances.

\subsection{Implications for the Formation History of the Galaxy}

Several authors have suggested that the ``young halo'' Galactic globular 
cluster system at least may have been formed from the accretion of dwarf 
galaxies resembling the present-day dwarf satellites of the Milky Way 
\citep[e.g.,][]{sz78,z93,mvdb05}. However, we have already seen 
(Sect.~\ref{sec:SYS}) that the 
distribution of RR Lyrae periods for the Galactic globulars, including 
the ``young halo'' component, is different from that of those systems. 
It is instructive to perform a similar quantitative comparison between 
the distribution of {\em HB types} 
for the Galactic and nearby extragalactic globular clusters. 

Using only the ``young halo'' component, we find a 
Kolmogorov-Smirnov value of $D_{\rm KS} = 0.5425$, implying a 
probability of only $P_{\rm KS} = 0.2\%$ that the ``young halo'' 
Galactic globulars and the nearby extragalactic globulars were drawn 
from the same parent population. If we remove the LMC globulars from 
the sample (given that many authors suggest that the protogalactic 
fragments that led to the formation of the Milky Way were actually 
dSph-like, not LMC-like), we find $D_{\rm KS} = 0.498$, implying a 
probability of only $P_{\rm KS} = 2.4\%$ that the samples were
drawn from the same parent distribution. Therefore,  
the Kolmogorov-Smirnov test rejects, with high statistical
significance, the hypothesis that the Galactic globular cluster 
system was assembled from the capture 
of protogalactic fragments that resembled the present-day dwarf 
satellite galaxies of the Milky Way---Fornax and Sagittarius in 
particular. 

This does not mean, of course, that there may 
not have been a (now extinct) 
primordial dwarf galaxy population with properties that 
were different from those of the surviving dwarf satellite galaxies 
of the Milky Way. We thus face the task of defining whether our current 
sample of dwarf galaxies is atypical in some respect. For instance, it 
is now well established that the detailed chemical patterns of even the 
more metal-poor populations of the Milky Way dwarf satellite galaxies 
bear little resemblance to that found among most stars in our galaxy 
\citep[e.g.,][]{kvea04}. Note, however, that our satellite galaxies do 
{\em not} appear to be anomalous in this respect, since \citet[][; see 
especially their Fig.~5]{pbea04}
have shown that their abundance patterns are consistent with those for 
objects which are supposed to have given rise to dSph galaxies---namely, 
the Damped Lyman~$\alpha$ (DLA) systems \citep*[e.g.,][]{fmea97,ahea05}. 
Also, the 
globular clusters which were unquestionably accreted by the Milky Way 
from dSph fragments and which have been carefully studied for 
chemical abundance patterns, such as M54 \citep*{jbea99}, Palomar~12 
\citep{jbea97,jc04}, and Terzan~7 \citep{gtea04,lsea05,lsea07}, show 
clearly different abundance patterns with respect to bona-fide Galactic 
globulars. A careful and 
detailed discussion of this point has recently been presented by 
\citet*{bpea05}. In fact, the presence of such peculiar abundance 
patterns, particularly a lowered abundance of the alpha-captured 
elements, has recently been viewed as a signature of a possible 
accretion origin; the reader is referred to the recent spectroscopic 
study of NGC~5694 by \citet*{jwlea06} for a recent example. 
Note, on the other hand, that there are at least 
some outer-halo globular clusters with completely normal abundance 
patterns (\citeauthor{iiea01} \citeyear{iiea01}; \citeauthor{cm05}
\citeyear{cm05}; see also footnote~\ref{foo:M5} below), 
thus suggesting that it is not only the inner 
Galactic halo that contains globular clusters that formed ``in situ.'' 
On the other hand, \citet*{mmea08} have recently claimed 
that Arp~2 and Terzan~8---which are also Sagittarius globulars---both 
have alpha-element abundances that are typical of Galactic halo clusters, 
which certainly confuses the chemical picture somewhat. As far as the 
newly discovered SDSS dSph galaxies \citep[e.g.,][]{vbea06,vbea07}, 
recent surveys have revealed that at least some of them are also 
Oosterhoff intermediate \citep[e.g.,][]{ckea08} and present unusual 
abundance patterns (Frebel et al. 2009, in preparation), although 
seemingly ``regular'' OoII dwarfs are also present 
(\citeauthor{ms06} \citeyear{ms06}; \citeauthor{mdoea06} \citeyear{mdoea06};
\citeauthor{cgea08} \citeyear{cgea08}; Musella et al. 2009, in preparation; 
see also \citeauthor{mc09} \citeyear{mc09}
for a critical discussion). 

In conclusion, 
most of the recent evidence suggests that, even though we seem 
to be witnessing mergers between dSph satellite galaxies and our own 
galaxy ``in real time'' (e.g., Sagittarius, CMa), these are not truly 
representative of the events that led to the formation of the Milky Way, 
and only a relatively minor fraction of the Galactic halo may have been 
assembled from dSph-like protogalactic fragments resembling the 
present-day Milky Way dSph satellites \citep*[see also][ for a recent
discussion]{dgea07}.

\subsubsection{RR Lyrae Stars in M31 and the Origin of the Galactic Halo}\label{sec:M31halo}

It has recently been suggested that the ``young'' second-parameter globular clusters 
in our galaxy (as well as at least some of the Milky Way dSph satellites) were 
accreted from M31 when the latter was forming its Population~II stars
\citep{vk02}. Given 
that our galaxy's globular clusters clearly present the Oosterhoff dichotomy, 
but not so its dSph satellite galaxies, the question naturally presents 
itself: how does M31 classify, in terms of Oosterhoff status? 

Of course, this is a very difficult question to answer, given that 
the detection of RR Lyrae variable stars at the distance of M31 is very difficult 
from the ground. It was originally attempted by \citet{pvdb87} using the CFH 3.6m
telescope. These authors claimed the detection of 30 probable RR Lyrae stars in
their surveyed field, thus suggesting that the M31 halo has a specific 
frequency of RR Lyrae variables comparable to that in RR Lyrae-rich Galactic 
globular clusters. They also reported a mean period for the ab-type 
variables of $\langle P_{\rm ab} \rangle = 0.548$~d, thus classifying the M31 
halo field as OoI. 

More recently, \citet{adea04} performed 
a variability survey, using the WIYN 3.5m telescope, of an M31 halo field that 
includes the \citet{pvdb87} field, finding a specific frequency of RR Lyrae 
variables dramatically lower than in the \citeauthor{pvdb87} study. Using  
time-series observations with ACS onboard the {\em HST}, \citet{tbea04a} 
have shown that, in fact, the specific RR Lyrae frequency of the M31 
halo is intermediate between the two quoted studies. Interestingly, they 
also find a $\langle P_{\rm ab} \rangle = 0.594$~d, which corresponds to the 
Oosterhoff-intermediate domain in Figure~\ref{fig:OO}. 
On the other hand, as we have seen, the dSph satellites of the Milky Way are 
indeed almost exclusively Oosterhoff-intermediate---and so are probably the 
dSph satellites of M31 as well \citep{bpea02a,bpea05a}. Therefore, the 
incorporation of dSph satellites from the M31 system into our own galaxy's 
would not fundamentally change the stellar pulsation properties of the 
Milky Way's dSph system. 

To be sure, extensive variability surveys of not only the M31 halo field
and its dwarf satellite galaxies, but also (and most challenging of all) 
of its {\em globular clusters}, 
are badly needed to place constraints on its old halo's (and indeed, 
as we have just seen, the whole Local Group's) formation history. It would 
be very important to establish whether the Oosterhoff dichotomy is present 
in the M31 globulars, as well as in M31's general field, since 
this would enable a systematic comparison between the very oldest 
stars in the M31 dSph satellites and their peers in the general 
M31 halo, thus posing constraints on the extent to which the latter may 
have been assembled from ``building blocks'' similar to the former.  
Unfortunately, such studies have so far remained rather limited 
\citep{gcea01,rocea08}, given that they require expensive time-series
observations from space. 
Ground-based surveys could in principle be carried out using large telescopes 
equipped with adaptive optics instruments, but unfortunately, this technique 
remains limited to near-infrared observations---a wavelength regime where not 
only the red giants are brightest but also the amplitudes of RR Lyrae stars 
are smallest \citep*[e.g.,][]{alea85,lj90,rjea96}, thus making their detection 
hardest.

\section{The Second-Parameter Problem}\label{sec:TPR}

The {\em second parameter problem} of globular cluster astronomy is commonly  
defined as the existence of globular clusters with very similar metallicity 
\citep[the ``first parameter'';][]{sw60} and yet different HB morphologies. 
It appears to have been first noted by \citeauthor{sw60} (see pages 607 
and 608 in their paper), and later also by 
\citet{sw67} and \citet{vdb67}. The phenomenon has now been recognized to 
exist also in the M31 halo \citep{dmea06}, as well as in the Fornax dSph 
galaxy \citep{rbea98,rbea99}. 

The second parameter candidate that 
was first noted in the literature seems to have been age. We quote 
\citet{sw60}: 

\begin{quotation}
... the character of the 
horizontal branch is spoiled by the two clusters M13 and M22. [Individual 
stars in] M13 appear to be metal-rich, whereas the character of the 
horizontal branch simulates that of the very weak-lined group (M15, M92, 
NGC~5897). [...] M13 is younger than M2 or M5 \citep{a59}. Consequently, 
in addition to chemical composition, the second parameter of age may be 
affecting the correlations...
\end{quotation}

\noindent (Note that the sense of the claimed age 
difference between M13 and other clusters of similar 
metallicity but redder HB type is the {\em opposite} of what 
modern stellar evolution theory indicates to be necessary to account 
for their differences in HB types.) 

The next candidate second parameter was the helium abundance 
\citep{sw67,vdb67}. 
Later on, with the advent of the \citet{sz78} 
scenario for the formation of the Galactic halo---which was based upon 
the hypothesis that age is the second parameter, outer-halo globular 
clusters with predominantly red HBs being younger than inner-halo 
globulars with bluer HBs (see their Fig.~10)---age has quickly become 
the most popular second parameter candidate. 
However, the age hypothesis has proven 
controversial, with different perspectives having been presented on both 
the {\em presence} of significant age differences among Galactic globular 
clusters \citep[e.g.,][; and references therein]{psea96,scd97} 
and the {\em amount} that may be required to account for different 
second-parameter pairs entirely in terms of age 
\citep[e.g.,][]{cfp93,ldz94,ffea97}. In addition, 
other candidate second parameters 
abound in the literature; these include (or are at least related to, and 
often involve combinations of) 
mass loss \citep{rp82,c00}, cluster ellipticity \citep{jn83,jn87}, stellar 
rotation \citep{mg76,fpr78,rp85}, magnetic fields \citep{rs81,vc83,cp84}, 
the cyanogen distribution among red giants \citep{jn81,jnea81,sn83}, 
[CNO/Fe] \citep{rs81}, super-oxygen-poor RGB stars \citep{cfp95}, the Na-O 
correlation among cluster giants \citep{cg96,ecea07}, the 
helium-core mass at the helium flash \citep*{pdea72,sc98}, helium mixing
\citep*{vs88,lh95,as97a,as97b,sc98,cn00,aea01,c01}, planetary systems 
\citep*{ns98,sl99,sh00,sh01,nsea01,ls02,sh07}, 
globular cluster core density or concentration \citep{fpea93,rbea97}, 
and cluster mass \citep{rbea06}. 
A very informative and relatively 
recent review of the several ``second parameters'' that have been 
proposed in the literature is provided by \citet{fpb97}. 

Recently, \citet*{csp04} have argued that neither the deep mixing nor 
the primordial channel are capable of satisfactorily accounting for the 
abundance patterns observed among globular cluster red giant stars. In 
the same vein, \citet{csea04} present an extremely interesting conundrum 
in regard to the extreme abundance anomalies seen in stars close to the 
RGB tip in M13, and whose solution may hold the key to the extent to 
which deep mixing and/or primordial contamination may affect the 
evolution of the stars on the HB: 

\begin{quotation}
The correlation of O and the Mg isotopic abundance ratio suggests
that M13 giants in a very small interval of initial mass preferentially 
underwent severe pollution or ablation or, alternatively, were 
preferentially formed from material secularly ejected earlier from more 
massive cluster giants. These stars only just at this moment arrived 
essentially at the red giant tip, exactly where the effect of deep
mixing, if it indeed exists, would likely be most easily manifest. It 
is a most remarkable coincidence, if indeed that's what it is.
\end{quotation}

\begin{figure}[t]
  \plotone{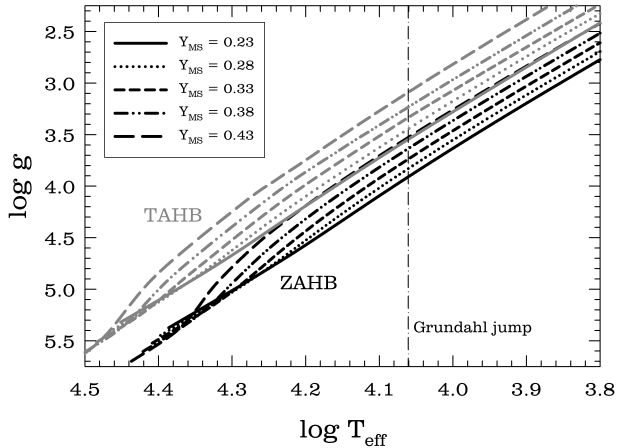} 
  \caption{Effect of an increased helium abundance upon the gravities of 
  HB stars. The lines labeled ``ZAHB'' correspond to the zero-age HB, 
  whereas the lines labeled ``TAHB'' correspond to the terminal-age HB
  (i.e., to core helium exhaustion). The vertical line indicates the 
  temperature of the \citet{gea99} jump.\label{fig:LOGG}
}  
\end{figure}

\subsection{Helium as a Second Parameter: Some Empirical Constraints}\label{sec:HSP}

As mentioned previously, the helium abundance was one of the first 
second parameter candidates to be proposed in the literature. Recently, 
with observations of faint stars in globular clusters having shown that 
many of the abundance patterns observed among bright giants that 
might have been ascribed to deep mixing are also present down to the 
main-sequence level 
\citep[see][ for an extensive review and references]{rgea04b},
thus severely constraining the extent to which 
deep mixing may contribute to the observed patterns among the brighter
giants, the helium abundance scenario has 
regained impetus, now under the label of 
{\em primordial contamination} by a previous generation of 
(likely massive AGB) stars 
\citep[e.g.,][; and references therein]{dea02,fdea05,dac04,dac08,jn04,cd05,cd07,lea05,gbea07,nt07,gpea07,sjyea08}. 
For critical discussions of different theoretical scenarios for the 
formation of helium-enhanced (sub-)populations in globular clusters, 
the reader is referred to the recent papers by 
\citet{bn06}, 
\citet{lc06}, 
\citet{mm06}, 
\citet{pc06}, 
\citet{kbea07}, 
\citet{dv07}, 
\citet{cy07,cy08}, and 
\citet*{tdea07}. 
The earlier paper by \citet[][, see its \S4]{xs95}, which unfortunately 
has received little attention, already contained an interesting discussion 
of several different scenarios that could generate helium-enhanced 
stars in globular clusters. 
In turn, modern views of the deep helium mixing scenario have recently 
been advanced as well, including papers by 
\citet{sf06}, 
\citet{tsea07}, and 
\citet*{syea08}. These authors propose that, 
in the dense environments of globular clusters, 
such deep mixing can indeed take place along the RGB, as triggered by 
tidal interactions. According to them, these phenomena could 
explain not only some level of He enrichment on the HB, but also the 
high rotation velocities that are observed among sufficiently cool 
blue HB stars. The same authors also suggest that cluster main sequence 
stars can likewise get polluted, by encounters with AGB stars 
\citep*{ttea07}. \citet{sh07} also speculate that He mixing can 
take place during the pre-ZAHB phase, triggered by the engulfment of 
low-mass companions and the resulting spin-up of the star's envelope.  

What are the constraints that may be posed on the enhanced helium 
hypothesis, using HB stars? In Figure~\ref{fig:LOGG}, we show the impact 
of an enhanced helium abundance upon the $\log g - \log T_{\rm eff}$ 
diagram for blue HB stars. 
These results are based on the evolutionary 
calculations by \citet{sc98}. As one can clearly see, a large 
increase in the helium abundance is expected to lead to a marked signature 
in this plane, with lower gravities resulting due to the increased 
luminosities that are brought about, with an increase in $Y$, for stars 
with efficient H-burning shells. For stars cooler than the \citet{gea99}
``jump'' (at about 11,500~K; see Sect.~\ref{sec:BHB}), 
comparison between the spectroscopically
derived gravities and the theoretical predictions should be fairly 
straightfoward, the only noteworthy complication being associated with 
the distortion of the Balmer lines that may be effected by the 
predicted presence of a stellar wind \citep{vc02}. 
In the case of M13, an increase in $Y$ up to 0.28 has been suggested 
\citep{cd05}. The spectroscopic results by \citet{smea03} do not 
rule out a $Y_{\rm MS} \simeq 0.28$ for these cooler stars (but note  
again that proper inclusion of stellar winds might bring the derived 
gravities up, and therefore the required $Y$ values down), but appear  
inconsistent with a $Y_{\rm MS} \approx 0.33$ or higher (see their Fig.~8b); 
in addition, they do not favor a difference in $Y$ between M13 and M3, 
although more data would certainly be needed to firmly establish this point.  
Note that the high helium abundance suggested for blue HB stars in NGC~2808 
could also be constrained in this way, since \citet{lea05} suggest that 
even the cooler blue HB stars in this cluster have helium abundances as 
high as $Y_{\rm MS} = 0.33$, which should leave an obvious mark in the 
$\log g - \log T_{\rm eff}$ plane (Fig.~\ref{fig:LOGG}). The helium enhancement 
proposed for the same cluster by \citet{dac04} is significantly smaller, 
being at the level of $Y_{\rm MS} = 0.27$ for the cooler blue HB stars; 
the difference should be testable with a sufficiently large sample of 
stars. The same technique can likely be used to constrain the suggestion 
by \citet{cd08} that even in the case of M3 the blue HB stars may present 
some degree of helium enhancement.  

The comparison between evolutionary model predictions and spectroscopically 
derived gravities for temperatures higher than the Grundahl jump is a much 
more complicated affair, since the 
presence of radiative levitation/gravitational diffusion effects in this 
region require the computation of detailed model atmospheres properly 
taking into account the  
{\em extremely complex} observed abundance patterns \citep*{pbea95,fcea97}
in order for reliable gravities, bolometric corrections and color
transformation to be derived. To the best of our knowledge, no such 
detailed calculations have been carried out yet. 
Accordingly, it remains unclear whether models with an {\em overall} 
enhancement in all metals to super-solar levels (which is not what 
the quoted abundance studies indicate), as commonly employed in the 
literature, are adequate to derive gravities 
of stars hotter than the Grundahl jump.

\subsection{Why Are the Most Metal-Poor Globular Clusters Not the Ones
            with the Bluest HBs?} 

The influence of the ``first parameter'' (metallicity) upon HB morphology can 
be summarized as the observed trend of HB type getting redder as metallicity 
increases \citep{sw60}.\footnote{\citet{fpea93} (see their Sect.~2.1) provide 
an enlightening discussion of the dependence of HB temperature on metallicity 
and RGB mass loss, showing how it can become very difficult for a
metal-rich system to produce blue HB stars, compared to metal-poor systems. 
Note, in addition, that the 
RGB mass increases with metallicity for a fixed age and helium 
abundance [see, e.g., eq.~(1) in \citeauthor{cfp93} \citeyear{cfp93}], 
which also favors the production of redder HB types at higher $Z$.} 
Interestingly though, the 
most metal-poor globular clusters are {\em not} the ones with the bluest 
observed HBs, contrary to what might be expected from basic theory (see 
the isochrones overplotted on Figure~\ref{fig:HBR}).
In fact, the trend of HB type becoming bluer with decreasing 
[Fe/H] appears to be reversed at a ${\rm [Fe/H]} \approx -1.8$ 
(see Fig.~\ref{fig:HBR}, {\em left panel}). The reason 
for such a reversal of trend is not entirely clear; while some have suggested 
that it could be due to a decrease in mass loss eficiency as metallicity 
decreases \citep{r73}, others have suggested instead 
that a majority of the metal-poor globulars may in fact be somewhat 
{\em younger} than Galactic globular clusters of intermediate metallicity 
\citep{yl02}. According to the isochrones shown in Figure~\ref{fig:HBR}, 
the required age difference between the ``old halo'' globular clusters and
the most metal-poor ones with $0.1 \la \LZ \la 0.7$ would have to be fairly 
large. On the other hand, the HB morphology-based age difference would 
decrease using mass loss formulae which imply a strong mass loss dependence 
on metallicity, as indeed favored by the mass formulae listed in 
Table~\ref{tb:DM} (see Fig.~\ref{fig:DM}) but contrary, it should be 
noted, to the recent results by \citet{loea02} (see Sect.~\ref{sec:DM} 
above). In any case, inspection 
of recent papers in which globular cluster ages have been computed provides 
little support for an age difference, except perhaps in the case of 
M68 (NGC~4590) \citep{area99,dv00,sw02,fdaea05}. Interestingly, M68 does 
have a redder HB type than most other Galactic globular clusters of similar 
[Fe/H]---and the possibility that it may be relatively young was raised 
early on by \citet{bcea96}.

\subsection{A Reanalysis of Specific Second-Parameter Pairs}

In what follows, we will readdress 
some classical second-parameter pairs, including  
NGC~288/NGC~362 and M13/M3, with the goal of answering the following 
question: are the measured turnoff age differences between these clusters 
sufficient to explain the observed difference in their HB types? We will 
also briefly discuss the case of the ``young'' outer-halo globular 
clusters Pal~3, Pal~4, and Eridanus, and provide a summary of recent
developments on the second-parameter effect in metal-rich globular 
clusters. Note that an RGB mass loss that presents no dependence on 
$L$, $g$, or $R$, as suggested by \citet{loea02}, would require 
larger age differences than computed in the subsections below in order 
to account for any given second-parameter pair in terms of age.

\subsubsection{The Pair NGC~288--NGC~362} 

This is probably the second-parameter pair which has attracted the most 
attention in the literature, given the very similar metallicities of the 
two clusters but their strikingly different HB types---NGC~362 with a 
predominantly red HB, and NGC~288 with an almost entirely blue HB.  
Reviews and extensive references to early 
work have been recently provided by \citet{mbea01} and \citet{cea01a}. 
Here we present new model calculations computed along the same veins as 
in \citeauthor{cea01a}, using the several different prescriptions for 
mass loss on the RGB summarized in Table~\ref{tb:DM} above, and compare the 
results with the age difference for this pair, as carefully derived by 
\citeauthor{mbea01} on the basis of the so-called ``bridge method'' 
\citep{psea96}. The essence of this method is that the HB of NGC~1851
looks very similar to the sum of the NGC~288 and NGC~362 HBs, so that, 
by carefully superposing the latter CMDs with that for NGC~1851, the 
relative positions of their turnoffs---and hence their relative 
ages---can be derived.\footnote{Underlying this method is of course the 
assumption that NGC~1851 is chemically similar to both NGC~288 and 
NGC~362, as discussed in detail by \citet{mbea01}. 
The reader should keep in mind the possibility that this may 
not be strictly true, given the abundance peculiarities recently detected 
in NGC~1851 by \citet{yg08}, the discovery of a bimodal subgiant branch 
in the cluster by \citet{apmea08}, and the results of the theoretical 
analyses by \citet{scea08} and \citet*{msea08}. Note, on the other hand, 
that fairly tight constraints 
on the possibility of a spread in $Y$ in the cluster were recently 
provided by \citet{mc09}.} 

\begin{figure}[t]
  \plotone{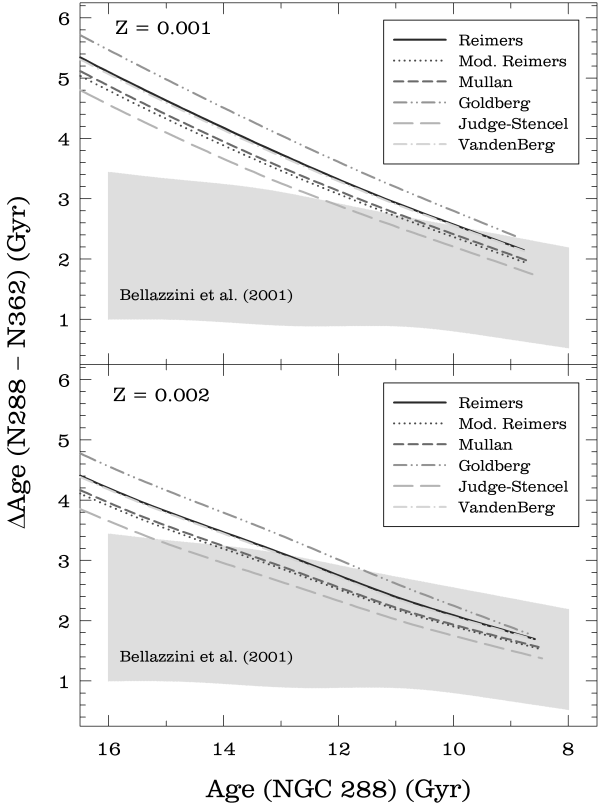}
  \caption{The relative age that is required to explain the difference in HB 
  type between NGC~288 and NGC~362 entirely in terms of age ({\em lines}) 
  is plotted as a 
  function of the NGC~288 absolute age for two different metallicities: 
  $Z = 0.001$ ({\em upper panel}) and $Z = 0.002$ ({\em lower panel}). For 
  each panel, each line corresponds to a different mass loss formula, as 
  indicated in the insets. The shaded area corresponds to the turnoff
  age difference for the pair, from \citet{mbea01}. (This plot 
  represents an update with respect to the similar one presented in
  \citeauthor{cea01a} 2001a.)\label{fig:288362}
}
\end{figure}

Figure~\ref{fig:288362} shows the result of this comparison. In this 
plot, the lines indicate the age difference that is required, from the 
standpoint of canonical stellar evolution theory, to account for the 
observed difference in HB morphology between the clusters, under the 
assumption that the second parameter is age, and for the different 
indicated recipes for the RGB mass loss. The gray bands illustrate,  
in turn, the measured age difference. (These bands are not horizontal 
because the age difference depends on the absolute age value itself,  
the {\em log} of the age difference remaining instead more 
approximately constant as a function of age.) The upper plot 
corresponds to an adopted metallicity $Z = 0.001$, whereas the bottom 
plot to a metallicity $Z = 0.002$. This figure is similar to Figure~7 in 
\citet{cea01a}, but one sees that the new results require slightly 
larger age differences between NGC~288 and NGC~362 to account for 
their relative HB types. The bottom plot clearly shows that, if the 
metallicity of these clusters is of order $Z = 0.002$ and NGC~288 is 
younger than about 12~Gyr, the pair can be accounted for in terms of 
age as the second parameter, irrespective of the mass loss formula 
used. Such a metallicity corresponds to an ${\rm [Fe/H]} \simeq -1.2$ 
for an $[\alpha/{\rm Fe}] = +0.3$ \citep*{msea93}. Such an [Fe/H] value 
is very similar to the values provided in the Feb.~2003 edition of the 
\citet{h96} catalog, and only slightly lower than obtained in the  
\citet{cg97} scale---namely, ${\rm [Fe/H]} = -1.07$ for NGC~288, and 
${\rm [Fe/H]} = -1.15$ for NGC~362. Therefore, it appears fair to say 
that, in the \citeauthor{cg97} scale, this pair is indeed consistent 
with age as the sole second parameter. 

On the other hand, the case $Z = 0.001$, which corresponds to an 
${\rm [Fe/H]} \simeq -1.5$ for an $[\alpha/{\rm Fe}] = +0.3$, does 
not provide equally satisfactory consistency with the hypothesis that 
age is the second parameter. As can be seen from the 
plot, only for NGC~288 ages lower than about 9~Gyr is it that 
consistency with the relative turnoff age difference is recovered, 
irrespective of the mass loss formulation used. For ages higher than 
10~Gyr, several mass loss formulae lead to results that allow one  
to question the hypothesis that age is the (only) second parameter 
for this pair. The best consistency with the age hypothesis obtains 
when using the ``Judge-Stencel'' formula; the worst, when using the 
``Goldberg'' formula (see Table~\ref{tb:DM}). 
Note that an ${\rm [Fe/H]} \simeq -1.5$ falls 
within the uncertainty range of the \citet{zw84} metallicity value, 
since these authors give ${\rm [Fe/H]} = -1.40 \pm 0.12$ for NGC~288, 
and ${\rm [Fe/H]} = -1.27 \pm 0.07$ for NGC~362; similar values are 
provided by \citet{ki03} and \citet{grea97} (in the \citeauthor{zw84}
scale). Therefore, we conclude that it is more difficult to account 
for the pair NGC~288/NGC~362 entirely in terms of age if the 
\citeauthor{zw84} scale better describes the actual abundances of 
globular cluster stars.

It should be noted, in this sense, that the recent results by 
\citet{maea04} \citep[see also][]{jm04}
for the solar metal abundance imply a major downward 
revision in $Z$ with respect to the canonical value (i.e., from 
$Z_{\odot} \approx 0.02$ down to $Z_{\odot} = 0.0126$). 
If this propagates to the 
metallicity scale used for metal-poor stars, a major downward revison 
in the [Fe/H] and $Z$ values for Galactic globular cluster stars may 
be in store as well. According to the above discussion, this downward 
revision would further complicate the explanation of 
second-parameter pairs in terms of age. The same applies, as 
already stated, if the mass loss results from \citet{loea02} are used, 
as opposed to the mass loss recipes given in Table~\ref{tb:DM} (see 
Sect.~\ref{sec:DM} above for a critical discussion). 

To close, we note that \citet{fg03} has recently
investigated the ages of NGC~288 and NGC~362 on the basis of Str\"omgren 
photometry, finding that the two clusters differ in age by less than 
1~Gyr. If so, Figure~\ref{fig:288362} clearly shows that age cannot 
be the (sole) second parameter for this pair.

\subsubsection{The Pair M13--M3}

This pair was first seriously and quantitatively considered in the context 
of the second-parameter phenomenon by \citet{cfp95}, who suggested 
that the age 
difference required to account for the difference in HB type between M3 
(uniformly populated HB) and M13 (blue HB with a long blue tail) 
exceeded the constraints from turnoff stars. A similar conclusion was 
later reached by \citet{ffea97}. More recently, \citet{screa01} have 
reanalyzed the problem, incorporating the \citet{r75a,r75b} mass loss 
``law'' into their analysis, and finding an age difference of 
$1.7\pm 0.7$~Gyr between the two clusters---which, according to them, 
``can produce the difference in HB morphology between the clusters.'' 
Here we 
revisit the problem, investigating the impact of the several different 
mass loss formulae for red giant stars summarized in Sect.~\ref{sec:DM}. 

First we proceed to compute synthetic HB models using the global sample 
of M3 HB stars discussed in \citet{cea01b} and \citet{c04a}, and assuming 
the canonical metallicity for the pair---namely, $Z = 0.001$. For M13, we 
have retrieved the {\em HST} photometry from \citet{gpea02} and performed 
number counts along the HB of the cluster. When we attempted to 
reproduce the observed M13 HB morphology in terms of canonical synthetic 
HBs, we quickly arrived at the conclusion that a unimodal distribution was 
inadequate, since it was unable to explain the large number of very low-mass
stars at the extreme hot end of the M13 HB. Therefore, we incorporated a 
second, low-mass mode to our simulations, thus obtaining a much better
agreement with the overall HB morphology of the cluster. It should be 
noted that the EHB of the cluster is extremely populous, containing 
about 30\% of all HB stars in M13. We then computed the average 
total mass loss required to explain the derived masses of the HB stars 
for each mass loss formula in Table~\ref{tb:DM} 
and over a range in ages, which then allowed 
us to compute the age difference that was needed, with respect to M3, to 
account for the observed difference in HB type.

\begin{figure}[t]
  \plotone{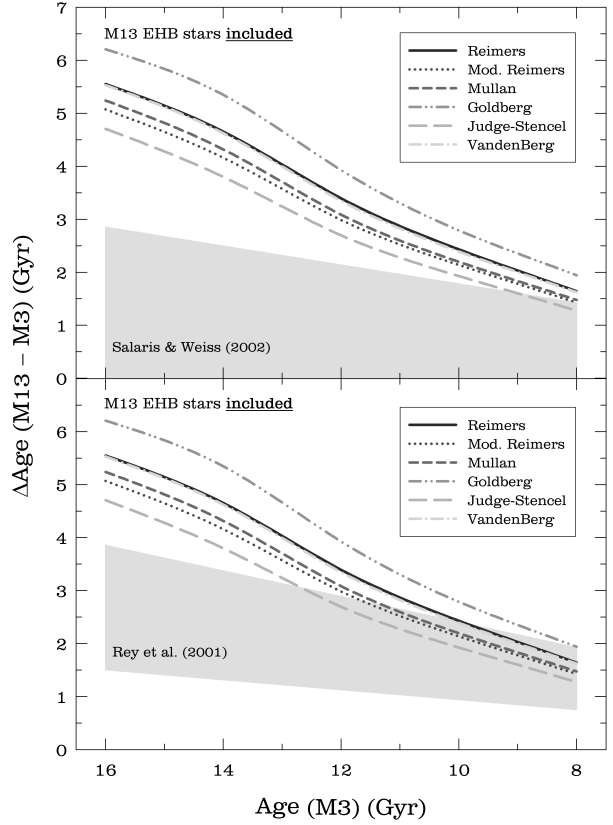}
  \caption{Same as in the previous plot, but for the pair M13--M3. The upper 
  panel shows the comparison between the theoretical results and the \citet{sw02}
  turnoff age difference, whereas the lower plot compares the theoretical 
  values with the \citet{screa01} result based on the 
  turnoff points.\label{fig:M3M13EHB} 
}
\end{figure}

\begin{figure}[t]
  \plotone{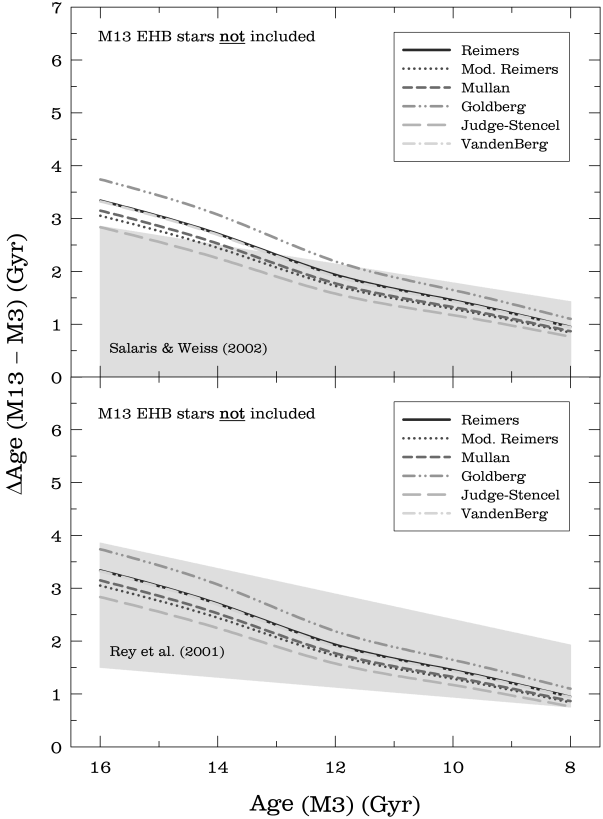}
  \caption{Same as in the previous plot, but now {\em ignoring} M13's EHB 
  component in the calculations.\label{fig:M3M13}  
}
\end{figure}

The result is shown in Figure~\ref{fig:M3M13EHB}. 
In the upper plot, we compare the 
theoretical results with the empirical age difference for the pair, as 
recently derived by \citet{sw02}. (Unless otherwise stated, in this 
section we use the \citeauthor{sw02} results obtained in the 
\citeauthor{zw84} \citeyear{zw84} scale.) 
In the lower plot, the same theoretical 
results are compared with the age difference derived by \citet{screa01}. 
We have also compared the model results with the \citet{dv00} age 
difference values, but these are intermediate between the other two 
studies so that we omit further discussion of this case in what follows. 
Note, in addition, that, within the errors, \citet{jb98} and \citet{area99}
find M13 and M3 to be essentially coeval, whereas \citet{ps98} claims 
that M13 may even have the younger turnoff of the two (see below). 
\citet{fg99} provides an impressive comparison between the two clusters 
in the Str\"omgren $u$, $(u\!-\!y)_0$ plane (his Fig.~2, left panel), 
where one can 
clearly see that there is not much room for a significant age difference 
between the two clusters---\citeauthor{fg99} himself finding an age 
difference of only $0.7\pm 0.2$~Gyr between M13 and M3, the former 
being older. 

As can clearly be seen, the required age difference appears too large in 
comparison with the \citet{sw02} age difference measurements, irrespective 
of the mass loss formula used. The larger age difference measured by 
\citet{screa01}, on the other hand, may be compatible with the turnoff
age difference, particularly for lower M3 ages and if the ``Goldberg'' 
formula provides a less reliable description of mass loss rates on the 
RGB than do the others. 

Some authors have suggested, on the other hand, that the origin of EHB 
stars may not be directly linked to the age of a globular cluster, being 
more likely instead to be due to other physical processes, including 
helium enrichment, binarity, and any other processes that may trigger 
enhanced mass loss on the RGB \citep[e.g.,][]{egea01,tbea01}. 
If so, it follows that at least 
30\% of the HB stars in M13 do not owe their present-day color to age, 
a different second parameter being required. In fact, \citet{screa01} 
favor this option, although this is in conflict with the scenario of 
\citet{pl97} and \citet{hclea02} for the origin of the ``UV upturn 
phenomenon'' and for the photometric evolution of galaxies, according 
to which even the hottest HB stars owe their existence to age (high 
ages naturally being implied for giant elliptical galaxies in this case). 
Note that this age scenario forms the basis upon which the recently 
measured GALEX ultraviolet spectra of early-type galaxies and 
extragalactic globular clusters is currently being interpreted 
\citep[e.g.,][]{ywlea05a,screa05}. On the other hand, in our own galaxy 
EHB stars are now known to be present even in {\em open} clusters 
\citep*{ku92,jlea94,egea97}, again reinforcing the impression that old 
ages cannot be solely responsible for the origin of EHB stars---and hence 
suggesting that age evolution alone cannot fully account for the UV 
upturn phenomenon and for the evolution of the photometric properties 
of galaxies as a  function of redshift.  

To check whether the pair M13/M3 might be explained by assuming a 
completely different formation channel for the M13 EHB stars, we have 
repeated the above exercise but now {\em removing} all M13 EHB stars 
from the sample. Since there are so many EHB stars in the cluster, this 
clearly leads to a significantly 
higher mean mass for the M13 HB stars, and therefore 
to a smaller expected age difference between this cluster and M3. 
Indeed, Figure~\ref{fig:M3M13} confirms that, if one admits that 
the EHB stars owe their 
origin to a different physical process, one is able to account for the 
different HB types of M13 and M3 entirely in terms of the reported 
turnoff age differences, irrespective of the adopted mass loss recipe, 
provided \citep[in the][ case]{sw02} M3 is younger than 12~Gyr. More 
stringent constraints on the M3 absolute age would derive in the 
case of using the \citet{loea02} results for the (lack of a) 
dependence of RGB mass loss rate on $L$, $g$, or $R$. 

Finally, it is important to note that, according to \citeauthor{ps98}'s 
(\citeyear{ps98}) report on ultra-precise photometry for M13 and M3 
obtained with the CFHT, {\em the intrinsic position 
of the turnoff point in M13 is actually bluer 
than in M3}, which is of course completely unexpected if the former is 
indeed older than the latter. The author strongly argues that there 
are very few (if any) sources of systematic errors that could have led 
him to underestimate the turnoff color for M13 compared to M3 in his 
study. One might suspect reddening uncertainties to be the culprit, but 
it should be noted that both clusters have very low reddening: the 
Feb.~2003 edition of the \citet{h96} catalog lists reddening values 
$E(\bv) = 0.02$~mag and 0.01~mag for M13 and M3, respectively (the 
same values as used by \citeauthor{ps98}). From \citeauthor{ps98}'s 
study, one finds that, for both clusters to have the same turnoff color
(and hence the {\em same} ages), the relative reddenings of the two 
clusters, compared with the \citeauthor{h96} catalog value, 
would have to be incorrect by $\Delta E(B\!-\!I) = 0.04$~mag, which 
amounts to $\Delta E(\bv) = 0.015$~mag according to the \citet{rl85}
standard extinction law---in the sense that the M13 reddening must have 
been overestimated, and/or the M3 reddening underestimated. While this 
may seem like a small change, it is worth noting that, on the 
basis of the \citet*{dsea98} dust maps, one finds $E(\bv) = 0.017$~mag 
for M13, and $E(\bv) = 0.013$~mag for M3: while the shifts are correct in 
sign, they are clearly insufficient in size to bring M13 to even the same 
turnoff age as M3. An even larger change would be needed, of course, to 
make M13 significantly older than M3. \citeauthor{ps98} argues, in fact, 
that a more likely explanation for the differences in CMD positions 
{\em and shapes} between the two clusters would be provided by a 
difference in helium abundance, as has indeed been suggested by other 
authors as well \citep[e.g.,][]{cd05,dcea05}. Even stronger constraints 
on the relative reddening values between the two clusters would certainly 
prove helpful in clarifying the situation.

\subsubsection{Red HB Globular Clusters in the Extreme Outer Halo}

It is worth revisiting the case of the outer halo globular clusters 
with predominantly red HB types which have had their ages measured with 
{\em HST}, such as Palomar~3, Palomar~4, and Eridanus. According to 
\citet{psea99} and \citet{dv00}, these clusters are slightly younger than 
the bulk of Galactic globular clusters with similar metallicity. Is the 
detected age difference between these clusters and inner halo clusters, 
such as M3 and M5,\footnote{In fact, proper motion studies 
\citep{rsea96,kc97} indicate that 
M5 is an {\em outer halo} globular which just happens to be close to its 
perigalacticon at this point in time, the cluster actually spending much 
of its life at galactocentric distances larger than 50~kpc.\label{foo:M5}}
consistent with the observed difference in HB morphology between them? 

\subsubsubsection{The Pair M3--Pal~3}
In terms of metallicity, Pal~3 provides a good match to M3. Accordingly, 
\citet{cea01b} performed a study of the age difference between Pal~3 and 
M3 that is required to account for their relative HB types, and found that 
the age difference based on analyses of the turnoff points could easily 
account for the difference in mean HB color between the two clusters. We 
have repeated their exercise, finding a slightly larger age difference 
being needed to account for their different HB types, 
but basically confirming their results. On the other 
hand, \citeauthor{cea01b} call attention to the fact that the mass 
dispersion along the HB of Pal~3 appears entirely consistent with zero, 
thus being significantly different from M3's. If differences in mass 
loss among individual red giants is responsible for the presence of mass 
dispersion along the HBs of globular clusters, then the mass loss process 
clearly operated in a different way in Pal~3 than it did in M3. Accordingly, 
\citet{cea01a} suggest that while age may be the ``global'' second 
parameter driving the {\em mean} HB color for at least some globular 
clusters, for many globular clusters environmental effects might be a 
``local'' second parameter responsible for generating a {\em dispersion 
in color} around this mean value.

\subsubsubsection{The ``Pair'' M5--Pal~4/Eridanus} 
In terms of chemical composition, both Pal~4 and Eridanus appear to have 
metallicities similar to M5's. In addition, since Pal~4 and Eridanus 
appear to have similar chemical composition and CMD morphology but 
coarsely populated CMDs, it seems 
reasonable to combine the data for the two and perform a single analysis 
of the ``pair'' comprised by M5, on the one hand, and Pal~4/Eridanus, on 
the other \citep{c00}.

\begin{figure}[t]
  \plotone{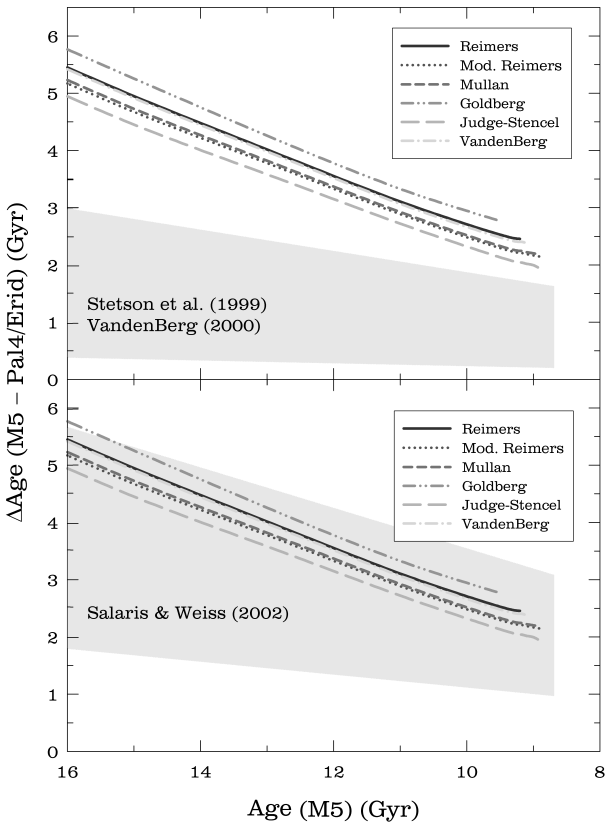}
  \caption{Same as in the previous plots, but for the case of M5 and the 
  outer-halo globulars Pal~4 and Eridanus. The latter are studied as a 
  single object, as explained in detail by \citet{c00}. The upper panel 
  compares the model predictions with the turnoff-based age 
  difference from \citet{psea99} and \citet{dv00}, whereas the bottom 
  panel shows a similar comparison against the turnoff ages by 
  \citet{sw02}.\label{fig:M5P4}  
}
\end{figure}

Again, we repeat the analysis carried out by \citet{c00}, but now 
comparing the theoretical results with the age differences derived on 
the basis of the turnoff points by \citet{psea99}, \citet{dv00}, and 
\citet{sw02}. The results are shown in Figure~\ref{fig:M5P4}, where the upper 
panel shows the comparison between the HB morphology-based analysis 
and the turnoff age difference from \citeauthor{psea99} and \citeauthor{dv00},
whereas the lower panel shows a comparison between the same results for 
the HB stars in these clusters and the turnoff ages from \citeauthor{sw02}.

These plots \citep[which again give slightly higher HB type-based age differences
between the clusters than originally reported by][]{c00} reveal that the 
turnoff age difference measured by \citet{psea99} and \citet{dv00} is 
insufficient to account for the difference in HB types between M5 and 
Pal~4/Eridanus, irrespective of the mass loss formula used. On the other 
hand, the turnoff age difference reported by \citeauthor{sw02} is clearly 
more consistent with the hypothesis that the age difference between M5 
and Pal~4/Eridanus is sufficient to account for their relative HB types. 
It is clearly very important to establish what the actual turnoff age 
difference between M5 and Pal~4/Eridanus is before we are in a position 
to decide whether age can be the (``global'') second parameter for this pair.

\subsection{The Second-Parameter Effect at High Metallicities}

While the second-parameter problem is traditionally thought to affect 
mostly intermediate-metallicity globular clusters, the fairly recent 
discovery of large (and peculiarly bright) RR Lyrae populations 
\citep{lea99,pea00}
and prominent blue HB tails \citep{gpea97,mrea97} 
in the moderately metal-rich Galactic 
globular clusters NGC~6388 and NGC~6441 has 
brought the phenomenon to the realm of ${\rm [Fe/H]} \sim -0.5$
globular clusters as well. 

Several hypotheses have been discussed in the literature to explain 
the observed HB morphology and peculiar RR Lyrae periods in these 
clusters. These include tidal collisions \citep{mrea97}, 
helium mixing on the RGB, a primordially increased helium abundance, 
and increased core masses at the RGB tip as a result of internal 
rotation \citep{sc98}. Other explanations include a large spread in 
metallicities \citep{gpea97,as02}, a selective metal depletion 
scenario \citep{as99}, and a range in internal helium abundances 
\citep{dac04}. In addition, and also as an attempt to explain their
peculiar HB morphologies, \citet{crea02} have suggested that 
NGC~6388 and NGC~6441 might be similar to $\omega$~Cen in nature, 
with a (small) internal metallicity spread and a (fairly large) 
internal range in ages; in their scenario, the blue HB and 
bright RR Lyrae components of NGC~6388 and NGC~6441 would be 
ascribed to a combination of lower metallicity and old 
ages, RR Lyrae stars being somewhat more metal-poor stars evolved 
away from a position on the blue ZAHB. 

Unfortunately, most---if not all---of these scenarios face strict 
observational and theoretical constraints. \citet{mrea97} show 
that tidal collisions cannot produce the observed HB 
morphology---and we add that it cannot lead to peculiarly 
bright RR Lyrae stars, either. 
A primordially increased helium abundance does not appear 
consistent with the position of the RGB bump in these clusters 
\citep{rea02}. A large spread in metallicities, as also pointed
out by \citeauthor{rea02}, is not supported by observations of 
the cluster CMDs in the RGB region, since the latter do not show 
the large spread in colors that would be expected in this case. 
Moreover, \citet{gcea05} have recently found, based on VLT spectra, 
only a small (though significant) spread in [Fe/H] among the RR Lyrae 
stars in NGC~6441---although this result could not be confirmed by 
\citet{rgea07} among the cluster's red giants. 
Evolution away from a position on the blue ZAHB 
does not produce enough bright RR Lyrae variables to explain the 
observed RR Lyrae period distributions, and neither is the sloping
nature of the HB quantitatively reproduced in the \citet{crea02} 
scenario \citep{pea02}. 

Most puzzling of all is the evidence that the 
blue HB stars in these clusters cooler than the \citet{gea99} 
jump do {\em not} appear to have peculiarly low 
gravities \citep*{smea99a}, which effectively rules out any scenario
that requires the blue HB and RR Lyrae components to be brighter 
than in canonical models. However, a brighter HB at both the blue 
HB and the RR Lyrae levels seem to be required by i)~The glaring
evidence that the tip of the blue HB is $\sim0.5$~mag brighter in 
$V$ than the red HB component; ii)~The very 
long periods of the RR Lyrae stars in both clusters. Given the
apparently irreconcilable evidence, we suggest 
that a reassessment of the spectroscopic gravities for a larger 
sample of blue HB stars in both clusters would prove well worth 
the effort.\footnote{After this paper had been completed, 
\citet{smas06} revisited the problem and concluded that those previous 
results were indeed incorrect, most likely due to problems with the 
background subtraction from the spectra (blends).} 

Assuming that the spectroscopic gravities can indeed be reconciled 
with the requirement that the blue HB component be unusually bright, 
we discuss the possibility that a internal spread in helium 
abundances \citep{dac04}
may help account for the HB morphology and
RR Lyrae pulsation properties in these clusters \citep[see][ for 
numerical simulations in the case of helium mixing]{sc98}. 

Consider the case
of NGC~2808, and assume that, as suggested by \citet{dac04}, there
is an internal helium abundance range among the stars in the cluster. 
In their scenario, NGC~2808's red HB and RR Lyrae components have a 
``normal'' helium abundance, blue HB stars at the ``horizontal'' 
level have a mild helium enhancement, and hotter blue HB stars 
have a much higher (initial) helium content, up to about 35\%. 
\citet{lea05} suggest a similar scenario for the cluster, only 
invoking an even higher helium enhancement for the blue HB stars. 
Now imagine how NGC~2808 might have looked like in the not too remote
past (say, a couple of Gyr ago): its present-day helium-enhanced blue
HB stars would necessarily be redder as a consequence of the younger 
age, and therefore {\em helium-enhanced stars would be present in 
relatively high numbers inside the RR Lyrae instability strip}---thus
leading to overluminous, long-period RR Lyrae stars. The 
stars at the tip of the blue HB would have an even higher helium 
abundance, 
leading to a marked {\em sloping HB} when compared with 
the red HB and RR Lyrae components in the same cluster. Clearly, 
the NGC~2808 envisaged by \citeauthor{dac04} and \citeauthor{lea05} 
would have looked a lot like the present-day NGC~6388 and NGC~6441
in the past! Accordingly, a possible explanation 
for the observed peculiarities in these clusters could involve 
both an enhanced helium abundance among a fraction of the cluster 
stars (note that the blue HB plus RR Lyrae components in both 
clusters constitute relatively small fractions of the overall HB 
populations, so that the RGB bump constraint might not be violated
in this case) and a younger age with respect to the bulk of the 
Galactic globular clusters---the latter not necessarily being
required, depending on metallicity (i.e., first parameter) and 
mass loss effects: indeed, the deep HST photometry presented by 
\citet{mcatea06} for NGC~6388 reveals that the cluster is comparable
in age to 47~Tuc. 

A potential problem with this scenario is the fact 
that, for every single helium abundance value, one should expect, 
by analogy with what is observed in ``single-population globular 
clusters,'' a spread in mass loss as well. Therefore, at any 
given color, a spread in helium abundance should also be present, 
and it is unclear whether the implied HB luminosity distribution 
would match in detail the observed one, which appears fairly tight. 
This problem is especially evident in the recent simulated CMD 
by \citet[][; see their Fig.~7]{gbea07}, but is not as apparent in 
the simulated CMDs of \citet{cd07}---which again may be due to 
the lack of a mass spread for the populations with enhanced $Y$ 
in their simulations. Note, in this sense, that the simulations 
presented by \citet{sc98} in their helium 
mixing and rotation scenarios also ignore the effect of a spread
in masses for each individual $Y$. In any case, 
\citet{gp08} has recently reported that NGC~6388 does show 
a double subgiant branch, thus reinforcing the evidence for 
the presence of more than one stellar population in the cluster.  

We conclude that stringent tests of such a scenario could 
be provided by deep {\em HST} photometry (so that the turnoff
ages, and possible splits indicative of multiple stellar populations, 
can be reliably established), spectroscopic gravities for 
a large sample of moderately cool blue HB stars in both clusters, 
and HB simulations in which both a spread in $Y$ and in RGB mass 
loss are simultaneously taken into account.

\section{On the RR Lyrae Luminosity-Metallicity Relation} 

Much has been written over the past several years in regard to the luminosities 
of HB stars at the RR Lyrae level, particularly in the $V$ band, and its 
dependence on metallicity; recent reviews of the subject include papers by 
\citet{bc99}, \citet{cc99,car03}, \citet{cc03}, and \citet{jst06}. \citet{gb03}
and \citet{sc05} have provided recent reviews in which theoretical uncertainties 
affecting the predicted properties of HB stars, including their luminosities, 
have been discussed in considerable detail. Several different theoretical 
results have recently been compared by 
\citet[][; see their Fig.~1]{cc03} and 
\citet*[][; see their Fig.~9]{gea05}. 
Accordingly, 
in what follows we will content ourselves with providing an updated estimate 
of the HB luminosity-metallicity relation as based on the trigonometric 
parallax of RR Lyrae itself, only briefly mentioning recent progress in 
understanding some discrepancies that have prevented the establishment of 
a universally accepted $M_V{\rm (RRL)}-{\rm [Fe/H]}$ relation. 

It should be 
noted that this crucial relation has traditionally provided the very  
basis for the Population~II distance scale, thereby constituting
one of the most important techniques used to help nail down the first 
step in the cosmological distance ladder---namely, the distance to the 
LMC \citep[e.g.,][]{bea02,gk03,jst06}. 
Moreover, its slope and zero point 
have long been recognized as crucial ingredients in the determination of 
the absolute ages of globular clusters and their variation with metallicity 
\citep[e.g.,][]{sc90,w92c}. On the other hand, it should also be noted that
this relation can only be used in a very approximate way to estimate the 
{\em average} absolute magnitudes of RR Lyrae stars of a given metallicity. 

In order to properly evaluate the absolute magnitudes of {\em individual} 
RR Lyrae stars, more precise techniques, frequently involving a 
{\em period-luminosity relation}, are required. Unlike the case of classical
Cepheids, however, good period-luminosity relations are {\em not} available
for RR Lyrae stars in the visual bandpasses, for reasons that have been 
discussed in detail by \citet*{cea04}. On the other hand,   
it has long been known that good RR Lyrae period-luminosity relationships 
are present in the near-infrared \citep*{alea86,alea90}. Empirical results
have recently been critically discussed by \citet*{asea06} and \citet{mfea08}, 
and \citet{asea08} have recently provided a detailed analysis of the $J$, $H$, 
$K$ light curves of RR Lyrae itself.  
Theoretical calibrations using the near-infrared bandpasses $J$, $H$, 
$K$ have been provided by \citet{scea04}, \citeauthor{cea04}, and \citet{mdpea06}, 
among others. As pointed out by \citet{isea03} and \citeauthor{cea04}, a 
period-luminosity relation is also present in $I$, although in this case one 
finds somewhat more scatter and a stronger metallicity dependence than in 
the near infrared. Finally, \citet{ccc08} and \citet{clcc08} have recently 
shown that very precise period-luminosity and period-color relations may also 
be defined for RR Lyrae stars in the Str\"omgren and SDSS filter systems, 
respectively.

\subsection{The Variable Star RR Lyr and the HB Luminosity-Metallicity
  Relation} 

The star RR Lyr is the closest of its class, and accordingly has proven
of great interest for trigonometric parallax studies. Its variability was 
noted by Williamina P. Fleming at the Harvard College Observatory prior 
to July 1889, but the discovery was not announced until a few years later 
by \citet{ep01}. 

\citet{bea02} obtained, using the {\em Hubble Space Telescope},  
a much more accurate (and significantly smaller) value for the absolute 
parallax of RR Lyrae than had previously been provided by {\em Hipparcos}, 
namely $\pi_{\rm abs} = 3.82\pm 0.20$~mas 
\citep[compared to $\pi_{\rm abs}^{\rm Hip} = 4.83\pm 0.59$~mas;][]{pea97}. 
Recently, \citet{vl07} revised 
the trigonometric parallaxes provided by Hipparcos, and arrived at a value 
$\pi_{\rm abs}^{\rm Hip} = 3.46\pm 0.64$~mas for RR Lyr. 
A weighted average of ground-based studies \citep*{valh95} indicated 
a parallax $\pi_{\rm abs} = 3.0\pm 1.9$~mas for RR Lyrae \citep[see Fig.~6 
in][]{bea02}. 
Taking a weighted average of these results, we obtain a 
final value of $\pi_{\rm abs} = 3.78\pm 0.19$~mas for RR Lyr.
This implies a revised distance modulus of $(m\!-\!M)_0 = 7.11\pm 0.11$~mag 
for the star.

\citet{bea02} argue in favor of a relatively low extinction towards RR Lyr, 
namely $A_V \simeq 0.07\pm 0.03$~mag. In recent work, an intensity-weighted 
mean magnitude of $\langle V \rangle = 7.76$~mag \citep{fea98}\footnote{This 
value is provided in Table~1 of their paper, which is only available in 
electronic format, from 
{\tiny\texttt{http://vizier.u-strasbg.fr/viz-bin/VizieR?-source=J/A+A/330/515}}.} 
has been adopted for RR Lyr. However, we note that 
this value is based on {\em Hipparcos} photometry, which may require a 
non-trivial transformation to the standard system. For comparison, 
\citet{l94} determines a $\langle V \rangle = 7.66$~mag, and \citet{lea96} 
find instead $\langle V \rangle = 7.74$~mag. On the other hand, \citet{gp98} 
argue strongly in favor of the {\em Hipparcos}-based magnitudes of 
\citeauthor{fea98}, only proposing an {\em additional}, reddening-related 
correction: $V_{\rm GP99} = V_{\rm F98} - 0.2 \, E(\bv)$. Using a standard 
extinction law, $A_V \simeq 3.1 \, E(\bv)$, and the reddening value obtained
by \citet{cc08} on the basis of Str\"omgren photometry, namely 
$E(\bv) = 0.015\pm 0.020$~mag, leads to a final,  
extinction-corrected RR Lyr mean magnitude of 
$\langle V_0 \rangle \simeq 7.71 \pm 0.06$~mag. 

It is important to note that the intensity-mean magnitude (and even more 
so the corresponding magnitude-weighted average) does not necessarily 
correspond to the magnitude of the ``equivalent static star'' (i.e., the 
magnitude 
the star would have if it were not pulsating): an amplitude-dependent 
correction has to be applied. Such a correction has been obtained, both 
for fundamental (RRab) and first-overtone (RRc) variables, by \citet*{bea95}
on the basis of detailed hydrodynamical pulsation models of RR Lyrae stars. 
RR Lyrae has long been known to present the \citet{b07} effect, and its 
$P_{\rm Blazhko}$ appears to fall in the range between 
40~d \citep{sea03} and 41~d \citep{sk00}, with an additional, longer-term 
periodicity ($P \simeq 4$~yr) also being present \citep{ds73}. 
Inspection of the light curves for RR Lyr presented by \citet{hs95}, 
\citeauthor{sk00}, and \citeauthor{sea03} suggest that the amplitude in 
$V$ oscilates in the range between 0.5~mag and 1.1~mag. According to 
Table~2 in \citeauthor{bea95}, this is precisely the amplitude range  
over which the intensity-weighted mean magnitude provides the most 
accurate description of the magnitude of the equivalente static star. 
Therefore, no amplitude corrections appear to be needed in the case of 
RR Lyrae. Taking, accordingly, a value 
$\langle V_0 \rangle = 7.71 \pm 0.06$~mag 
and a distance modulus $(m\!-\!M)_0 = 7.11\pm 0.11$~mag, one finds an 
absolute magnitude for the star of $M_V = 0.60 \pm 0.13$~mag. 

In regard to the star's metallicity, \citet{cea95} obtain 
${\rm [Fe/H]} = -1.39$~dex and $[\alpha/{\rm Fe}] = 0.31$~dex, which are 
quite typical values for Galactic halo stars. Previous metallicity 
measurements for this star had provided values in the range between 
${\rm [Fe/H]} = -1.14$ and $-1.21$~dex \citep[see Table~6 in][]{cea95}, 
so that the \citeauthor{cea95} result represented a significant downward 
revision. According to \citet{bea01}, the \citeauthor{cea95} measurements 
are in a scale that closely mimicks the \citet{zw84} scale. We accordingly 
adopt a metallicity value ${\rm [Fe/H]} = -1.39 \pm 0.10$~dex for RR Lyr. 

Note that the analysis of \citet{cc08} suggests that RR Lyr is a somewhat
evolved star, with an overluminosity of $\simeq 0.06\pm 0.01$~mag in $V$
with respect to the mean for other RR Lyrae stars with similar metallicity
\citep[see also][]{mfea08}. To within the errors, this result is not 
inconsistent with the recent near-infrared study by \citet{asea08}. 
Therefore, the average luminosity of RR Lyrae variables with metallicity 
similar to that of RR Lyr itself should be around $0.66\pm 0.13$~mag. 

What do these results imply, in terms of the traditionally employed RR Lyrae  
luminosity-metallicity relation in the $V$ band, usually taken in the 
form $M_V({\rm RRL}) = \alpha \, {\rm [Fe/H]} + \beta$? 

To answer this question, we shall first adopt a slope $\alpha = 0.23\pm 0.04$ 
for this relation, as found and/or favored in several recent reviews of the 
subject, including \citet{bc99}, \citet{cc99, car03}, and \citet{cc03}.  
Several recent analyses do provide additional 
support for this result: for instance, 
\citet{cea03} and \citet{gea04a} obtain $\alpha = 0.214\pm 0.047$ from 
analysis of RR Lyrae variables in the LMC, whereas 
\citet{oea03} obtained $\alpha = 0.21-0.28$, depending 
on their treatment of presumably well-evolved RR Lyrae variables with 
periods around 0.7~d, from analysis of the RRab stars in $\omega$~Cen. 

Using the \citet{cea95} metallicity for RR Lyr 
in the \citet{zw84} scale, one then finds a value 
of $\beta = 0.98 \pm 0.13$, implying a final relationship of the 
following form: 

\begin{equation}
   M_V({\rm RRL}) = (0.23 \pm 0.04) \, {\rm [Fe/H]}_{\rm ZW} + (0.98 \pm 0.13). 
   \label{eq:RRL}
\end{equation}

\noindent If one transforms the \citeauthor{cea95} metallicity to the 
\citet{cg97} scale and then repeats the analysis, one finds instead

\begin{equation}
   M_V({\rm RRL}) = (0.23 \pm 0.04) \, {\rm [Fe/H]}_{\rm CG} + (0.93 \pm 0.13). 
   \label{eq:RRLCG97}
\end{equation}

These relations, while based on a detailed reassessment of the 
absolute magnitude and evolutionary status 
of RR Lyr, turn out to be similar to 
the relation derived in the recent review papers by \citet{bc99}, 
\citet{cc99,car03}, and \citet{cc03}, based on a critical analysis 
of several calibration techniques (but ignoring the evolutionary status
of the star). This notwithstanding, some methods 
have provided somewhat discrepant slopes and/or zero points, and we  
shall momentarily address two such cases. Before doing so, however, we 
will immediately proceed to deriving the all-important distance modulus 
of the LMC that is implied by these relations
\citep[see also][ for a recent review]{dra04}.

\subsection{The Distance Modulus of the LMC} 

Eq.~(\ref{eq:RRL}) implies an absolute magnitude 
$M_V({\rm RRL}) = 0.64\pm 0.14$~mag at ${\rm [Fe/H]} = -1.48\pm 0.07$. 
The latter is the 
mean metallicity derived for LMC RR Lyrae variables by \citet{gea04a}, 
in the \citet{zw84} scale. Using a value 
$\langle V_0 \rangle = 19.068\pm 0.102$~mag from \citeauthor*{gea04a}, 
one then 
finds an updated true distance modulus for the LMC of 
$(m\!-\!M)_0^{\rm LMC} = 18.42\pm 0.17$.
If one uses instead the average values for LMC RR Lyrae stars independently 
determined by \citet{jbea04}, namely ${\rm [Fe/H]} = -1.46\pm 0.09$~dex and 
$\langle V \rangle = 19.45\pm 0.04$~mag, with their favored reddening of 
$E(\bv) = 0.11$~mag, one finds $\langle V_0 \rangle = 19.11\pm 0.04$~mag 
for a $M_V = 0.65\pm 0.14$~mag, thus implying a true distance modulus
$(m\!-\!M)_0^{\rm LMC} = 18.46\pm 0.15$.
Taking a weighted average over these two results, we arrive at 
the following distance modulus for the LMC, based on our updated analysis 
of the star RR Lyr:

\begin{equation}
   (m\!-\!M)_0^{\rm LMC} = 18.44\pm 0.11. 
   \label{eq:LMC3}
\end{equation}

\subsection{Are We Converging on a $M_V{\rm (RRL)}-{\rm [Fe/H]}$ 
            Relation Yet?} 

In what follows, we discuss a few discrepant calibrations of the HB 
luminosity-metallicity relation, and describe how the problem has recently 
been solved or what suggestions may have been advanced to reconcile the 
discrepant calibrations. 

\subsubsection{The Shallow Slope Obtained from HST Photometry of M31 Globular 
   Clusters} 

Almost a decade ago, a very shallow slope, $\alpha = 0.13\pm 0.07$, was 
derived by \citet{fpea96} on the basis of {\em HST} observations of M31 
globular clusters. However, the determination of the HB level in their 
CMDs was far from straightforward, since instead of seeing a {\em horizontal} 
branch at the RR Lyrae level, whenever a blue or intermediate HB component 
was present their CMDs revealed instead surprisingly {\em sloped} ``horizontal'' 
branches, not unlike what one sees when plotting an $I, \,\, (V\!\!-\!\!I)$ 
CMD for Galactic globulars. Such sloping HBs are not seen among
Galactic globular clusters, and even though differences between Galactic 
and extragalactic globulars (related, for instance, to the different 
chemical evolution histories of the different galaxies, or to the amount
of angular momentum available in the different protogalactic clouds) 
may indeed exist, theoretical models of HB stars do 
not predict similar CMD morphologies.   

A possible solution to this puzzle has recently been provided by \citet{rea05}, 
who found that the \citet{fpea96} CMDs were strongly affected by photometric 
blends, not accounted for in their original analysis. Investigating the impact
of these blends upon the morphology of the HB in their CMDs, \citeauthor{rea05}
have found that the strange sloping nature of the HB in \citeauthor{fpea96} 
is likely due to the presence of photometric blends. Accordingly, 
\citeauthor{rea05} derived new CMDs for an enlarged sample of 
M31 GCs in which the effects of blends were taken into account. As a 
result, they provide a revised slope of $\alpha \simeq 0.20\pm 0.09$, 
clearly in much better agreement with eq.~(\ref{eq:RRL}). It should be 
noted, however, that careful inspection of the CMDs published by these 
authors still reveal unrealistic HB shapes, thus raising the possibility 
that additional corrections will be needed before a final relation between
HB magnitude and metallicity can be derived on the basis of observations 
of M31 globulars.

\subsubsection{The Faint Zero Point of the Method of Statistical Parallaxes}

As far as the zero point of eq.~(\ref{eq:RRL}) is concerned, the recent 
review by \citet{cc03} 
shows that there is reasonable agreement among the several different methods 
that are used to infer it. Importantly, the Baade-Wesselink method, which 
used to favor a faint zero point \citep[i.e., fainter than provided by the 
above calibration by at least 0.1~mag; see review of earlier work by][]{fea98b}, 
is now seen to be in good agreement with eq.~(\ref{eq:RRL}) and with a brighter zero point 
for the HB luminosity-metallicity relation. Indeed, \citet{gk03}, by applying 
the same Baade-Wesselink algorithm to both Galactic RR Lyrae and Cepheid 
variables, has recently shown that a consistent, ``long'' distance 
modulus for the LMC obtains, namely $(m\!-\!M)_0 = 18.55$~mag---about 0.05~mag
{\em brighter} than implied by eq.~(\ref{eq:RRL}). 

This notwithstanding, at least one method remains that does keep 
repeatedly providing a faint zero point to the LMC: the {\em statistical 
parallaxes} method. For instance, 
\citet{gp98} favor a $M_V = 0.77$~mag at ${\rm [Fe/H]} = -1.6$~dex from 
analysis 
of a sample of 147 RR Lyrae stars, which is 0.21~mag fainter than implied 
by eq.~\ref{eq:RRL}; or $M_V = 0.80$~mag at ${\rm [Fe/H]} = -1.7$~dex after adding 
to the sample 716 non-variable metal-poor stars, which is 0.26~mag fainter 
than eq.~(\ref{eq:RRL}). While \citet{cc03} hint that the presence of disk stars 
may affect the \citeauthor{gp98} results, \citet{dr01} recently provided 
a new application of the method in which thick disk stars were carefully 
separated from halo stars using both kinematic and metallicity criteria, 
but essentially confirming the \citeauthor{gp98} results. \citet{pg98a,pg98b}
have provided a very careful and detailed analysis of the possible 
systematic uncertainties affecting the method, without succeeding in 
identifying a likely cause for the difference with respect to several 
other methods---in fact, they hint that the problem lies with 
the latter \citep[see also][]{pg99}. 
It is indeed unclear what the solution to this problem will be, 
but \citeauthor{cc03} suggest that the \citeauthor{dr01} results may be 
affected by inhomogeneities in the distribution of their {\em halo} stars. 
Indeed, that the distribution of Galactic halo stars is not quite uniform 
has recently been noted, using blue HB stars, by \citet{aea05} and 
\citet{lcea05} \citep[but 
see also][]{wbea04}; using RR Lyrae stars, by 
\citet{iea00}, \citet{vea01}, \citet{vz03,vz06}, \citet{kea04}, 
\citet{zea04}, and \citet{skea08}; 
and, using several other types of tracers, by, for instance,  
\citet{nea02}, 
\citet{iea03}, \citet{mea03,mea04}, \citet{mdea04}, \citet{ny05}, and 
\citet{rpea06}, 
among many others.\footnote{The widespread use of HB stars, whether 
variable or not, in many such studies clearly shows their importance  
as probes of the Galaxy's structure, formation, and evolution. Other 
recent examples include \citet{esea04}, \citet{ctea05}, 
\citet{wbea05,wbea08}, \citet*{tkea05}, and \citet{ck06}---among others.}

\subsubsection{Differences between HB Stars in Globular Clusters and in 
the Field?} 

The similarity between metal-poor field and cluster RR Lyrae stars 
appears reasonably established \citep{c98,dsc99,ecea00}, 
but there may be differences in regard to at least the more extreme 
HB stars, both in regard to their luminosity distribution 
\citep[][ claim that the field contains significantly fewer stars at the 
blue end than do such globular clusters as M15 and NGC~6229]{wbea05} 
and their physical origin: \citet{mbea05,mbea06} and 
\citet*{cmbea08} find that the EHB stars in NGC~6752 and 
M80 (NGC~6093) do {\em not} appear to be associated to binary systems, 
unlike what seems to happen most frequently among field sdB 
stars \citep{egea01}. \citet{cmbea08} interpret this phenomenon as 
likely due to a binary fraction-age relation (globular cluster stars
being on average much older than field stars), a hypothesis
which has recently been supported by \citet{zh08}. In addition, the 
smaller envelope masses of the older cluster red giants make it
much easier for the single-star channel to produce EHB stars, as 
also noted by \citet{cmbea08}.

It should also be noted that 
the abundance anomalies which are commonly seen in globular cluster red 
giants \citep[e.g.,][; and references therein]{csea04,rgea04b} 
appear {\em not} to be present in field red giants \citep[][; see also
Sect.~5.2 in Grundahl et al. 1999 for extensive references to work prior 
to the year 2000]{rgea00}.  
Therefore, if these abundance anomalies affect somehow the evolution of 
RGB and HB stars (through the ``primordial'' and/or ``deep mixing'' 
channel), as has indeed been often suggested in the literature 
(see \S\ref{sec:HSP} for extensive references), 
they should naturally be expected to have an impact upon the observed 
properties of HB stars---including color distribution, 
luminosities, gravities, and pulsation characteristics. While 
these differences have not been frequently detected, it is worth recalling 
the cases of NGC~6388 and NGC~6441, whose RR Lyrae stars are {\em very} 
different from field RR Lyrae stars with similar metallicity 
\citep[][; see also \citeauthor{al95} \citeyear{al95} and \citeauthor{c04b}
\citeyear{c04b} for additional references to metal-rich globular clusters 
harboring relatively long-period RR Lyrae 
stars]{sc98,lea99,pea00,pea01,pea02,gcea05,mcatea06,nmea06,nm07}.

\subsubsection{The $M_V{\rm (RRL)}-{\rm [Fe/H]}$ Relation: Linear, 
               Quadratic, or Even More Complex?}

{\em Steeper} slopes than provided by eq.~(\ref{eq:RRL}) may result from the 
inclusion of RR Lyrae stars more metal-rich than ${\rm [Fe/H]} \simeq -1$, 
as originally suggested by \citet*{cea91} from theoretical computations. 
In fact, for field stars in
our galaxy, a {\em quadratic} relation between $M_V{\rm (RRL)}$ and [Fe/H] 
is likely to be superior to a linear approximation 
when such metal-rich variables are included, in the sense that the metal-rich 
variables are fainter than would be expected using a linear fit to the 
more metal-poor data---a result which is strongly supported by the 
theoretical models (e.g., 
\citeauthor{cea91} \citeyear{cea91}; \citeauthor{fcea00} \citeyear{fcea00}; 
\citeauthor{bea03} \citeyear{bea03}; 
\citeauthor{cea04} \citeyear{cea04}; and references therein). 

On the other hand, it should also be noted that the {\em helium abundance} 
is also expected to increase with $Z$, by amounts which may differ in different 
environments as a consequence of different chemical enrichment laws 
\citep[as early noted by][]{vdb67}. Accordingly, it is extremely 
important to realize that there is no strong physical basis for a universal 
$Y-{\rm [Fe/H]}$ relation. As an example, \citet{cfp96} analyzed three 
different chemical enrichment scenarios which, though providing essentially 
the same helium abundance at low metallicities (${\rm [Fe/H]} < -1$), led 
to differences in $Y$ by $\simeq 0.035$ at ${\rm [Fe/H]} \approx -0.5$, 
and by up to 0.1 at solar metallicity \citep[see also][]{mc08b}. 
Indeed, 
it would perhaps be surprising if galactic disks and spheroids presented 
precisely the same $\Delta Y - \Delta Z$ ``law,'' given the evidence that 
they present different $\alpha$-element enrichment patterns---variations 
in the latter being expected to be accompanied by variations in the  
$\Delta Y - \Delta Z$ relation as well 
\citep[see Figs.~1 and 2 in][]{mc08b}.

This can impact HB luminosities in a quite dramatic way. 
Since the HB luminosity depends on 
the helium abundance as 

\begin{equation}
  \frac{d M_{\rm bol}^{\rm HB}}{d Y} \approx 4.5
  \label{eq:BOLHB}
\end{equation}

\noindent \citep{c96},\footnote{This is a little higher than what obtains 
from ZAHB models \citep*[see, e.g., Table~1 in][]{sea87}, due to the 
fact that high-$Y$ evolutionary tracks present much more important luminosity 
evolution than do models for lower $Y$ \citep[see, e.g.,][]{sg76,s87}.} 
the quoted differences in helium abundance may lead 
to changes in the HB luminosity, from one chemical enrichment scenario to 
the next, by as much as 0.15~mag at ${\rm [Fe/H]} = -0.5$, and by 
0.45~mag (!) at ${\rm [Fe/H]} = 0$. The helium enrichment law clearly plays 
a crucial role in defining the slope of the HB luminosity-metallicity relation 
at high metallicities, and this should be properly taken into account when 
studying different stellar populations.\footnote{We note, in passing, that 
possible differences in the helium enrichment law should also be taken into 
account when studying classical Cepheids in external galaxies. In particular,  
\citet{fea02} show that the Cepheid distance scale also presents an important,  
intrinsic dependence on $Y$. Note that the usual approach, in Cepheid-based
studies of the extragalactic distance scale, is to take into account 
statistical uncertainties in a ``universal'' $\Delta Y - \Delta Z$ 
enrichment law, whereas our point is that such a relation need not be 
universal. \citeauthor{fea02} already point out that the galaxy M101 may 
require different enrichment laws in its inner and outer parts in order to 
account for observed differences in their dependence on metallicity.} 

In the same token, we believe it is a risky procedure to derive relative 
globular clusters ages, using the HB luminosity level as a standard candle, 
for globular clusters of different metallicities belonging to different 
populations, as has been done by \citet{oea95} in the case of 47~Tuc
and NGC~6528, since they may have undergone difference helium
enrichment laws and therefore have significantly different HB luminosities.  

Another important implication of this result is that the production of hot 
HB sources at high metallicities, which are believed to be the main sources 
of the ``UV upturn'' (or UVX) light coming from elliptical galaxies and the 
bulges of spirals \citep[see][ for a review and extensive references]{oc99}, 
is importantly affected by their precise helium enrichment laws 
\citep[e.g.,][]{syea97}. In the Galactic bulge, evidence of the presence of
hot HB stars has been provided and discussed by \citet{gbea96}, 
\citet{dtea99,dtea04}, \citet{rpea01}, \citet{gbea05}, and \citet{bm08}. 
It should also be noted that 
the precise RGB mass loss recipe adopted may also significantly impact 
the predicted production of UV sources (see Sect.~\ref{sec:DM}). 

Therefore, a calibration such as the one provided by eq.~(\ref{eq:RRL}) 
may not be straightforwardly applicable to just any galaxy, especially when 
more metal-rich components are considered. Also, from \citeauthor{r1879}'s 
(\citeyear{r1879})
relation, one finds that the pulsation properties of RR Lyrae stars are 
directly inherited from, or rather reflected upon, their luminosities. 
However, Figure~\ref{fig:OO} clearly shows that the RR Lyrae that belong 
to our galaxy 
present a rather peculiar behavior, at least compared to those belonging 
to its neighboring galaxies. But if this is indeed so, what guarantees do 
we have that $M_V{\rm (RRL)}-{\rm [Fe/H]}$ relations based upon Galactic 
halo RR Lyrae stars are universally applicable to other galaxies? Arguably, 
one should make sure that, if not the Oosterhoff dichotomy itself, at least 
the Oosterhoff-Arp-Preston-Sandage period-metallicity {\em progression} 
\citep[][; and references therein]{o39,o44,a55,p59,as81,s04} is verified 
in the sample under scrutiny \citep{c04b}---which is only possible by using 
time-series observations. 

Even for bona-fide Galactic RR Lyrae stars, one should recall that the 
$M_V{\rm (RRL)}-{\rm [Fe/H]}$ relation can be strongly affected by 
evolutionary effects \citep[e.g.,][]{l90,dea00}; these are more properly avoided 
when using the RR Lyrae {\em period-luminosity relation} to infer distances. 
In any case, one must keep in mind that there appear to be 
``pathological outliers'' for which the available calibrations of any of 
these relations may not be straightforwardly applied, 
due to the fact that they may not be properly described by canonical 
models \citep{sc98,pea01,pea02,rea02}. Also, if the helium 
abundance is enhanced in at least a fraction of the stars in some globular 
clusters, whether due to primordial or to evolutionary effects, models 
assuming $Y \simeq 0.23-0.25$ 
will not provide a correct distance to the RR Lyrae 
stars in the systems which partook a primordial enhancement, or in which 
some RGB stars may have undergone helium enrichment and ended up as 
overluminous RR Lyrae. For critical discussions of several different 
channels for the production of high-$Y$ stars in globular clusters, 
the reader is referred to the extensive list of references provided in 
\S\ref{sec:HSP}.

Even if present-day Galactic globular clusters 
have only their blue HB components affected by a high $Y$, {\em a few Gyr 
ago these blue HB components were actually (overluminous) RR Lyrae stars}. 
Therefore, even if not in our present-day galaxy, there is no guarantee
that in other (somewhat younger) galaxies and their respective globular 
cluster systems we might not run into overluminous RR Lyrae that are their 
analogues of the present-day, helium-enhanced, Galactic blue HB stars.   
The effect of a change in $Y$ upon the RR Lyrae 
period-luminosity relation has been recently 
discussed by \citet{cea04}. Again, over/underluminosity should be 
reflected upon the pulsation periods of RR Lyrae stars, so that proper 
monitoring of their pulsation cycles remains the only way to safely apply 
local (or standard theoretical) HB luminosity calibrations to external 
systems. For nonvariable HB stars, unfortunately, such a consistency 
diagnostic is not available. 

For the sake of argument, consider the \citet{yl02} scenario, whereby 
the Galactic OoII component is not a genuine component of our galaxy, having 
instead been accreted from an external galaxy at some point in the 
past.\footnote{In fact, \citet{vdb93} and \citet{lc99b}, among others, 
suggest precisely the opposite, namely, that it is the 
{\em OoI} component of the Galaxy that is of external origin, the OoII 
clusters having formed very early in the Galaxy's infancy.}
In this scenario, had such an accretion event not taken place, our Galactic 
globular cluster system would be characterized by the OoI component shown 
in Figure~\ref{fig:OO} ({\em left panel}) plus the OoIII component shown 
in the same panel. 
In this case, one would find for our Galactic globulars an {\em opposite} 
trend to what is seen in the Oosterhoff-Arp-Preston-Sandage progression, 
with the more metal-rich globulars presenting {\em much longer periods}---and 
therefore presumably having brighter HBs---than the OoI clusters. There is no 
guarantee that in external galaxies similar to our own but which have not 
(again in the \citeauthor{yl02} scenario)
incorporated an ``OoII protogalactic 
fragment'' in the course of its history, such a seemingly far-fetched 
behavior is not precisely what happened in practice.

\subsection{A Caveat: The Evolutionary Status of the Star RR Lyr}

One possible caveat with employing the star RR Lyr to constrain the 
average HB luminosity-metallicity relation at the instability strip level 
is the fact that we do not know 
{\em a priori} the evolutionary status of the star. Therefore, it could be 
significantly brighter (or fainter) than the average field RR Lyrae at the 
same metallicity, in which case our results could be {\em systematically} 
off by 0.1~mag or more, given the intrinsic spread in RR Lyrae luminosities
\citep[e.g.,][]{s90}. Indeed, \citet{cc08} have recently argued, based on 
Str\"omgren photometry and theoretical modelling, that RR Lyr may be 
brighter than other field RR Lyrae stars of similar metallicity by 
$0.064\pm 0.013$~mag in $V$, whereas \citet{mfea08}, comparing distances 
derived using type II Cepheids and RR Lyrae stars, similarly advocate 
that RR Lyr is overluminous by $0.16\pm 0.12$~mag in $V$. Therefore, 
it may not be entirely safe to assume, as frequently done, that RR Lyr is 
truly representative of other RR Lyrae-type stars of similar metallicity. 
A definitive solution to this problem will probably 
have to wait for new, accurate parallax determinations for large numbers 
of RR Lyrae stars, such as will be afforded by the GAIA and SIM missions.

\begin{figure}[t]
  \plotone{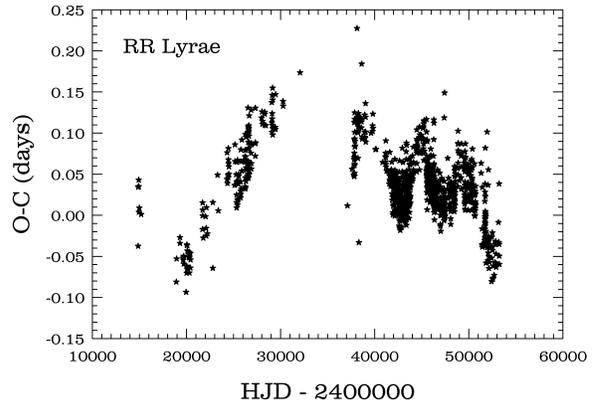}
  \caption{{\em Observed minus computed} (O-C) phase of maximum light (in days) 
    as a function of heliocentric Julian day for the variable star 
    RR Lyrae.\label{fig:OC}
}
\end{figure}

\begin{figure}[t]
  \plotone{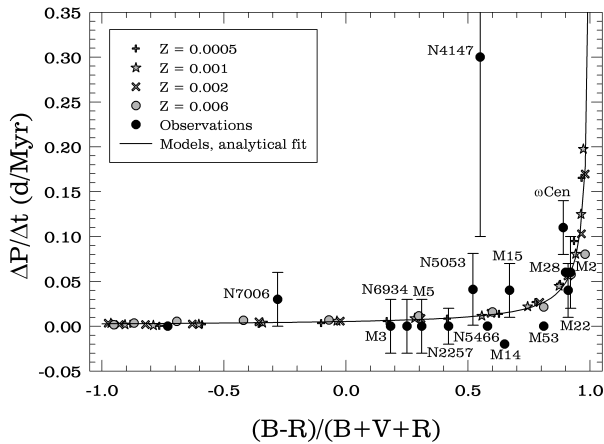}
  \caption{Average period change rate for RR Lyrae stars in Galactic 
    globular clusters. The plusses, stars, crosses, and filled gray symbols 
    represent new theoretical results for four different metallicity values
    (as shown in the inset); 
    these can be very well represented by the analytical formula provided in 
    eq.~(\ref{eq:PCR}) and indicated by a solid line in this plot.\label{fig:PCR}
}
\end{figure}

In the meantime, however, it could also be useful to look further into 
another diagnostic that is available to us at this point in time: the 
{\em period change rate} of RR Lyr. More specifically, a large, positive 
period change rate for the star could raise a yellow flag indicating the 
possibility of its being in a fast, advanced evolutionary stage, and 
therefore likely brighter than most other stars of similar metallicity. 
The presence of a near-zero or negative period change rate would be 
more in line with the star being found around the slowest part of the 
HB evolution. Of course, the caveat should be kept in mind that it is 
very often the case that RR Lyrae stars are seen to exhibit erratic period 
changes which are in some cases even orders of magnitude different from 
what is predicted by stellar evolution theory (see \citeauthor{hs95} 
\citeyear{hs95} for a review 
and discussions), so that the present argument should be taken with 
due caution.   

Automatically computed and updated values of {\em observed minus computed}
(O-C) times of maxima for RR Lyr can be found 
in the {\em GEOS RR Lyr Database} 
web page\footnote{\tiny\texttt{http://dbrr.ast.obs-mip.fr/}}
\citep*[see][]{lbkb08}. In 
Figure~\ref{fig:OC} we show one such curve, obtained with respect to a reference 
period of 0.5668378~d. As recently reviewed by \citet{s05}, a variable 
star with a period that is linearly changing with time will show a 
quadratic curve in the period-epoch diagram, with the value of the 
quadratic term given by 

\begin{equation}
  \beta^{\prime} = \frac{1}{2} \frac{dP}{dt} \langle P\rangle E^2, 
  \label{eq:DPDR} 
\end{equation}

\noindent where $\langle P\rangle$ is the average period value during the 
timespan of the observations, and $E$ is the epoch. A {\em formal} 
least-squares fit to the data shown in Figure~\ref{fig:OC} provides a period change 
rate of $\sim -0.1$~d/Myr---although it should be emphasized that there is no 
a priori reason why the period should change linearly with time, and 
therefore no strong reason to believe that the assumed quadratic law 
provides an adequate representation of the data for RR Lyrae stars 
\citep[e.g.,][ and references therein]{rs97}. As a matter of fact, 
according to this diagram and the similar one presented by \citet{sk00}, 
the rate of period change in RR Lyr has not at all been constant; one 
sees instead an apparent increase in the period at JD~2,418,000 or 
JD~2,420,000, only to be followed by a sudden decrease in the period 
that took place some time around JD~2,432,000, with a seemingly oscillatory 
behavior thereafter (interpreted by Szeidl \& Koll\'ath as being due to the 
accumulation of random fluctuations in the period). 
One way or another, there is 
certainly no strong evidence of a sharply increasing period, as might be 
expected in the case of a very bright, extremely evolved RR Lyrae star 
that is not strongly affected by random period variations. 
While this by no means represents conclusive evidence that RR Lyr is 
not well evolved and therefore overluminous, it is at least consistent 
with the hypothesis that the star presents a luminosity that is not 
dramatically different from that of most other RR Lyrae stars of 
similar metallicity.

\begin{table*}[t]
\caption{Period Change Rates for RR Lyrae Stars in Globular Clusters\label{tb:PCR}}
\smallskip
\begin{center}
{\small
\begin{tabular}{cccccc}
\tableline
\noalign{\smallskip}
Name   &   Other         & [Fe/H]  & $\langle \frac{\Delta P}{\Delta t}\rangle$  & $\LZ$ & Population \\
       &                 &         &          (d/Myr)                &       & \\
\noalign{\smallskip}
\tableline
\noalign{\smallskip}
NGC 2257 &               & $-1.63$ & $+0.00\pm 0.02$ & $+0.42$ & LMC \\
NGC 4147 &               & $-1.83$ & $+0.30\pm 0.20$ & $+0.55$ & Sag? \\
NGC 5024 & M53           & $-1.99$ & $+0.00$         & $+0.81$ & OH \\
NGC 5053 &               & $-2.29$ & $+0.04\pm 0.04$ & $+0.52$ & YH \\
NGC 5139 & $\omega$ Cen  & $-1.62$ & $+0.11\pm 0.03$ & $+0.89$ & $\omega$C \\
NGC 5272 & M3            & $-1.57$ & $+0.00\pm 0.03$ & $+0.18$ & YH \\
NGC 5466 &               & $-2.22$ & $+0.00$         & $+0.58$ & YH \\
NGC 5904 & M5            & $-1.27$ & $+0.00\pm 0.03$ & $+0.31$ & OH \\
NGC 6171 & M107          & $-1.04$ & $+0.00$         & $-0.73$ & OH \\
NGC 6402 & M14           & $-1.39$ & $-0.02$         & $+0.65$ & OH \\
NGC 6626 & M28           & $-1.45$ & $+0.06$         & $+0.90$ & OH \\
NGC 6656 & M22           & $-1.64$ & $+0.04\pm 0.03$ & $+0.91$ & OH \\
NGC 6934 &               & $-1.54$ & $+0.00\pm 0.03$ & $+0.25$ & YH \\
NGC 7006 &               & $-1.63$ & $+0.03\pm 0.03$ & $-0.28$ & YH \\
NGC 7078 & M15           & $-2.26$ & $+0.04\pm 0.03$ & $+0.67$ & YH \\
NGC 7089 & M2            & $-1.62$ & $+0.06\pm 0.04$ & $+0.92$ & OH \\
\tableline
\end{tabular}
}
\end{center}
\end{table*}

\section{Period Change Rates of RR Lyrae Stars: The Evolutionary Connection}

As we have just seen in the case of RR Lyrae (Fig.~\ref{fig:OC}), more often than 
not RR Lyrae variables show erratic rates of period change, presumably 
due to mixing events in the stellar core \citep{sr79} or to the presence 
of hydromagnetic effects related to the conjectured existence of a 
magnetic cycle similar to the Sun's \citep{s80}. Interestingly, 
transient magnetic fields in the H and He ionization zones have recently 
been suggested to be responsible for the Blazhko effect as well
\citep{rs06}. Note that the presence 
of a strong magnetic field in RR Lyr (as also required in other 
theoretical scenarios for the \citeauthor{b07} effect) has recently been 
ruled out by the high-precision spectropolarimetric observations (carried 
out over an almost 4-yr timespan) by \citet{mchea04}. 
\citet{rs06} speculates, again drawing an analogy with the Sun, that 
these measurements may simply indicate that a strong magnetic 
field may be deeply embedded in the star, becoming rather weak only 
at its surface (see \S3 of his paper for extensive references to earlier 
work on this subject). 
Relatively recent reviews of the Blazhko effect in RR Lyrae stars have 
been provided by \citet{s97,hs05} and \citet{kk02}. 

This notwithstanding, given sufficient time coverage and large enough 
samples of stars, one may be able to detect, superimposed on the random 
period changes that characterize many RR Lyrae variables, the slow period 
changes that are due to the star's evolution accross the CMD. That the 
period of the star should change with time is a very basic prediction of 
stellar evolution and pulsation theory: as a star slowly changes its 
luminosity and temperature, its mean density algo changes; 
according to \citeauthor{r1879}'s \citeyearpar{r1879} 
period-mean density relation of pulsation 
theory, $P \sqrt{\langle \rho\rangle} \approx {\rm const.}$, the period 
should accordingly change in inverse proportion with the density, implying 
that a star with decreasing density---which may be due either to a 
decreasing temperature and/or an increasing luminosity, assuming the stellar 
mass is constant---should have an increasing period. Since stars 
evolving to the red and towards higher luminosities on the HR diagram 
are thought to be present mainly in globular clusters with blue HB types, 
it follows that positive period change rates that can be ascribed to 
evolutionary effects are expected mostly on blue HB clusters. Clusters 
with predominantly red or intermediate HBs, on the other hand, should 
show, on average, period change rates around zero, or even negative. 

Some models which clearly show this predicted trend of increasing period 
change rate for bluer HB types have been presented by \citet{l91}.  
Recently, \citet{cea04} computed a new, extensive grid of synthetic HB models, 
which are also suitable for a reassessment of these canonical theoretical 
predictions. Based on those simulations, which we augmented with an 
additional set of simulations for both extremely blue and extremely red 
HBs, period change rates were computed for the metallicity values 
$Z = 0.0005$, $0.001$, $0.002$, and $0.006$, as shown in Figure~\ref{fig:PCR}. 
Note that these theoretical results can be extremely well described by an 
exceedingly simple analytical formula (solid line in Fig.~\ref{fig:PCR}), 
namely 

\begin{equation}
   \left\langle \frac{\Delta P}{\Delta t}\right\rangle = \frac{0.0053}{1 + 0.99 \, 
        \LZ} \,\,\, {\rm (d/Myr)},
   \label{eq:PCR}
\end{equation}

\noindent where $\LZ = \BVR$.

\begin{figure*}[ht]
  \includegraphics[scale=0.545]{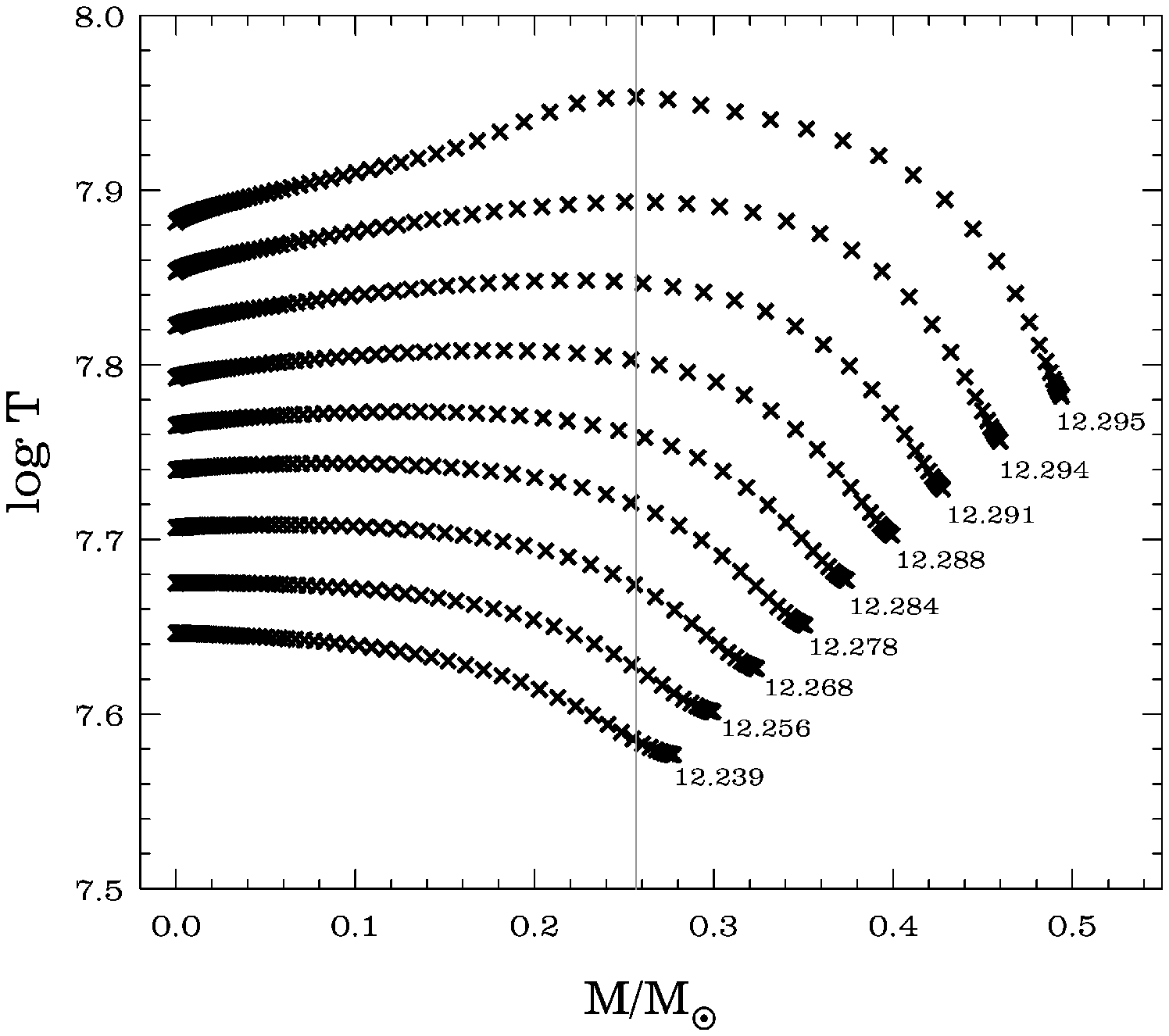} 
  \includegraphics[scale=0.545]{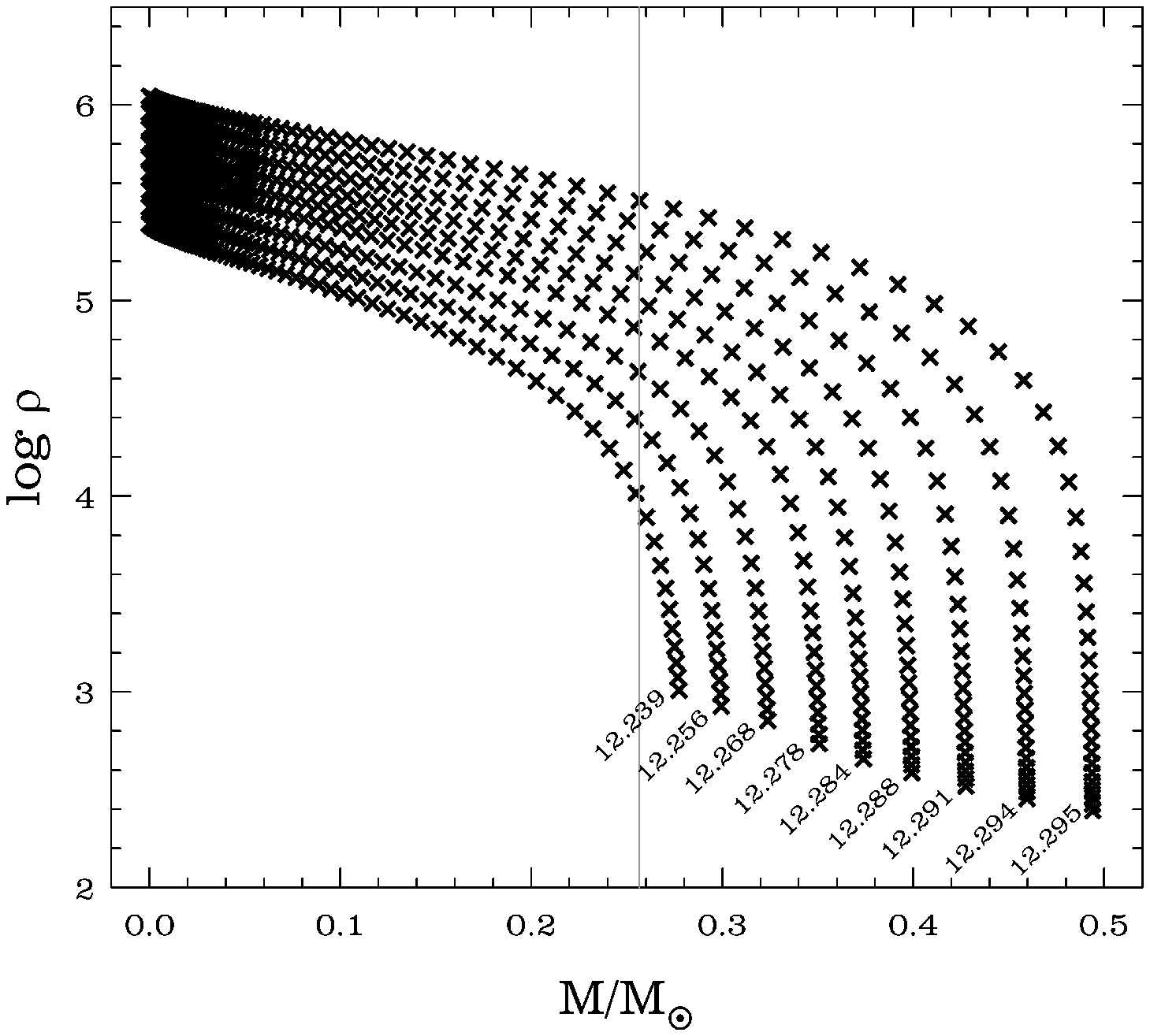} 
  \caption{Run of temperature ({\em left panel}) and density ({\em right panel}) as 
    a function of the Lagrangian coordinate mass in the core of selected RGB models. 
    Labels next to each temperature or density profile indicate the time (in Gyr) 
    since the star arrived at the zero-age main sequence. The top model profile in 
    each panel shows the core profile at the onset of the He-flash. The vertical 
    line accordingly shows the mass coordinate where He ignition first takes 
    place.\label{fig:TRHOM}
} 
\end{figure*}

The empirical data are also shown in Figure~\ref{fig:PCR}, overplotted on 
the theoretical results. Using the recent \citet{n04} compilation as a 
starting point, we have added to the sample several additional globular 
clusters, including M2 \citep{lc99} and 
NGC~4147 \citep{sea05}, as well as earlier data for several globulars 
as summarized in Table~5.1 of \citet{hs95}---the latter corresponding to 
the datapoints without error bars. These data are listed in 
Table~\ref{tb:PCR}. As can be seen, there 
is good agreement between these evolutionary models and the observations, 
within the empirical error bars. In the case of NGC~4147, the timespan of 
the available observations has presumably been insufficient to reliably 
detect the evolutionary changes in the pulsation periods.

\section{Conductive Opacities and the He-Core Mass at the Helium Flash}\label{sec:MC} 
In an earlier review, \citet*{cea96} 
argued that uncertainties in the {\em conductive opacities}  
remained that might still lead to significant changes in the computed 
properties of HB stars, especially their luminosities. Since over 10 years 
have passed since the \citeauthor{cea96} study, a new look at the problem 
seems especially worthwhile---and this is the purpose 
of the present section. 

To be sure, 
our approach does {\em not} mean that there are not other physical processes 
which are uncertain enough as to potentially lead to important changes in our
understanding of HB evolution; quite the contrary, in fact. For instance, the 
treatments of diffusion,  
convection, and mixing, as well as such a key nuclear reaction rate as 
$^{12}{\rm C} (\alpha, \gamma) ^{16}{\rm O}$, remain subject to considerable
uncertainty, and future developments may accordingly affect HB evolutionary 
predictions in quite significant ways. 
Unfortunately, it would not be practical for us here to provide an 
extensive review of all of the many different physical processes 
that play an important role in defining the predicted properties of 
HB stars. The interested reader is referred to the recent 
reviews by \citet{msea02}, \citet{sc05}, and \citet{mc07} for critical 
discussions of several of the more salient uncertainties affecting the 
computation of HB models. 

\subsection{Overview} 
One of the main ingredients affecting the 
luminosity level of HB stars are the {\em conductive opacities} in the 
cores of their progenitor RGB stars. The cores of RGB stars are  
electron-degenerate, so that electron conduction 
is the main form of energy transport. Naturally, the more efficient 
the transport of energy away from the He-core, the more difficult it becomes 
for the core to reach high enough temperatures for the onset of helium 
burning, with the end result that the He-flash is delayed for higher 
electron conductivities (i.e., smaller conductive opacities), the star 
reaching a higher He-core mass and a brighter location on the RGB 
tip---thus also being expected to lose more mass during its first ascent 
of the RGB and thereby settle on a bluer post-flash location on the 
ZAHB.

\begin{figure}[t]
  \plotone{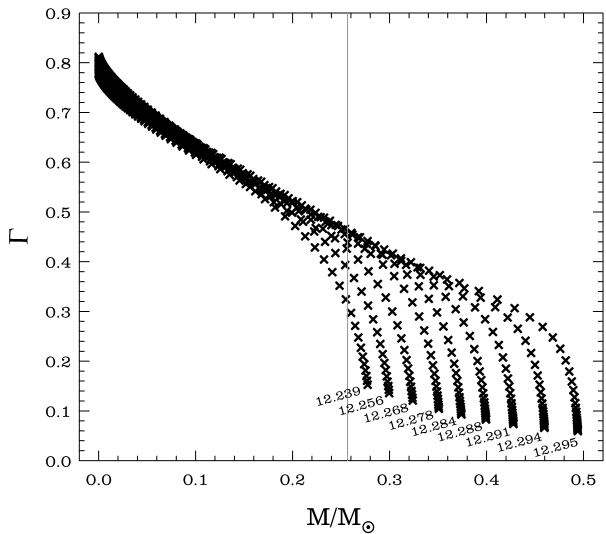}
  \caption{As in Figure~\ref{fig:TRHOM}, but for the 
    run of the Coulomb $\Gamma$ parameter as a function of mass in the RGB 
    core. The vertical line shows the mass coordinate where He ignition first 
    takes place.\label{fig:GAMMA}
} 
\end{figure}

The He-core mass $M_{\rm c}$ is also known to have a marked, {\em direct} 
impact upon the HB structures. The shapes of HB evolutionary tracks are 
strongly affected by the precise value of $M_{\rm c}$, with a larger $M_{\rm c}$
leading to less marked {\em blueward loops} on the HB phase \citep{sg76}
and thereby affecting the predicted fraction of well-evolved RR Lyrae stars 
in globular clusters with predominanty blue HBs \citep{c92}. Most importantly, 
higher $M_{\rm c}$ values lead to brighter HBs as 

\begin{equation}
  \frac{d M_{\rm bol}^{\rm HB}}{d M_{\rm c}} \approx 7.3-9.0\,\, ({\rm mag}/M_{\odot})
  \label{eq:MC}  
\end{equation}

\noindent (\citeauthor{sea87} \citeyear{sea87}; \citeauthor{c96} \citeyear{c96};
\citeauthor{cea96} \citeyear{cea96}); 
therefore, an uncertainty in the value of 
$M_{\rm c}$ by only $\pm 0.01 \, M_{\odot}$ (or about $\pm 2\%$) is capable 
of affecting the HB luminosity level by $\pm 0.07-0.09$~mag. The 
numerical experiments by \citet{sg78} indicate that a change in $M_{\rm c}$ 
by about $0.01 \, M_{\odot}$ obtains when one reduces the conductive opacities 
used in their calculations by a (uniform) factor of 2. 

The implied level of uncertainty in the HB models 
is important not only for the sake of comparing the predicted 
and empirically calibrated HB luminosity-metallicity relationships, but also 
(and perhaps most importantly) in the context of the self-consistent 
determination of globular cluster ages from evolutionary models of globular 
cluster stars; of the establishment of the RGB tip distance scale; of the 
determination of the helium abundance in globular clusters using 
Iben's (1968) $R$-method; and of the placement of astrophysical 
constraints on fundamental (especially particle) physics parameters. As 
to the latter, constraints on the neutrino magnetic moment using bright 
RGB and HB stars were discussed by \citet{rw92} and \citet{cdi93}; 
those on the axion mass, 
by \citet{rd87}; and those on the number of extra dimensions in the Universe 
by \citet{cea00}. Reviews of the subject and extensive additional references
have been provided by \citet{r96,r00} and \citet{cea96}. 

Until the mid-90's, the two main sources of conductive opacities 
were the calculations 
by \citet{hl69} and \citet{iea83}. To date, most authors still use one of 
these two, as can be seen from the recent 
summary of physical ingredients in different evolutionary models compiled
by \citet{gea05}. Yet, as pointed out by \citet{cea96}, the theoretical 
calculations of \citeauthor{iea83} in particular are not applicable to 
the physical conditions characterizing the interiors of RGB stars, since 
they are restricted to the regime of relatively strong Coulomb intreactions
(Coulomb parameter 
$\Gamma > 2$), while in the RGB core $\Gamma < 0.81$---an intermediate 
regime between the liquid and crystal phases 
considered by \citeauthor{iea83} and an
ideal gas. Since then, however, new conductive opacity calculations have 
been presented by \citet{p99} and \citet{pea99}. 

Figure~\ref{fig:TRHOM} shows the run of temperature 
({\em left panel}) and density ({\em right panel}) 
in the interior of a metal-poor, low-mass star as it climbs up the 
RGB. These models, which refer to a $0.8\, M_{\odot}$ star with 
${\rm [Fe/H]} = -2.3$, $[\alpha/{\rm Fe}] = 0.3$, and $Y = 0.2484$, 
were kindly provided by D. A. VandenBerg (2005, priv. comm.). 
Nine ``snapshots'' are taken over an interval covering the final 
$56 \times 10^6$~yr of the star's evolution in the RGB phase; as can be 
seen from this plot, during this time interval the mass of the He core 
increases from about $0.28 \, M_{\odot}$ all the way up to its final 
value, of order $0.5 \, M_{\odot}$. Note that the He-flash actually 
takes place at a mass value just below $0.26 \, M_{\odot}$, due to the 
fact that the density-dependent neutrino energy losses lead to an  
efficient cooling of the innermost regions and therefore to a 
temperature inversion in the RGB core, as first shown by \citet{t67}. 
For the sake of simplicity, in what follows we shall assume a pure helium 
plasma.

Figure~\ref{fig:GAMMA} shows the variations in the Coulomb $\Gamma$ parameter [see, e.g., 
eq.~(21) in Catelan et al. 1996], which measures the strength of the 
electrostatic interactions in the plasma, throughout the He core as a 
function of the star's evolutionary status. Clearly, this diagram fully 
supports the assertion that $\Gamma \la 0.81$ in the RGB interior. 
This accordingly 
confirms the caveat that one should avoid using the \citet{iea83} 
results in evolutionary calculations of low-mass stars.

Since the more recent \citet{p99} and \citet{pea99} calculations 
have superseded the \citet{iea83} 
results, the preceding argument leads naturally to the following
question: are the \citeauthor{p99} results themselves fully applicable 
to the conditions characterizing RGB interiors? 

In fact, not quite: \citet{pea99} explain that a fit to the static structure 
factor that is valid over the range $1 \leq \Gamma \leq 225$ was employed 
in their calculations. Therefore, the possibility that a problem similar 
to the one affecting the \citet{iea83} results is present also in the 
case of \citeauthor{p99}'s calculations cannot be excluded; a fully 
self-consistent calculation for $0.1 \la \Gamma \la 0.8$, which is the 
relevant range for the typical RGB interior (Fig.~\ref{fig:GAMMA}), would prove of 
great interest in the context of precision models of RGB stars and their 
HB progenitors. \citeauthor{pea99} do note that Coulomb logarithms in the 
transition domain from weak ($\Gamma \ll 1$) to strong ($\Gamma \ge 1$) 
ion coupling could be calculated using the \citet*{bea82} formalism, but 
unfortunately such a formalism was not applied in their calculations. 
In fact, it appears that the reason why such a range of $\Gamma$ values 
has not been given great emphasis in such calculations as those by 
\citeauthor{iea83} or \citeauthor{pea99} is that such studies are 
generally carried out primarily with applications to the more extreme 
conditions characterizing neutron stars and white dwarf cores in mind, 
the interiors of RGB stars having received comparatively little attention.

\begin{figure}[t]
  \includegraphics[scale=0.52]{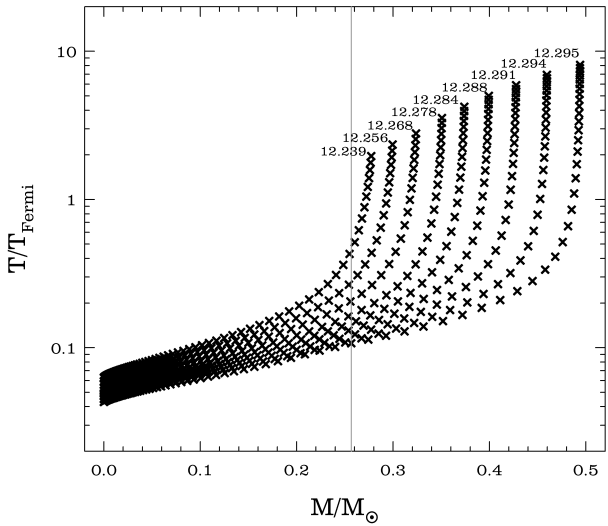} 
  \caption{As in Figure~\ref{fig:TRHOM}, but for the 
    ratio between actual temperature and the Fermi temperature in the RGB 
    core.
	\label{fig:FERMI}
}
\end{figure}

This, in fact, also brings to mind the question whether other approximations 
to physical situations that are rather more extreme in neutron stars and 
white dwarf cores may not have been adopted in these latest calculations that 
could render uncertain their application to the conditions characterizing 
RGB cores. In this sense, the most obvious ingredient is the 
{\em electron degeneracy level}. 
Figure~\ref{fig:FERMI} shows the run of the 
ratio between the actual temperature 
and the Fermi temperature for the same models shown in the previous plots. 
As well known, for $T \ll T_{\rm Fermi}$, strong degeneracy is present, 
whereas $T \gg T_{\rm Fermi}$ implies a classical regime where degeneracy
effects are unimportant. This plot indicates that 
RGB interiors fall in an intermediate regime between strongly degenerate
and classical. In other words, {\em partial 
degeneracy} better describes the status of the RGB interior. Note, in 
particular, that at the mass coordinate where the He-flash takes place, 
the temperature ranges from about 10\% to 45\% the Fermi temperature. 
Clearly, the effects of partial degeneracy should be fully taken into 
account for conductive opacities that are applicable to the conditions 
characterizing RGB cores to obtain. How does this compare with the 
\citet{p99} and \citet{pea99} analyses? 

As clearly stated by \citet{pea99}, their analysis ``is limited by the 
condition $T \ll T_{\rm Fermi}$''---in other words, strong degeneracy is 
assumed in their calculations. Unfortunately, this cannot be expected 
to provide particularly reliable results for the RGB 
interior.\footnote{On the other hand, Potekhin (2005, priv. comm.) 
points out that the technique of ``Fermi-Dirac averaging'' has been 
adopted in the region of partial degeneracy, which at least improves 
somewhat the situation compared to what would obtain from straight 
application of the strongly degenerate results to partially degenerate 
conditions \citep[see][ for more details]{p99}.}

\begin{figure}[t]
  \plotone{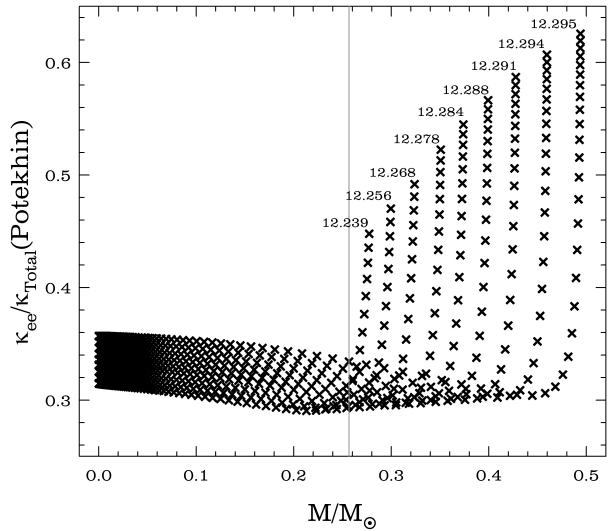}
  \caption{As in Figure~\ref{fig:TRHOM}, but for the 
    ratio between the 
    component of the conductive opacity due to electron-electron 
    interactions and the total conductive opacity, according to the calculations by 
    \citet{p99} and \citet{pea99}, in the aforementioned RGB core models. The 
    electron-electron contribution is clearly a major component of the conductive
    opacity for the conditions characterizing the cores of 
    RGB stars.\label{fig:EE}
} 
\end{figure}

\begin{figure}[ht]
  \plotone{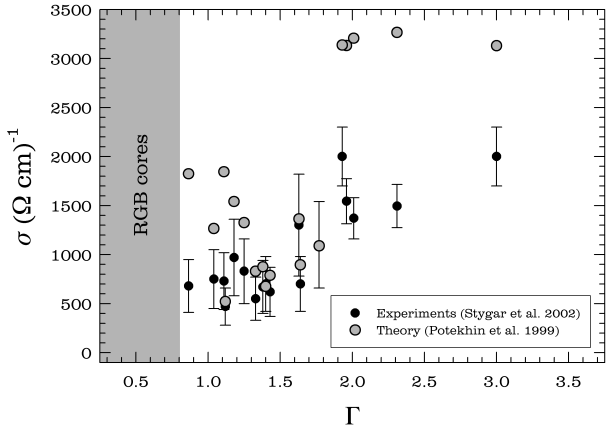}
  \caption{Electric conductivities: comparison between laboratory experiments 
    ({\em black symbols with error bars}, from Stygar et al. 2002) and the 
    expected theoretical values according to \citet{p99} and 
    \citet{pea99} ({\em gray symbols}).\label{fig:LAB}
} 
\end{figure}

\begin{figure}[ht]
  \plotone{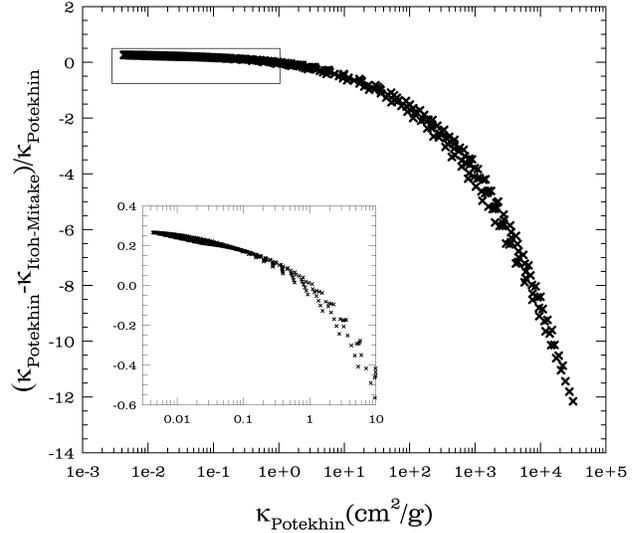}
  \caption{Relative difference between the 
    conductive opacity computed by \citet{iea83}, 
    with quantum corrections by \citet{mea84}, 
    and by \citet{p99} and \citet{pea99} 
    in the aforementioned RGB core models.
    The inset shows a blowup of the region 
    indicated by a rectangle in the upper left 
    of the plot.\label{fig:IP}
} 
\end{figure}

\begin{figure}[ht]
  \plotone{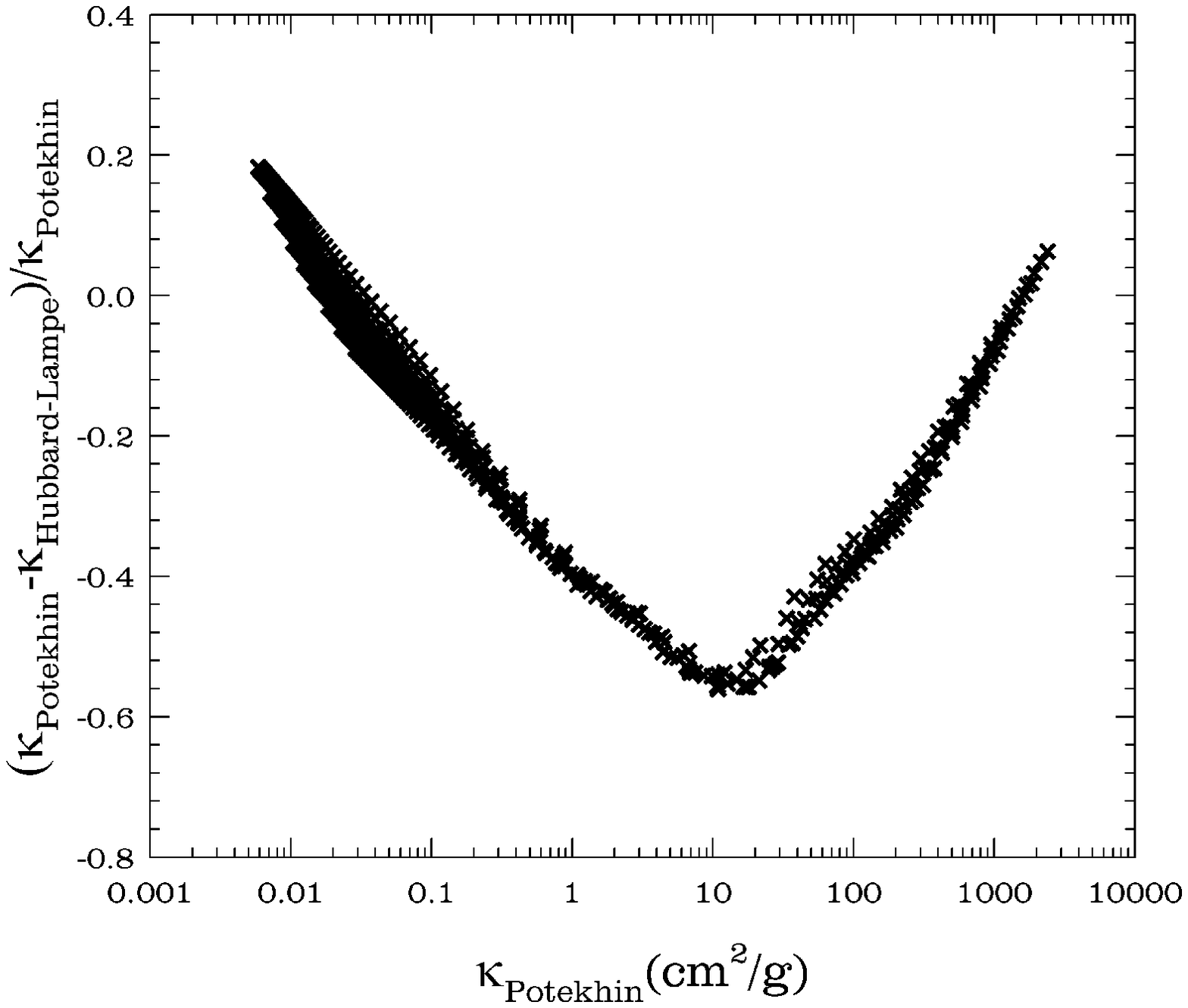}
  \caption{Relative difference between the 
    conductive opacity computed by \citet{hl69} 
    and by \citet{p99} and \citet{pea99} in the aforementioned RGB core 
    models.\label{fig:HLP}
} 
\end{figure}

The above discussion refers to the contribution of electron-ion ({\em ei}) 
interactions to the conductive opacity. An apparently more subtle effect 
is related to the {\em electron-electron (ee) interactions}. Since these are 
usually of little importance in the more extreme conditions characterizing 
neutron stars and white dwarfs, the \citet{p99} and \citet{pea99} calculations 
have not treated this component in as much detail as in the case of 
the {\em ei} interactions. The {\em ee} component was indeed computed assuming 
complete degeneracy, with no corrections whatsoever 
due to partial degeneracy 
effects---and are accordingly expected to be accurate only to within 
a factor of $\sim 2$ (Potekhin 2005, priv. comm.). However, it must 
be noted that {\em ee} effects are a {\em major} component of the 
conductive opacity in a He plasma 
in the conditions characterizing the RGB interior \citep{hl69,cea96}. 
This is shown in Figure~\ref{fig:EE}, where the ratio between the 
component of the conductive opacity due to {\em ee} interactions 
and the total conductive opacity, according to the \citeauthor{p99} 
calculations, is displayed. One sees that the {\em ee} contribution reaches 
a {\em minimum} of about 28\% of the total conductive opacity close to 
the mass coordinate where the He-flash takes place. The {\em ee} contribution 
increases further as the He-flash is approached. Evidently, an error by 
a factor of two in the {\em ee} contribution to the conductive opacity implies 
that the total opacity is uncertain by a comparable factor. Accordingly, 
greater attention should be devoted to this crucial physical ingredient 
in future calculations of conductive opacities for RGB stars.

\subsection{Laboratory Experiments}
It is obviously of great interest to check how the \citet{p99} and \citet{pea99} 
opacities compare with experiments conducted in the laboratory. Such a 
comparison, in the case of electrical conduction, has recently been carried out 
by \citet*{sea02}, using laboratory data from \citet*{mea80}, 
\citet*{bea99}, and \citet{b00}. 
Their results are shown in Figure~\ref{fig:LAB}, where the electrical 
conductivity is plotted as a function of the Coulomb $\Gamma$ parameter. 
These comparisons clearly reveal a very complex pattern, with the models 
systematically overestimating the conductivity (i.e., underestimating 
the conductive opacity) for $\Gamma \ga 1.9$; presenting good agreement 
with the empirical results for $1.3 \la \Gamma \la 1.7$; and again 
overestimating the conductivity for $0.85 \la \Gamma \la 1.25$. It is 
unclear at present what the reasons for this behavior are, though Potekhin 
(2005, priv. comm.) points out that the laboratory plasmas, when 
heavy elements are present (the above experiments were carried out using 
xenon and aluminum), are almost always partially ionized, and it is 
very difficult to determine, and therefore account for, the actual degree of 
ionization, because of the plasma nonideality. Accordingly, more empirical 
tests of the \citeauthor{p99} results using low-$Z$ gases, especially He, 
would be highly desirable.

\subsection{Comparison between Different Prescriptions}
It is instructive to compare the \citet{p99} and 
\citet{pea99} 
results with those from the \citet{iea83} and \citet{hl69} analyses. This 
is done in Figures~\ref{fig:IP} and \ref{fig:HLP}, respectively, which show 
the relative 
differences between the conductive opacities computed by \citeauthor{iea83}
and by \citeauthor{hl69} with respect to the \citeauthor{p99} results for 
the physical conditions characterizing the RGB interior calculations. In 
the \citeauthor{iea83} case, we have included the quantum corrections by 
\citet*{mea84}, which however are rather small over the region of interest.  
The agreement between the Itoh-Mitake prescriptions 
and \citeauthor{p99}'s is 
quite poor, with the former systematically overestimating the conductive 
opacity except very close to the core, where the conductive opacity is 
instead underestimated by about 25\%.\footnote{According to Potekhin (2005, 
priv. comm.), this discrepancy can largely be ascribed to the fact that 
the \citet{iea83} calculations are limited not only to strong coupling, 
but also to strong electron degeneracy---compare with the discussion in 
the previous footnote.} On the other hand, the 
\citeauthor{hl69} calculations reveal a better agreement with the 
\citeauthor{p99} results, the maximum overestimate of the conductive 
opacity by \citeauthor{hl69} compared to \citeauthor{p99} not exceeding 
about 55\%. In the central core, again the older calculations underestimate 
the conductive opacity compared to the more recent results, albeit by 
a maximum of less than 20\%. We conclude that 
the \citeauthor{p99} results are in much better agreement with the canonical 
prescriptions by \citeauthor{hl69} than they are with the results by 
\citeauthor{iea83}. 

While updated conductive opacities that are entirely 
suitable for the conditions characterizing RGB interiors are not available, 
it is clear that one should avoid using the \citet{iea83} results in 
calculations of low-mass stellar evolution.

\subsection{Latest Developments} 
After the first version of this paper had been completed, new conductive
opacities were published by \citet{scea07}---in essence, a revised and 
updated installment of the \citet{p99} opacities. These calculations, 
while still not perfect, 
already do take several of the shortcomings noted in the previous sections 
into due account, and should accordingly be strongly preferred in 
state-of-the-art evolutionary computations. 

\S2.4, and especially Figures~3 and 4, in 
\citeauthor{scea07} provide the results of a comparison between the new 
conductive opacities computed by \citeauthor{scea07} and the earlier studies 
by \citet{hl69}, \citet{iea83}, and \citet{p99}. Important differences
are found with respect to the results from all previous studies, especially 
those from the \citeauthor{iea83} team. According 
to \citeauthor{scea07}, the differences are mainly due 
to the following: 
i)~Deficiencies in the earlier treatment of strongly coupled and
relativistic plasmas (\citeauthor{hl69}); ii)~An inadequate extension 
towards the $T > T_{\rm F}$ regime (\citeauthor{iea83}); 
iii)~A neglect of {\em ee} scattering (\citeauthor{p99}). 

When 
the new conductive opacities are used in evolutionary computations, 
\citeauthor{scea07} find that the changes, with respect to the results 
based on the earlier \citet{p99} prescriptions, to be relatively modest. 
More specifically, they find: i)~A reduction in the core mass at the He-flash 
by about $0.006 \, M_{\odot}$, irrespective of metallicity; ii)~An increase 
in $M_V({\rm ZAHB})$ at the RR Lyrae level, by an amount ranging from 
0.06~mag at $Z = 0.0001$ down to 0.04~mag at solar metallicity; 
iii)~An increase in the HB lifetime, by about 5-6\%. While relatively small, 
these systematic effects must obviously be properly taken into account in 
future precision studies involving HB stars.

\section{Epilogue} 

The study of horizontal branch stars encompasses and/or has direct implications
for many different branches of astrophysics. Without first looking into 
several of these areas, one might rush into the conclusion that the study of 
HB stars is a frustratingly limited affair. Nothing could be farther from the 
truth. As we have seen, in order to properly understand HB stars and 
appreciate their 
astrophysical implications, one must dwell on the physical 
processes (both canonical and non-canonical) that control the helium core 
flash in RGB stars; the physical processes that lead to mass loss in RGB
stars; whether (and how) deep mixing takes place in RGB stars, both before 
and during the helium flash; how angular momentum evolves with time in 
low-mass stars; where, and how fast, radiative levitation and diffusion 
effects in moderately high-gravity stars ``kick in''; how dramatically 
non-solar abundance ratios (again as a result of radiative levitation and 
gravitational settling) may affect model atmospheres and the photon output 
as a function of wavelength; how both radial and non-radial (both p and g) 
pulsation modes can be excited, and then evolve with time; how theory and 
observations of radial pulsators may place constraints on the physical 
parameters of stars, and thereby help determine the formation history of 
galaxies and the extragalactic distance scale; how asteroseismology of 
non-radial pulsators may help reveal the interior structure and thereby 
constrain the evolutionary history of stars; how the ultraviolet flux from 
giant elliptical galaxies and bulges of spirals come to being; how the stellar 
populations of resolved and unresolved galaxies evolve with time; and so on 
and so forth. In summary, the study of HB stars is a challenging and far-reaching  
intellectual adventure, and we hope that the present review will have helped 
unveil some of the reasons why this is so.

\acknowledgements   
The author warmly thanks A. Bonacic, G. Clementini, C. Moni Bidin, B. J. 
Pritzl, A. Reisenegger, and M. Zoccali, and most especially S. Cassisi, 
A. Y. Potekhin, H. A. Smith, A. V. Sweigart, and D. A. VandenBerg, for 
very helpful discussions and information. The staff and personnel of the 
Bologna Observatory and University, where part of this text was written, is 
warmly thanked for their hospitality. 
Support for this work was provided by Proyectos FONDECYT Regulares No.~1030954
and 1071002, by Proyecto Basal PFB-06/2007, by the FONDAP Center for Astrophysics
15010003, and by a John Simon Guggenheim Memorial Foundation Fellowship. 
This research has made use of the {\em GEOS RR Lyr Database}.


\end{document}